\documentclass[reprint, nofootinbib, amsmath,amssymb, aps]{revtex4-1}

\usepackage[colorlinks=true,citecolor=blue,linkcolor=blue,urlcolor=blue, backref=false,pdfborder={0 0 0}]{hyperref}

\usepackage{aas_macros} 
\usepackage{soul}

\newcommand{\rvline}{\hspace*{-\arraycolsep}\vline\hspace*{-\arraycolsep}}
\newcommand{\bigzero}{\mbox{\normalfont\Large\bfseries 0}}
\newcommand{\bigone}{\mbox{\normalfont\Large\bfseries 1}}

\newcommand{\threej}[6]{{\begin{pmatrix}
#1 & #2 & #3 \\
#4 & #5 & #6
\end{pmatrix}}}

\newcommand{\Lagr}{\mathcal{L}}

\newcommand{\deriv}[2]{\frac{\textrm{d} #1}{\textrm{d} #2}}

\newcommand{\beq}{\begin{equation}}
\newcommand{\eeq}{\end{equation}}
\newcommand{\barr}{\begin{eqnarray}}
\newcommand{\earr}{\end{eqnarray}}
\newcommand{\ball}{\begin{align}}
\newcommand{\eall}{\end{align}}
\newcommand{\bs}{\boldsymbol}

\newcommand{\hn}{\hat{n}}
\newcommand{\hk}{\hat{k}}
\newcommand{\bx}{\bs{x}}
\newcommand{\fbh}{f_{\rm pbh}}
\newcommand{\lsim}{\mathrel{\hbox{\rlap{\lower.55ex\hbox{$\sim$}} \kern-.3em \raise.4ex \hbox{$<$}}}}
\newcommand{\gsim}{\mathrel{\hbox{\rlap{\lower.55ex\hbox{$\sim$}} \kern-.3em \raise.4ex \hbox{$>$}}}}
\usepackage{cancel}
\newcommand{\sld}{\cancel{\delta}}
\newcommand{\bk}{\boldsymbol{k}}

\usepackage{enumitem}
\usepackage{cancel}
\usepackage{graphicx}
\usepackage{dcolumn}
\usepackage{bm}

\usepackage{braket}
\usepackage{placeins}
\usepackage{color,soul}
\usepackage[toc,page]{appendix}

\begin{document}
\title{CMB temperature trispectrum from accreting primordial black holes}
\author{Trey W. Jensen}

\author{Yacine Ali-Ha{\"i}moud}%
\affiliation{Center for Cosmology and Particle Physics \\
Department of Physics, New York University \\
New York, NY 10003, USA}%

\date{\today}

\begin{abstract}
It is known that Primordial Black Holes (PBHs) can leave an imprint on Cosmic Microwave Background (CMB) anisotropy power spectra, due to their accretion-powered injection of energy into the recombining plasma. Here we study a qualitatively new CMB observable sourced by accreting PBHs: the temperature trispectrum or connected 4-point function. This non-Gaussian signature is due to the strong spatial modulation of the PBH accretion luminosity, thus ionization perturbations, by large-scale supersonic relative velocities between PBHs and the accreted baryons. We first derive a factorizable quadratic transfer function for free-electron fraction inhomogeneities induced by accreting PBHs. We then compute the perturbation to the CMB temperature anisotropy due to a general modification of recombination, and apply our results to accreting PBHs. We calculate a new contribution to the temperature power spectrum due to the spatial fluctuations of the ionization perturbation induced by accreting PBHs, going beyond past studies which only accounted for its homogeneous part. While these contributions are formally comparable, we find the new part to be subdominant, due to the poor correlation of the perturbed temperature field with the standard CMB anisotropy. For the first time, we compute the temperature trispectrum due to accreting PBHs. This trispectrum is weakly correlated with the local-type primordial non-Gaussianity trispectrum, hence constraints on the latter do not lead to competitive bounds on accreting PBHs. We also forecast Planck's sensitivity to the temperature trispectrum sourced by accreting PBHs. Excitingly, we find it to be more sensitive to PBHs under $\sim 10^3 M_{\odot}$ than current temperature-only power spectrum constraints. This result motivates our future work extending this study to temperature and polarization trispectra induced by inhomogeneously-accreting PBHs.
\end{abstract}

\maketitle

\section{Introduction}\label{sec:intro}

Not only are primordial black holes (PBHs) a probe of very early-Universe physics, but they could also be the culprit behind several cosmological and astrophysical mysteries. For instance, even if they constitute only a small fraction of cold dark matter (CDM), intermediate-mass PBHs (1-$10^4$ $M_{\odot}$) could be the seed for supermassive black holes \citep{bean02a} or account for recent LIGO/Virgo gravitational wave observations \citep{bird16a}. Thus, even if the abundance of PBHs in this mass range is heavily constrained \citep{carr21a}, it proves invaluable to inspect further. 

The intermediate-mass range is where PBH accretion may leave a non-negligible signature on the cosmic microwave background (CMB). The underlying physical phenomena that lead to an indirect signal are the following. PBHs accrete primordial plasma throughout cosmic time; some fraction of the in-falling material is converted into radiation; this radiation propagates and deposits energy into the background recombining plasma, heating and ionizing it; finally, this change to the ionization history perturbs the last-scattering surface, ultimately altering the observed CMB temperature and polarization anisotropy. In fact, the strongest constraints on the abundance of PBHs in this mass range come from this effect \cite{carr21a}; however, previous literature have only looked for a signal in 2-point CMB anisotropy statistics \citep{ricotti08a, yacine17a,poulin17a}.

One avenue that has not been inspected is the non-Gaussianity that is induced in the CMB by accreting PBHs. Although the PBH accretion rate and radiation power are largely uncertain, they necessarily depend on the magnitude of the local relative velocity between the accreted matter (baryons) and PBHs, which behave as CDM on large scales \citep{ricotti08a, yacine17a,poulin17a}. This dependence implies a spatial modulation of the luminosity of accreting PBHs and thus inhomogeneities in their perturbation to recombination. It is known that inhomogeneous recombination generates non-Gaussian signatures in CMB anisotropies \citep{senatore09a,khatri09a, dvorkin13a}. The goal of this paper is to quantify this \emph{qualitatively different} CMB signature of accreting PBHs, for the first time. 

The effect considered here is similar in spirit to that studied in Ref.~\cite{dvorkin13a} in the context of dark matter (DM) annihilation, with, however, two major differences. First, since the PBH luminosity depends on the relative velocity squared, the lowest-order non-Gaussian statistics induced by accreting PBHs is the \emph{trispectrum}, or connected 4-point function. This is to be contrasted with the bispectrum (3-point function) sourced by energy injection from inhomogeneous annihilating DM \cite{dvorkin13a}. Second, in the case of annihilating DM, the inhomogeneity in energy injection is of order the DM density fluctuation around recombination, that is of order $\sim 10^{-3}$ on scales $k \sim 0.1$ Mpc$^{-1}$. In contrast, the PBH luminosity has \emph{order-unity} fluctuations on the same scales \cite{jensen21a}, as it is strongly modulated by supersonic relative velocities \cite{Tseliakhovich_10}. This implies that the inhomogeneities in the free-electron fraction sourced by accreting PBHs are comparable to its mean enhancement, as we demonstrated explicitly in Ref.~\cite{jensen21a}, hereafter Paper I. We therefore expect the amplitude of the non-Gaussian signature of accreting PBHs to be $\sim 10^3$ times larger than that of inhomogeneously annihilating DM, at equal amplitudes of the 2-point function perturbation.

In Paper I, we found that, for a PBH abundance saturating CMB power-spectra limits, the free-electron perturbation is of order $\delta_e \sim 10^{-3}$ around $z \sim 10^3$, both in mean and in root-mean-square (see Fig.~14 of Paper I). This relatively large effect implies that the CMB trispectrum could be significantly more sensitive to PBHs than CMB power spectra, as we now show with two simple order-of-magnitude estimates. First, without any exotic energy injection nor primordial non-Gaussianity, recombination is intrinsically inhomogeneous, with perturbations $\delta_{e, \rm std} \sim 10^{-4}$ \cite{Novosyadlyj_06, senatore09a}. This leads to non-Gaussianities with an amplitude just below detectability threshold for Planck \cite{senatore09a, Huang_13}. This suggests that an inhomogeneity of order $\delta_e \sim 10^{-3}$ would lead to a non-Gaussian signal with a signal-to-noise ratio (SNR) of order 10. Second, in the presence of a perturbation $\delta_e$ to the free-electron fraction, the CMB temperature anisotropy $\Theta = \Theta^{(0)} + \Theta^{(1)}$ is displaced from its standard value $\Theta^{(0)} \sim \zeta$, where $\zeta \sim 10^{-4.5}$ is the primordial curvature perturbation, by an amount $\Theta^{(1)} \sim \delta_e \zeta$. We therefore expect the connected 4-point function to be of order $\langle \Theta \Theta \Theta \Theta \rangle_{\rm c} =\langle \Theta^{(0)} \Theta^{(0)} \Theta^{(0)} \Theta^{(1)} \rangle  \sim 10^{-3} \langle \zeta^2 \rangle^2$ for a PBH abundance saturating CMB power-spectra limits. In comparison, primordial trispectra lead to 4-point functions of order $\langle \Theta \Theta \Theta \Theta \rangle_{\rm c} \sim g_{\rm NL} \langle \zeta^2\rangle^3 \sim 10^{-9} g_{\rm NL} \langle \zeta^2 \rangle^2$. Planck's upper limits on the amplitude of local-type primordial non-Gaussianity is $|g_{\rm NL}| \lesssim 10^5$ \cite{planck20c}, implying that Planck is sensitive to a 4-point function of order $\langle \Theta \Theta \Theta \Theta \rangle_{\rm c} \sim 10^{-4} \langle \zeta^2 \rangle^2$. Here again, this estimates indicates that PBHs saturating CMB power spectra limits could lead to a trispectrum detectable with SNR $\sim 10$. Put differently, the trispectrum could be sensitive to PBH abundances an order of magnitude below current CMB power-spectra limits. As an ancillary effect, the perturbation of CMB power spectra induced by accreting PBHs ought to be modified by order unity when properly accounting for the inhomogeneities in $\delta_e$, which were neglected in past works \cite{ricotti08a, yacine17a, poulin17a}. 

These promising estimates warrant a detailed calculation of the effects of inhomogeneously-accreting PBHs on CMB power spectra and trispectra. In this work, we take the first step in this program by computing the temperature-only 2-point and 4-point functions. We moreover forecast Planck's sensitivity to PBHs from the temperature trispectrum. We find that the inhomogeneity in recombination only leads to a $\lesssim 10\%$ correction to the effect of accreting PBHs on the temperature power spectrum. We also find that the temperature trispectrum is approximately as sensitive to accreting PBHs as the temperature power spectrum is, and is thus not quite as powerful a probe as our simple order-of-magnitude estimates indicated. This is likely due to the imperfect correlation between the standard temperature anisotropy $\Theta^{(0)}$ and the perturbation $\Theta^{(1)}$ sourced by inhomogeneous ionization fluctuations. Still, we find that, for $M_{\rm pbh} \lesssim 10^3 M_{\odot}$, the CMB temperature trispectrum would be a more sensitive probe of accreting PBHs than the temperature power spectrum is. This result motivates exploring the full temperature and polarization trispectrum, which we take up in future work.

The remainder of this paper is organized as follows. In Section \ref{sec:paperI} we begin by briefly reviewing accreting PBHs as a source of inhomogeneous recombination. By assuming spherical accretion and taking the luminosity prescription from Ref.~\cite{yacine17a} (hereafter AK17), we derive a quadratic transfer function for the perturbed free-electron fraction. This transfer function incorporates the radiation transport simulation and perturbed recombination calculation from Paper~I. We are able to make the transfer function factorizable with some justified approximations specific to accreting PBHs, which tremendously reduces the computational cost of calculating the high-dimensional trispectrum.

In Section \ref{sec:temp_ani} we derive general equations for the perturbed temperature anisotropy at first order in free-electron fraction perturbations, starting from the Boltzmann-Einstein system, and using the line-of-sight method \citep{seljak96a}. The results of this section are general and not limited to perturbations from accreting PBHs. As in previous works \citep{dvorkin13a,khatri10a} we neglect ``feedback" terms in the first-order perturbation. However, for the first time we quantify the error induced by this approximation in the case of the power-spectrum perturbation induced by homogeneous free-electron perturbations.

In Section~\ref{sec:tempstats} we apply these results to recombination perturbations due to accreting PBHs. We compute the perturbation to the temperature anisotropy power spectrum sourced by the \emph{inhomogeneous part} of free-electron fraction perturbations, which we find to be more than an order of magnitude smaller than its counterpart induced by the homogeneous effect on the ionization history. We moreover compute the temperature trispectrum induced by accreting PBHs, given in Eq.~\eqref{eq:Trispec-final}, which is one of the main results of this work.

In Section \ref{sec:forecast} we extract new limits on PBH abundance from Planck upper bounds on the local-shape primordial trispectrum \cite{planck20c}, which indirectly constrains the PBH-induced trispectrum with which it partially overlaps. But due to a poor correlation between the two trispectra, the constraints are an order of magnitude weaker than the constraints from the power spectra analysis. We also forecast the sensitivity of Planck to the temperature 4-point function induced by accreting PBHs, based on the optimal trispectrum estimator of Ref.~\cite{smith15a}. We are able to make these computations efficiently by pre-computing purely geometric rotational-invariant coefficients. We find that the temperature trispectrum could probe PBH abundances lower than current temperature-only power-spectrum limits for $M_{\rm pbh} \lesssim 10^3 M_\odot$. We conclude and discuss future work in Section \ref{sec:conc}.

We discuss a few points in more detail in the Appendices. In Appendix~\ref{app:vbcsq}, we justify the approximation of general non-linear functions of $v_{\rm bc}$ by a biased tracer of $v_{\rm bc}^2$. We describe our numerical resolution and convergence tests in Appendix~\ref{app:conv}. We review a few useful properties of spin-weighted spherical harmonics in Appendix~\ref{app:spin}, which we then use in Appendix~\ref{app:Q-sums} to derive simple expressions for the rotational-invariant quantities involved in the trispectrum sensitivity forecast calculation. In Appendix \ref{app:auto} we compute the auto-power spectrum of the temperature perturbation induced by accreting PBHs and its correlation coefficient with the standard temperature anisotropy. Latsly, in Appendix~\ref{app:slope}, we inspect the redshift dependence of the signal-to-noise ratio of the PBH-induced trispectrum.

\section{Perturbed recombination from accreting PBHs}\label{sec:paperI}

In this section we briefly review the effect of accreting PBHs on the ionization history. We derive an approximate factorized form for the free-electron fraction fluctuations, quadratic in the initial perturbations, which will help simplify our trispectrum calculations later on.

\subsection{Effect of accreting PBHs on the ionization history: general expressions}

If present in the early Universe, PBHs would accrete baryons which would power some radiation---at minimum, the heated, compressed and eventually ionized accreted gas would emit free-free radiation. The PBH luminosity $L$ is a function of the baryon sound speed $c_s$ and of the magnitude of the local relative velocity between baryons and dark matter $\bs{v}_{\rm bc}(\bs{r})$ (both evaluated far from the accretion region). The detailed dependence is estimated in AK17, accounting for Compton heating and Compton drag, and in two limiting regimes for the ionization structure of the accretion flow; throughout this paper, and unless otherwise stated, we will assume the most conservative ``collisionally-ionized" limit. Following AK17, we approximate the effect of relative velocities by adding them in quadrature to the baryon sound speed $c_s$, i.e.~approximating $L(c_s; v_{\rm bc} \neq 0) \approx L(\sqrt{c_s^2 + v_{\rm bc}^2}; 0)$. While the baryon sound speed is very nearly homogeneous near recombination, relative velocities have large-scale fluctuations, with rms values of order five times the sound speed \cite{Tseliakhovich_10}; as a consequence, the PBH luminosity $L(\bs{r}) = \overline{L} (1 + \delta_L(z, \bs{r}))$ is strongly inhomogeneous, tracing the large-scale fluctuations of relative velocities.

Assuming, to simplify, that PBHs all have the same mass $M_{\rm pbh}$ and make a fraction $f_{\rm pbh}$ of the dark matter, their accretion-powered luminosity leads to a volumetric energy injection rate 
\begin{align}
\dot{\rho}_{\rm inj}(z, \bs{r}) &= \overline{\dot{\rho}}_{\rm inj}(z)\left(1 + \delta_L(z, \bs{r})\right),\nonumber\\
\overline{\dot{\rho}}_{\rm inj}(z) &\equiv f_{\rm pbh}\frac{\overline{\rho}_c(z)}{M_{\rm pbh}} \overline{L}(z)
\end{align}
where $z$ is the redshift and $\overline{\rho}_c$ is the mean dark matter mass density. Note that this equation is trivially generalizable to an extended mass distribution. This inhomogeneously-injected energy is partially deposited at some later time, and some distance away from the injection site. Some of this energy is deposited in the form of extra ionizations, leading to a perturbation $\Delta x_e(z, \bs{r})$ to the free-electron fraction. The latter is a convolution of the  volumetric energy injection rate with a dimensionless injection-to-ionization Green's function. In Fourier space, this convolution is a simple product:
\beq
\Delta x_e(z, \bs{k}) = \int_z^{\infty} \frac{d z'}{1+z'}  G_{x_e}^{\rm inj}(z, z', k) \frac{\overline{\dot{\rho}}_{\rm inj}}{n_{\rm H} H E_I}\Big{|}_{z'} \delta_L(z', \bk), \label{eq:Dxe(k)}
\eeq
where $n_{\rm H}$ is the mean number density of hydrogen, $H$ is the Hubble rate, and $E_I \equiv 13.6$ eV is hydrogen's ionization energy. The homogeneous part of the ionization-fraction perturbation is obtained from a similar time integral, involving the homogeneous part of the Green's function:
\beq
\overline{\Delta x_e}(z) = \int_0^a \frac{d z'}{1+z'}  G_{x_e}^{\rm inj}(z, z', 0) \frac{\overline{\dot{\rho}}_{\rm inj}}{n_{\rm H} H E_I}\Big{|}_{z'}.
\eeq

In Paper I, we computed the Green's function $G_{x_e}^{\rm inj}(z, z', k)$ numerically, by convolving the injection-to-deposition Green's function obtained from a radiative transfer code with the deposition-to-ionization Green's functions computed with a modified \texttt{HYREC}-2 \cite{hyrec2, yacine11a, YAH_10}.

\subsection{Quadratic transfer function of ionization perturbations}

\begin{figure}[htb]
\includegraphics[width=.95\columnwidth,trim={0cm 0.5cm 0.5cm 0.25cm}]{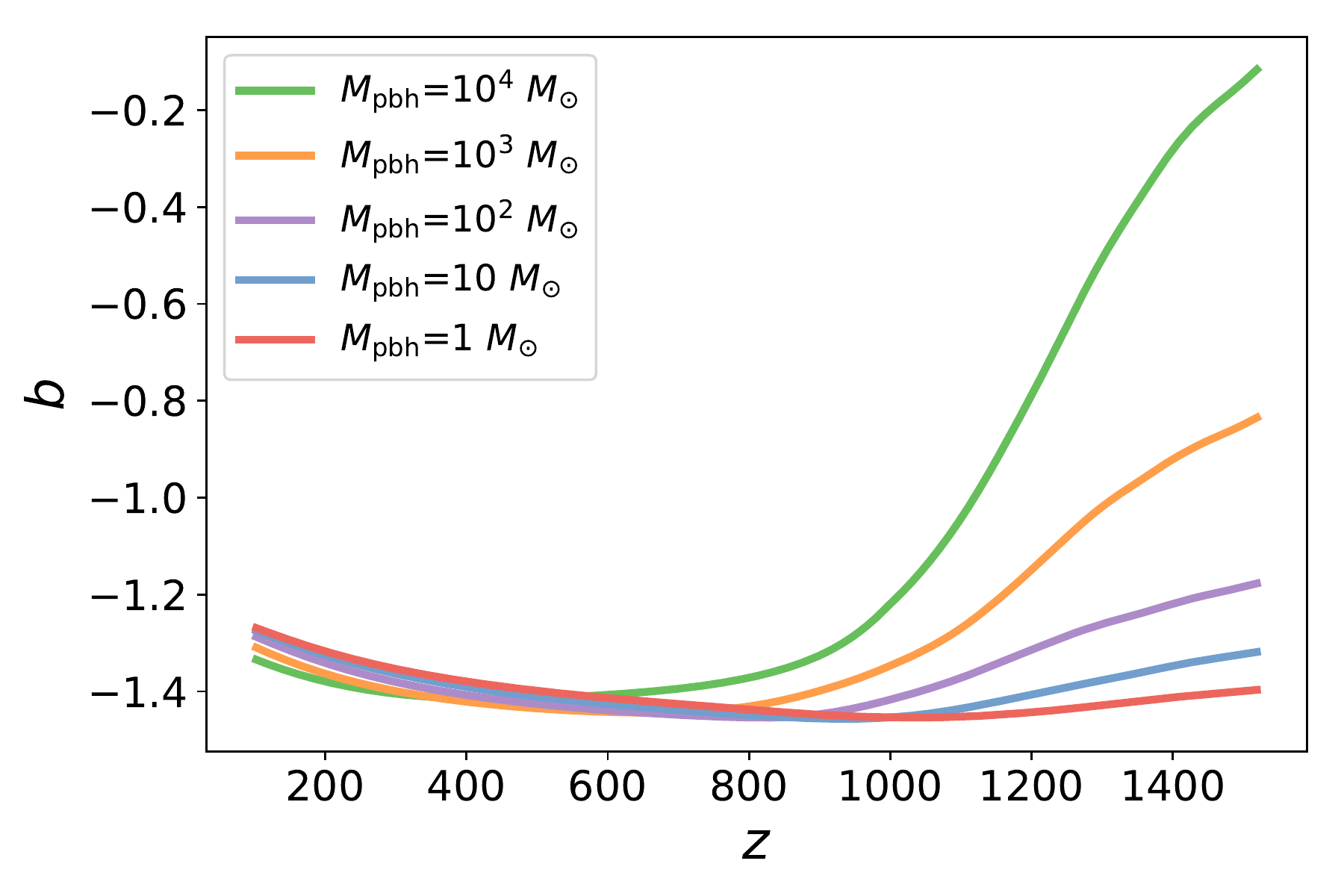}
\caption{\label{fig:b} Bias parameter $b(z)$ of the PBH accretion luminosity, approximated as a biased tracer of $v_{\rm bc}^2$ (see precise definition in Eq.~\ref{eq:b}). The bias is shown as a function of redshift and for several PBH masses $1M_{\odot} \leq M_{\rm pbh} \leq 10^4 M_{\odot}$.}
\end{figure}

The scale dependence of the luminosity perturbations $\delta_L$ is non-trivial, as the PBH luminosity is a nonlinear function of $v_{\rm bc}^2$. However, as we will see below, at lowest order $\Delta x_e$ affects CMB anisotropy statistics only through cross-correlations with other fields. As we demonstrate in Appendix \ref{app:vbcsq}, to a good approximation these cross-correlations can be obtained by approximating the full function by a biased tracer of $v_{\rm bc}^2$, with the same first moment:
\beq\label{eq:b}
\delta_L(z,\bs{r}) \approx b(z) \left(\frac{v_{\rm bc}^2(z,\bs{r})}{\langle v_{\rm bc}^2 \rangle(z)} -1 \right), \ \ \ b \equiv \frac32 \frac{\langle v_{\rm bc}^2 \delta_L\rangle}{\langle v_{\rm bc}^2\rangle}.
\eeq
This approximation is most accurate in both the large-scale and small-scale regimes, and as a consequence is reasonably accurate at all scales. We show the bias parameter $b$ as a function of redshift in Fig.~\ref{fig:b} for several black hole masses, for the AK17 accretion luminosity model. It is systematically negative, reflecting the suppression of accretion rate and luminosity in regions of large relative velocity, and its absolute value is roughly of order unity across a broad range of masses and redshifts. Although the accretion model is highly uncertain, we expect that these qualitative features should be robust, and hold even for very different accretion models, such as disk-like accretion \cite{poulin17a}.

Assuming scalar initial conditions and linear evolution, the relative velocity field is purely longitudinal, and we denote its transfer function by $\widetilde{v}_{\rm bc}(z, k)$ defined such that 
\beq
\bs{v}_{\rm bc}(z, \bk) = -i \hat{k} ~\widetilde{v}_{\rm bc}(z, k) \zeta(\bk), 
\eeq
where $\zeta(\bk)$ is the primordial curvature perturbation. We then have
\begin{align}
v_{bc}^2(z, \bk) =& (\bs{v}_{bc} \cdot \bs{v}_{bc})(z, \bk)\nonumber\\
=&- \int D(k_1 k_2)\sld(\bk_1 + \bk_2 - \bk) (\hat{k}_1 \cdot \hat{k}_2) \nonumber\\
&\quad\quad\times \widetilde{v}_{\rm bc}(z, k_1)  \widetilde{v}_{\rm bc}(z, k_2) \zeta(\bk_1) \zeta(\bk_2),
\end{align}
where from here on we denote $D(k_1 \cdots k_N) \equiv d^3 k_1/(2 \pi)^3  \cdots ~ d^3 k_N/(2 \pi)^3$ and $\sld(\bk) \equiv (2 \pi)^3 \delta_{\rm D}(\bk)$.

We denote by $\delta_e \equiv \Delta x_e/x_e^{(0)} = \overline{\delta}_e + \delta_{e, \rm inh}$ the total fractional perturbation to the standard (and homogeneous) ionization history $x_e^{(0)}$. The first part, $\overline{\delta}_e$, is the homogeneous contribution, and the second part, $\delta_{e, \rm inh}$, is the inhomogeneity, which has zero mean, $\langle \delta_{e, \rm inh} \rangle = 0$.

Inserting Eq.~\eqref{eq:b} into Eq.~\eqref{eq:Dxe(k)}, we obtain the Fourier transform of $\delta_{e, \rm inh}$ for $\bk \neq 0$:
\begin{align}
\delta_{e, \rm inh}(z, \bk) \equiv \frac{\Delta x_e(z, \bk)}{x_e^{(0)}(z)} &\approx  \fbh \int\!\! D(k_1 k_2)\sld(\bk_1 + \bk_2 - \bk) \nonumber\\
&\quad\times T_e(z, \bk_1, \bk_2)  \zeta(\bk_1) \zeta(\bk_2),  \label{eq:Te-def}
\end{align}
where, for $\bk_1 + \bk_2 \neq \bs{0}$, the ionization-perturbation quadratic transfer function $T_e$ is defined as
\begin{align}
T_e(z, \bk_1, \bk_2) &\equiv - \frac{\hat{k}_1 \cdot \hat{k}_2}{x_e^{(0)}(z)} \int_z^{\infty} \frac{d z'}{1+z'} G_{x_e}^{\rm inj}(z, z', |\bk_1 + \bk_2|) \nonumber\\
&~~~~~~~ \times \frac{\overline{\rho}_c \overline{L}~ b}{M_{\rm pbh} n_{\rm H} H E_I}\Big{|}_{z'} \frac{\widetilde{v}_{\rm bc}(k_1)\widetilde{v}_{\rm bc}(k_2)}{ \langle v_{\rm bc}^2\rangle}\Big{|}_{z'}. \label{eq:Te-general}
\end{align}
We moreover define
\begin{align}
T_e(z, \bk_1, -\bk_1) = 0, 
\end{align}
so that we may use Eq.~\eqref{eq:Te-def} even for $\bk = \bs{0}$, in which case it gives $\delta_{e, \rm inh}(\bk = \bs{0}) = 0$, as it should since $\delta_{e, \rm inh}$ is defined to have a vanishing spatial average\footnote{A more rigorous approach would be to keep track of the term proportional to $\sld(\bk)$ in $\delta_{e, \rm inh}(\bk)$; upon cross-correlating with other fields, this approach would give the same results as using Eq.~\eqref{eq:Te-def} for all $\bk$ and imposing $T_e(\bk_1, - \bk_1) = 0$.}.

\subsection{Factorized approximation of the quadratic ionization transfer function}\label{sec:approx}

We now derive an approximate, factorized form for $T_e$, which will tremendously simplify our subsequent calculations of CMB power spectra and trispectra. We do so by making two approximations.

$(i)$ For $z \gtrsim 10^3$, the Green's function $G_{x_e}^{\rm inj}(z, z')$ is peaked at $z' \approx z$ (see Fig.~9 in Paper I). We may therefore approximate the last ratio in Eq.~\eqref{eq:Te-general} by its value at $z' = z$. For $z \lesssim 10^3$, the Green's function is increasingly broad; however, after kinematic decoupling at $z_{\rm dec} \approx 1020$, relative velocities redshift as $\widetilde{v}_{\rm bc}(z, k) \propto (1+z)$, independently of scale \cite{Tseliakhovich_10}. Therefore, the last term in Eq.~\eqref{eq:Te-general} is independent of redshift for $z' \lesssim z_{\rm dec}$. We therefore make the following approximation in Eq.~\eqref{eq:Te-general}, which we expect to be accurate at all redshifts: 
\beq
\frac{\widetilde{v}_{\rm bc}(k_1)\widetilde{v}_{\rm bc}(k_2)}{ \langle v_{\rm bc}^2\rangle}\Big{|}_{z'} \approx \frac{\widetilde{v}_{\rm bc}(k_1)\widetilde{v}_{\rm bc}(k_2)}{ \langle v_{\rm bc}^2\rangle}\Big{|}_{z}.
\eeq
This approximation implies the following simplification:
\begin{align}
T_e(z, \bk_1, \bk_2) &=(\hat{k}_1 \cdot \hat{k}_2) G_e(z, |\bk_1 + \bk_2|) \nonumber\\
&\quad\quad\quad\quad\times \frac{\widetilde{v}_{\rm bc}(z, k_1) \widetilde{v}_{\rm bc}(z, k_2)}{\langle v_{\rm bc}^2 \rangle_z},\\
G_e(z, k) \equiv& -\int_z^{\infty}\frac{d z'}{1+z'} \frac{G_{x_e}^{\rm inj}(z, z', k)}{x_e^{(0)}(z)} \frac{\overline{\rho}_c \overline{L}~ b}{M_{\rm pbh} n_{\rm H} H E_I}\Big{|}_{z'}.\label{eq:G_e}
\end{align}

\begin{figure}[htb]
\includegraphics[width=.95\columnwidth,trim={0cm 0.5cm 0.5cm 0.25cm}]{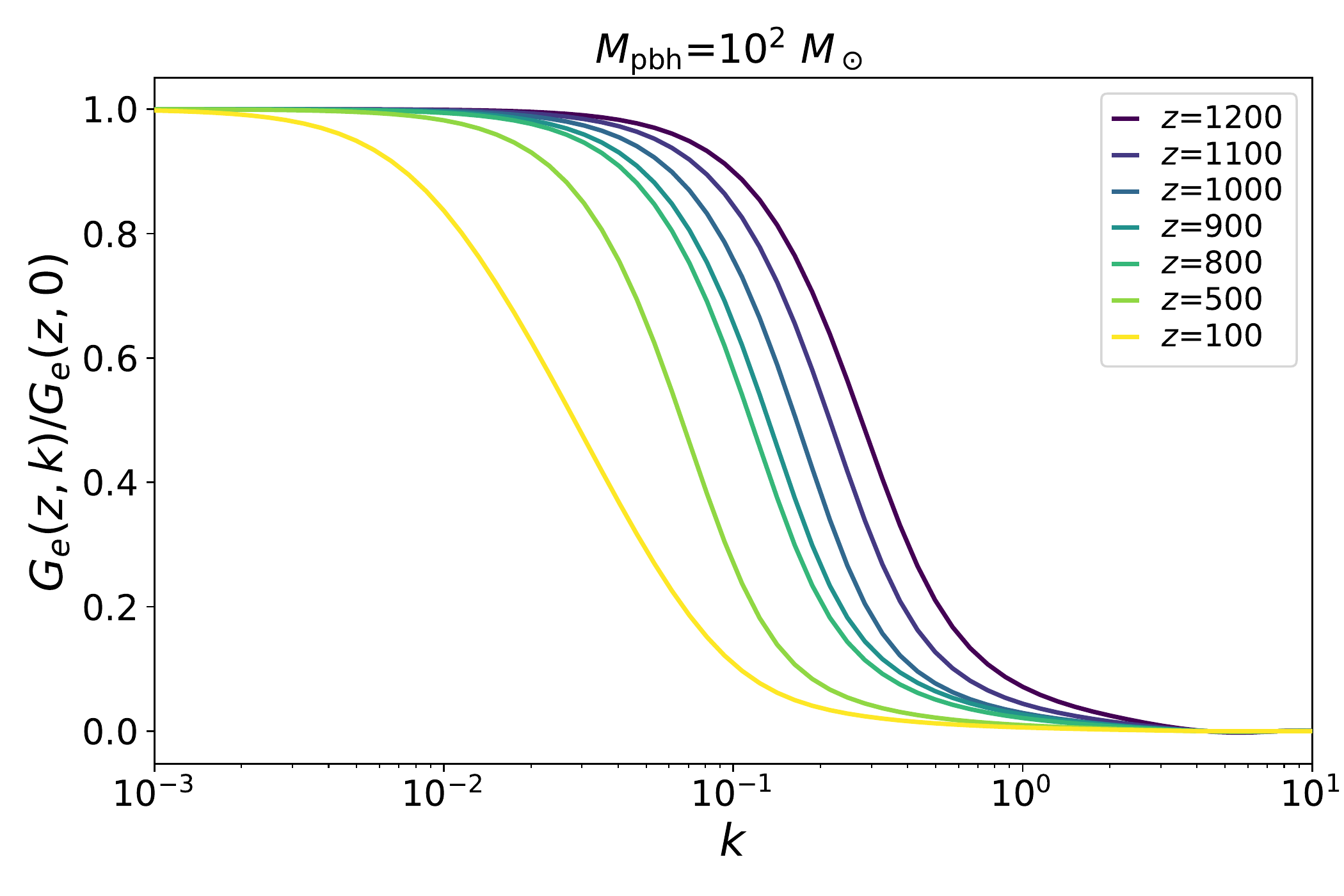}
\caption{\label{fig:G_e} Normalized injection-integrated Green's function defined in Eq.~\eqref{eq:G_e}, at various redshifts, for 100-$M_\odot$ PBHs. This function is approximately Gaussian with a characteristic cutoff $k_*(z)$, beyond which ionization inhomogeneities are suppressed due to finite propagation of injected photons.}
\end{figure}

$(ii)$ As illustrated in Fig.~\ref{fig:G_e}, we find that $G_e(z, k)$ is an approximately Gaussian function of wavenumber, with a characteristic cutoff at a redshift-dependent scale $k_*(z)$:
\beq
G_e(z, k) \approx G_e(z, 0) e^{- k^2/k_*^2(z)}.
\eeq
Given that $(\bk_1 + \bk_2)^2 \leq 2k_1^2 + 2 k_2^2$, we may therefore approximately bracket $G_e$ as follows: 
\beq
\frac{G_e(z, \sqrt{2} k_1)~G_e(z, \sqrt{2} k_2) }{G_e(z, 0)} \leq G_e(z, |\bk_1 + \bk_2|) \leq G_e(z, 0).
\eeq
By default, we will conservatively approximate $G_e$ by the lower bound of this range. This approximation is accurate at large scales $k_1,k_2 \lesssim k_*(z)$, at which propagation effects are not relevant to energy deposition. 

With these two approximations, the quadratic ionization transfer function takes on the factorized form
\begin{align}
T_e(z, \bk_1,\bk_2) &\approx (\hat{k}_1 \cdot \hat{k}_2) \Delta_e(z, k_1) \Delta_e(z, k_2), \label{eq:dele_pbh}\\
\Delta_e(z, k) &\equiv \frac{G_e(z, \sqrt{2} k)}{\sqrt{G_e(z, 0)}} \frac{\widetilde{v}_{\rm bc}(z, k)}{\langle v_{\rm bc}^2 \rangle_z^{1/2}}, \label{eq:Delta_e-def}
\end{align}
where we recall that this expression holds for $\bk_1 + \bk_2 \neq 0$ only, and that $T_e(z, \bk_1, - \bk_1) = 0$. 

\begin{figure*}[htp]
\includegraphics[trim={0.7 0 0.4cm 0},width=\columnwidth]{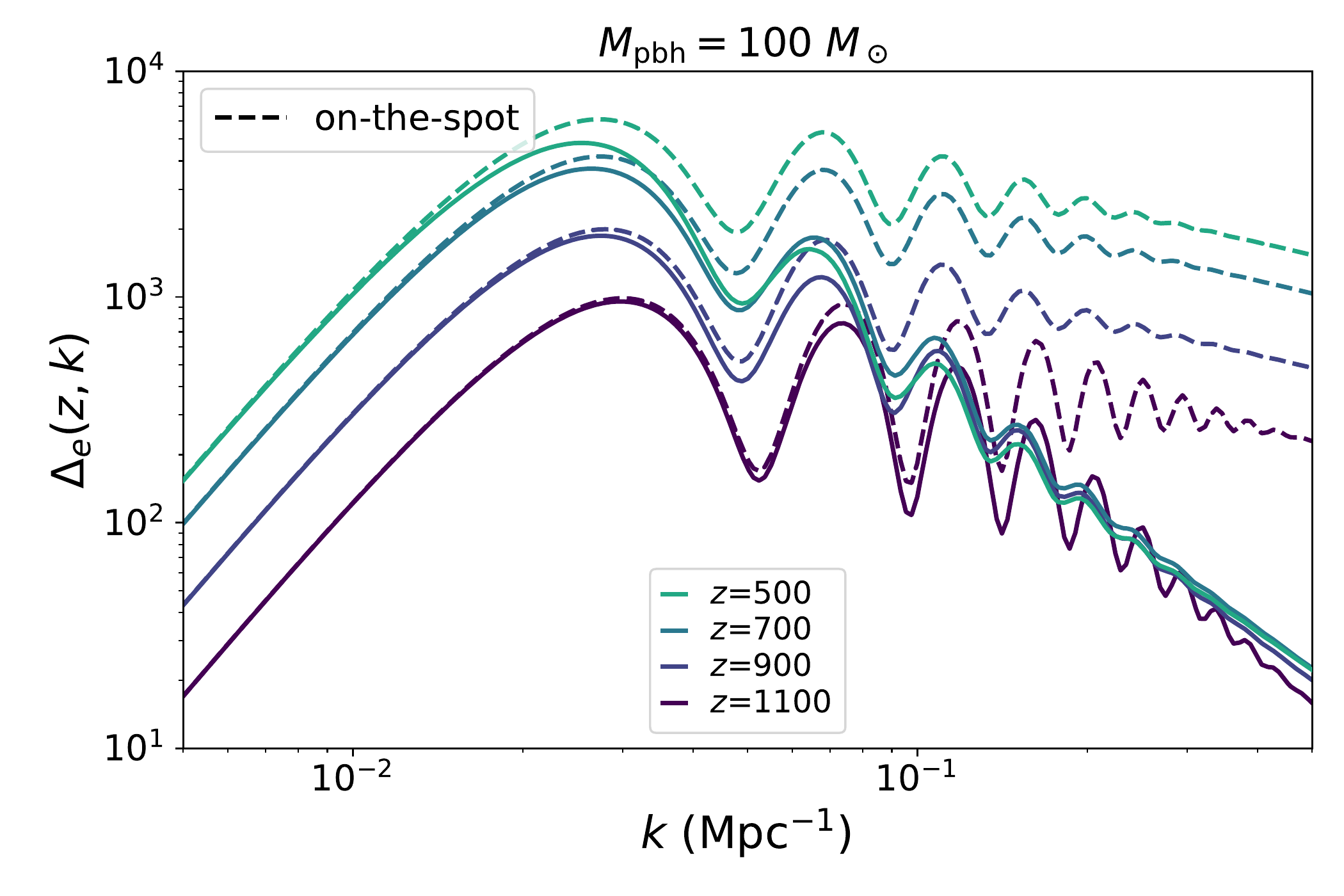}
\includegraphics[trim={0.7 0 0.4cm 0},width=\columnwidth]{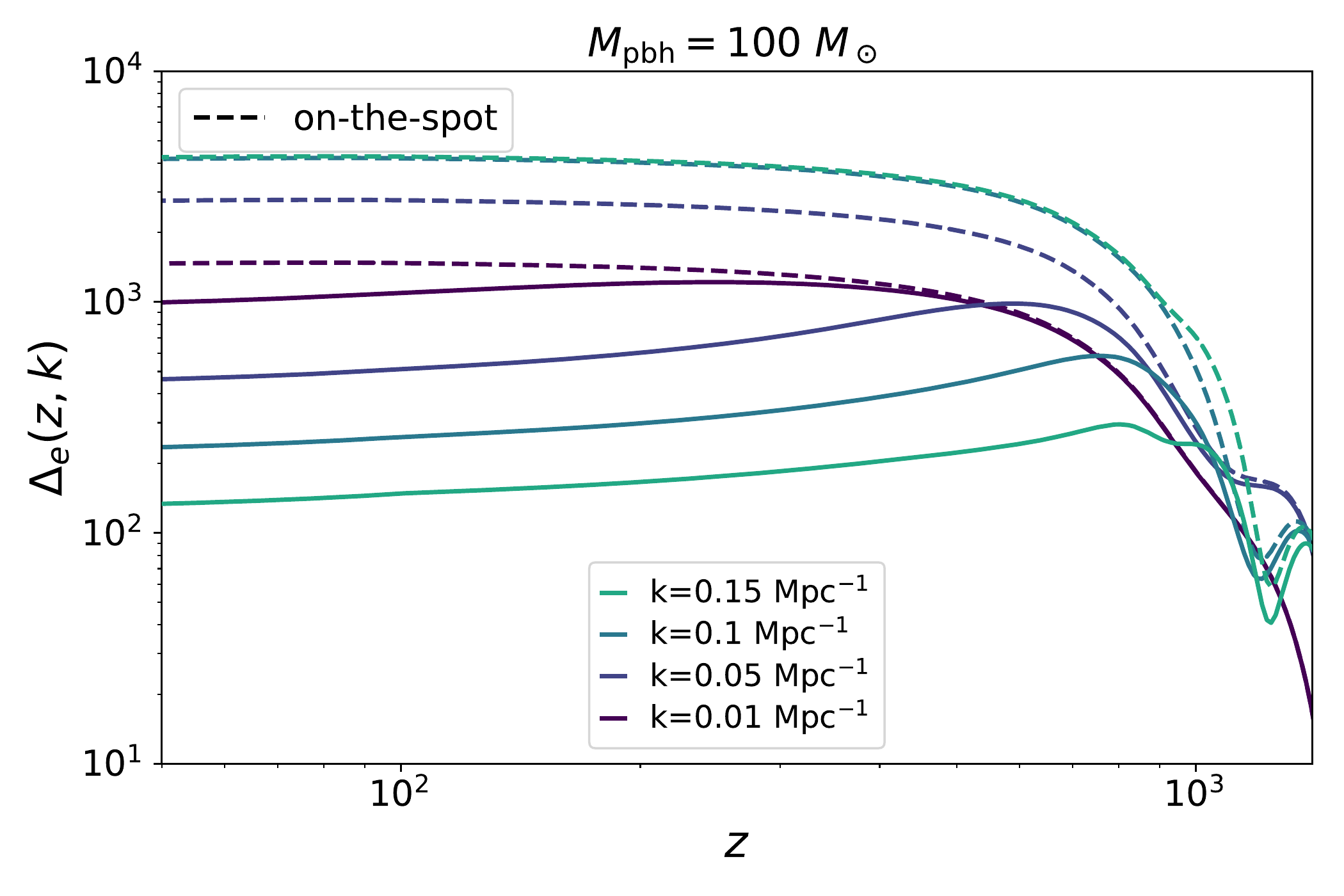}
\caption{\label{fig:Del_e} $\Delta_e(z,k)$ as defined in Eq.~\eqref{eq:Delta_e-def} for PBHs of 100 $M_\odot$ plotted for various redshifts as a function of scale (left) and for various scales as a function of redshift (right). We also show this function if we ignore photon propagation and consider energy deposition as spatially on-the-spot (dashed lines). These plots reveal the shape of the free-electron perturbations induced by $v_{\rm bc}^2$ as well as the amplitude suppression when considering nonlocal energy deposition from accreting PBHs.}
\end{figure*}

Our approximation for $G_e(z, |\bk_1 + \bk_2|)$ can significantly underestimate the true signal at small scales, in particular for $\bk_1 \approx -\bk_2$ or $k_1 \ll k_2$. Moreover, it modifies the geometric dependence of the signal. In order to estimate the error that this approximation induces, we will also show our results in the spatially-on-the-spot approximation $G_e(z, k) \approx G_e(z, 0)$, which systematically over-estimates the signal. In that case, the quadratic ionization transfer function still takes the form \eqref{eq:dele_pbh}, but with $\Delta_e(z, k) = \sqrt{G_e(z, 0)} \frac{\widetilde{v}_{\rm bc}(z, k)}{\langle v_{\rm bc}^2 \rangle_z^{1/2}}$. We show $\Delta_e(z, k)$ as a function of wavenumber and redshift in Fig.~\ref{fig:Del_e}, both for our default approximation, and in the on-the-spot limit.

\section{Temperature anisotropy from perturbed recombination: general equations}\label{sec:temp_ani}

We turn to computing the temperature anisotropy in the presence of a general deviation from the standard free-electron fraction evolution, including spatial variations. Because observed CMB anisotropies are consistent with the standard $\Lambda$CDM prediction and canonical homogeneous recombination, this deviation is necessarily small, allowing for a perturbative treatment.

\subsection{Temperature Boltzmann equation}

The evolution of the phase-space distribution of photons is governed by the Boltzmann-Einstein differential system. CMB photons follow geodesics in an expanding universe subject to Thomson scattering off free electrons. Provided photons remain thermal, they are described entirely by their temperature fluctuations $\Theta(\eta,{\bm x},\hat{n})$ and transverse, symmetric trace-free $3 \times 3$ polarization tensor $P_{ab}(\eta, \bm x, \hat{n})$, where $\hat{n}$ is the propagation direction and $\eta$ is the conformal time. In the conformal Newtonian gauge, the Boltzmann-Einstein equation for the temperature perturbation is \citep{ma95a} 
\begin{align}\label{eq:be_ODE}
    \deriv{\Theta}{\eta}\equiv \dot{\Theta}+\hat{n}\cdot\nabla\Theta+\hat{n}\cdot\nabla\psi-\dot{\phi}&=\dot{\tau}~ \mathcal{C}[\Theta, P_{ab}, {\bm v}_b],
\end{align}
where overdots denote partial derivatives with respect to $\eta$. 
In this equation, ${\bm v}_b$ is the baryon velocity, and $\dot{\tau}\equiv a n_e\sigma_T $ is the conformal scattering rate by free electrons with number density $n_e$, where $a$ is the scale factor, and $\sigma_{\rm T}$ is the Thomson cross section. $\mathcal{C}$ is a linear operator encapsulating the geometry of Thomson scattering:
\begin{align}
\mathcal{C}[\Theta, P_{ab}, {\bm v}_b](\hn)&\equiv \Lagr[\Theta, P_{ab}, {\bm v}_b](\hn) - \Theta(\hn), \label{eq:thomS_ang_avg}\\[5pt]
\Lagr[\Theta, P_{ab}, {\bm v}_b](\hn)&\equiv \Theta_0 + \hn \cdot \bs{v}_b + \hat{n}_a \hat{n}_b ~\Pi_{ab}, \label{eq:thoms_piece}
\end{align}
where $\Theta_0$ is the photon temperature monopole, 
\begin{align}
\Theta_0 \equiv \int \frac{d^2 \hn}{4 \pi} \Theta(\hat{n}),
\end{align}
and the symmetric trace-free tensor $\Pi_{ab}$ is a linear combination of the photon quadrupole moment and the angle-averaged polarization tensor:
\begin{align}
\Pi_{ab} \equiv  \int \frac{d^2 \hn}{4 \pi} \left[\frac14 (3 \hn_a \hn_b - \delta_{ab}) \Theta(\hat{n}) + \frac32 P_{ab}(\hat{n}) \right].
\end{align}

To close the Boltzmann-Einstein differential system, the evolution equations for baryons, cold dark matter, neutrinos, and photon polarization are needed \citep{ma95a}, but we do not explicitly list them here.

\subsection{Standard solution}\label{sec:canon}

We now briefly review the standard solution obtained with scalar initial conditions and for linear evolution, and given the standard, homogeneous free-electron fraction $x_e^{(0)}$. The notation and expressions derived here will be useful in the following section dealing with the perturbation to CMB anisotropies induced by modified recombination. We denote all standard variables by a superscript $(0)$, e.g.~$\Theta^{(0)}, \psi^{(0)}$, etc... For short, we also denote $\mathcal{C}^{(0)} \equiv \mathcal{C}[\Theta^{(0)}, P_{ab}^{(0)}, \bm v_b^{(0)}]$, and similarly for $\Lagr^{(0)}$.

The Boltzmann equation \eqref{eq:be_ODE} is most easily solved in terms of the variable $\Theta_{\rm eff} \equiv \Theta + \psi$, and in Fourier space. For the standard case, it takes the form
\begin{align}
\dot{\Theta}_{\rm eff}^{(0)} + i \bk \cdot \hat{n} \Theta_{\rm eff}^{(0)} + \dot{\tau}^{(0)} \Theta_{\rm eff}^{(0)} = \dot{\tau}^{(0)} S^{(0)},\\
S^{(0)} \equiv \Lagr^{(0)} + \psi^{(0)} + \frac1{\dot{\tau}^{(0)}} (\dot{\psi}^{(0)} + \dot{\phi}^{(0)}),\label{eq:S0-def}
\end{align}
The solution at an arbitrary conformal time is given by the line-of-sight solution \citep{seljak96a},
\begin{align}\label{eq:Thetaeff0_sol}
    \Theta^{(0)}_{\rm eff}(\eta, \bk, \hn) &= \exp\left(\int_\eta^{\eta_0}d \eta''~\dot{\tau}^{(0)}(\eta'')\right) \nonumber\\
    &\times \int_0^{\eta} d \eta' ~g(\eta')S^{(0)}(\eta',{\bm k},\hat{n}) e^{i \bs{k} \cdot \hn (\eta' - \eta)},
\end{align}
where $\eta_0$ is the conformal time today, and $g(\eta)$ is the standard visibility function, 
\begin{align}
g(\eta) \equiv \dot{\tau}^{(0)}(\eta) \exp\left(- \int_\eta^{\eta_0}d \eta'~\dot{\tau}^{(0)}(\eta')\right),  
\end{align}
In particular, the line-of-sight solution today, and at the spatial origin ${\bm x}=0$, is given by \citep{seljak96a}
\begin{align}\label{eq:los}
    \Theta^{(0)}_{\rm eff}(\eta_0,{\bm x}=0,\hat{n})=\int Dk \int_0^{\eta_0}\!\!d \eta\, g(\eta) \nonumber\\
    S^{(0)}(\eta, \bk, \hn) ~ e^{- i \bk\cdot\hat{n}\chi},
\end{align}
where from here on we denote $\chi \equiv \eta_0 - \eta$.

Under scalar adiabatic initial conditions, the baryon velocity is purely longitudinal, i.e.~in Fourier space, $\bm v_b^{(0)}(\bk) = - i \bk \theta_b^{(0)}/k^2$. Moreover, we have $\hat{n}_a \hat{n}_b \Pi_{ab}^{(0)}(\bk) = - \Pi^{(0)} P_2(\hat{k} \cdot \hn)$, where $\Pi^{(0)}\equiv (F_{\gamma 2}+G_{\gamma 0}+G_{\gamma 2})/8$ is a combination of the photon temperature quadrupole and polarization monopole and quadrupole moments (here we used the notation of Ref.~\cite{ma95a}). We thus have
\begin{align}\label{eq:S0}
S^{(0)}(\bk, \hn) = \Theta_0^{(0)} + \psi^{(0)} + \frac1{\dot{\tau}^{(0)}} (\dot{\psi}^{(0)} + \dot{\phi}^{(0)})\nonumber\\
- \frac{i }{k}  (\hat{k} \cdot \hn) \theta_b^{(0)}- P_2(\hat{k} \cdot \hat{n}) \Pi^{(0)} ,
\end{align}
where the dependence of $\Theta_0^{(0)}, \bm v_b^{(0)}$, etc... on $\bk$ is implicit.
 
To simplify this expression, note that $- i \hat{k} \cdot \hn ~e^{- i \bk \cdot \hn \chi }  = \partial_{k\chi} e^{- i \bk \cdot \hn \chi}$, where $\partial_{k \chi} \equiv \frac1{k} \frac{\partial}{\partial \chi}$. In Eq.~\eqref{eq:los} we may then conveniently substitute $S^{(0)}$ by an angle-independent differential operator acting on the geometric exponential term,
\begin{align}\label{eq:S_part}
S^{(0)}(\bk, \hn) \rightarrow S_{\partial}^{(0)}(\bk) \equiv \Theta_0^{(0)} + \psi^{(0)} + \frac1{\dot{\tau}^{(0)}} (\dot{\psi}^{(0)} + \dot{\phi}^{(0)})\nonumber\\
+ \frac {\theta_b^{(0)}}{k}  \partial_{k\chi} + \Pi^{(0)} \left( \frac3{2} \partial_{k\chi}^2 + \frac12 \right).
\end{align}
Using this substitution, and from the Rayleigh formula,
\begin{align}
e^{-i \bs{k} \cdot \hn\chi} &= \sum_{\ell} (-i)^\ell (2 \ell +1) j_\ell(k \chi) P_\ell(\hn \cdot \hk) \label{eq:plane-wave1}\\
&= 4\pi \sum_{\ell m} (-i)^{\ell} j_{\ell}(k \chi) Y_{\ell m}(\hn) Y_{\ell m}^*(\hk),\label{eq:plane-wave}
\end{align} 
where $j_\ell$ are the spherical Bessel functions, we may directly read off the harmonic multipoles (for $\ell > 0$) of the standard temperature anisotropy from Eq.~\eqref{eq:los}:
\begin{align}\label{eq:thet0}
    \Theta^{(0)}_{\ell m}&=4\pi (-i)^\ell \!\int\! Dk\, Y^*_{\ell m}({\hat k})\int_0^{\eta_0}\!\!
    d \eta \,g(\eta) S^{(0)}_\partial(\bk, \eta) j_{\ell}(k\chi),
\end{align}
where the operator $S^{(0)}_{\partial}$ now acts on the Bessel function. 

Lastly, we denote by $\widetilde{S}^{(0)}_{\partial}(k, \eta)$ the transfer function of $S^{(0)}_{\partial}$, defined such that
\begin{align}
S^{(0)}_{\partial}(\bk, \eta) = \widetilde{S}^{(0)}_{\partial}(k, \eta) \zeta(\bk),
\end{align}
where $\zeta(\bk)$ is the primordial curvature perturbation. We thus obtain
\begin{align}
   \Theta^{(0)}_{\ell m} &= 4\pi (-i)^\ell \int Dk~Y^*_{\ell m}({\hat k})\Delta_\ell(k) \zeta(\bk), \label{eq:Theta_lm^0}\\
   \Delta_\ell(k) &\equiv \int_0^{\eta_0} d \eta ~g(\eta) ~\widetilde{S}^{(0)}_{\partial}(k, \eta) j_{\ell}(k \chi).\label{eq:Deltal-def}
\end{align}

Assuming the primordial curvature perturbation is Gaussian, with power spectrum $P_{\zeta}(k)$, i.e.~such that 
\begin{align}
\langle \zeta(\bk) \zeta(\bk') \rangle = (2 \pi)^3 \delta_{\rm D}(\bk + \bk') P_{\zeta}(k) \equiv \sld(\bk + \bk') P_{\zeta}(k),
\end{align}
the canonical temperature anisotropy angular power spectrum, $\braket{\Theta^{(0)}_{\ell m}\Theta^{*(0)}_{\ell'm'}}\equiv \delta_{\ell \ell'}\delta_{m m'}C_\ell^{(0)}$, is then given by
\begin{align}\label{eq:C_l}
C_{\ell}^{(0)}= 4 \pi \int Dk ~[\Delta_\ell(k)]^2 P_{\zeta}(k).
\end{align}
The $\Delta_\ell(k)$ are the temperature fluctuation multipole transfer functions that can be extracted from cosmological codes such as \texttt{CLASS} \citep{CLASS}. In practice, since we will need to compute similar integrals later on, we compute the conformal time integral in Eq.~\eqref{eq:Deltal-def} ourselves, using only the source term transfer functions in Eq.~\eqref{eq:S_part} from \texttt{CLASS}. We also compute the $k$-integral in Eq.~\eqref{eq:C_l} ourselves, and checked that our results match those of \texttt{CLASS} to high accuracy. We discuss our numerical resolution and convergence tests in Appendix~\ref{app:conv}.

\subsection{Temperature anisotropy due to perturbed recombination }\label{sec:perturb}

We now suppose the free-electron fraction is perturbed, $x_e = x_e^{(0)} (1 + \delta_e)$. Importantly, we make no assumption about the spatial dependence of $\delta_e$, which in general has both a homogeneous and an inhomogeneous piece. As a result of the modified Thomson scattering rate $\dot{\tau} = \dot{\tau}^{(0)}(1 + \delta_e) \equiv \dot{\tau}^{(0)} + \dot{\tau}^{(1)}$, all matter and metric fields also get altered: $\Theta = \Theta^{(0)} + \Theta^{(1)}$, $\psi = \psi^{(0)} + \psi^{(1)}$, etc... For short, we again denote $\mathcal{C}^{(1)} \equiv \mathcal{C}[\Theta^{(1)}, P_{ab}^{(1)}, \bm{v}_b^{(1)}]$, and similarly for $\Lagr^{(1)}$.

In general, matter and metric fields depend nonlinearly on $\delta_e$; however, in the limit of small $\delta_e$, we may solve them with a perturbative expansion in $\delta_e \ll 1$. The zero-th order equation is the canonical Boltzmann-Einstein system discussed in Sec.~\ref{sec:canon}. At first order in $\delta_e \ll 1$, the photon temperature Boltzmann equation is
\begin{align}\label{eq:pertthet}    \dot{\Theta}^{(1)}+\hat{n}\cdot\nabla\Theta^{(1)}+\hat{n}\cdot\nabla&\psi^{(1)}-\dot{\phi}^{(1)}= \dot{\tau}^{(0)}\mathcal{C}^{(1)}+\dot{\tau}^{(1)}\mathcal{C}^{(0)}.
\end{align}

It is convenient to rewrite this equations in terms of the variable $\Theta_{\rm eff}^{(1)} \equiv \Theta^{(1)} + \psi^{(1)}$, as follows:
\begin{align}
\dot{\Theta}_{\rm eff}^{(1)} + \hat{n} \cdot \nabla \Theta_{\rm eff}^{(1)} + \dot{\tau}^{(0)} \Theta_{\rm eff}^{(1)} &= \dot{\tau}^{(0)} S^{(1)},\label{eq:Theta_eff1}
\end{align}
where the source term $S^{(1)}$ will be discussed shortly. Again, Eq.~\eqref{eq:Theta_eff1} can be easily solved in Fourier space, with the familiar line-of-sight solution. In particular, the order-one photon temperature perturbation at present time $\eta_0$, and at the spatial origin, takes the form
\begin{align}\label{eq:los_source}
    \Theta^{(1)}_{\rm eff}&(\eta_0,{\bm x}=0,\hat{n})=\nonumber\\
    &\int Dk\int_0^{\eta_0}\!\!d \eta\, g(\eta)S^{(1)}(\bm k, \eta, \hat{n}) e^{-i{\bm k}\cdot\hat{n}\chi}.
\end{align}
The first-order source term $S^{(1)}$ contains two pieces:
\begin{align}
S^{(1)} &= S^{(1) \rm d} + S^{(1) \rm f},\\ 
    S^{(1) \rm d} &\equiv \delta_e*\mathcal{C}^{(0)}, \label{eq:S1direct}\\
    S^{(1) \rm f}&\equiv\Lagr^{(1)}+ \psi^{(1)} + \frac{1}{\dot{\tau}^{(0)}}\left(\dot{\psi}^{(1)} + \dot{\phi}^{(1)}\right).
\end{align}
The first piece $S^{(1) \rm d}$, we coin as the ``direct" term, as it depends directly on the perturbed free-electron fraction $\delta_e$, and otherwise on zero-th order terms through $\mathcal{C}^{(0)}$, which can thus be extracted in a relatively straightforward fashion from \texttt{CLASS}. Note that $\delta_e * \mathcal{C}^{(0)}$ denotes a multiplication in real space, or a convolution in Fourier space. The second piece $S^{(1) \rm f}$, we dub the ``feedback" term, as it depends on first-order terms; it thus requires solving explicitly for the infinite Boltzmann hierarchy similar to that solved at zeroth order, but with an additional source term, containing wavemode mixing due to convolutions in Fourier space \cite{khatri10a}. 

As in previous studies \cite{khatri09a,dvorkin13a}, we will not solve for the feedback term in this work. However, we now quantify its magnitude for the first time, in the limit of \emph{homogeneous} perturbations to recombination.

\subsection{Magnitude of the feedback term for homogeneous \texorpdfstring{$\delta_e$}{de}}\label{sec:homo_de}

We consider the limiting case where $\delta_e(\eta, \bs{x}) = \overline{\delta}_e(\eta)$ is homogeneous. Our perturbative expansion in $\delta_e$ applies just as well in this case, as long as $\overline{\delta}_e \ll 1$. We shall only include the ``direct" source term, and then explicitly check our results against the exact output from \texttt{CLASS}, which can handle arbitrary homogeneous perturbations to the recombination history, thus effectively account for both ``direct" and ``feedback" sources (although the calculation is not split this way in \texttt{CLASS}).

Let us rewrite the direct source term as
\begin{align}
S^{(1) \rm d}_{\rm hom} = \overline{\delta}_e \mathcal{C}^{(0)} = \overline{\delta}_e \left(\Lagr^{(0)} + \psi^{(0)}\right) - \overline{\delta}_e \Theta_{\rm eff}^{(0)},
\end{align}
where the subscript ``hom" is there to remind the reader that we are considering a homogeneous free-electron fraction in this section.

The contribution of the second term to the innermost integral of Eq.~\eqref{eq:los_source} can be rewritten in the form 
\begin{align}
\int_0^{\eta_0} d \eta ~g(\eta) ~ \overline{\delta}_e(\eta) \Theta_{\rm eff}^{(0)}(\eta, \bk, \hn) e^{- i \bk \cdot \hn \chi}\nonumber\\
= \int_0^{\eta_0} d\eta ~g(\eta) ~ \overline{\mathcal{D}}_e(\eta) S^{(0)}(\eta, \bk, \hn)e^{- i \bk \cdot \hn \chi},
\end{align}
where 
\begin{align}
\overline{\mathcal{D}}_e(\eta)\equiv\int_{\eta}^{\eta_0} d \eta'~ \dot{\tau}^{(0)}(\eta') \overline{\delta}_e(\eta').
\end{align}
To obtain this result, we inserted the arbitrary-time line-of-sight solution \eqref{eq:Thetaeff0_sol} for $\Theta_{\rm eff}^{(0)}$, and switched the order of integration. We therefore arrive at the following expression for the direct contribution to the first-order temperature perturbation in the homogeneous case:
\begin{align}
{\Theta}^{(1) \rm d}_{\rm hom}(\eta_0,{\bm x}=0,\hat{n})=\int Dk\int_0^{\eta_0}\!\!d \eta\, g(\eta) \nonumber\\
\left[\overline{\delta}_e(\eta) (\Lagr^{(0)}_{\partial} + \psi^{(0)}) - \overline{\mathcal{D}}_e(\eta) S^{(0)}_{\partial} \right] e^{-i{\bm k}\cdot\hat{n}\chi},
\end{align}
where $\Lagr^{(0)}_{\partial}(\bk, \eta)$ is the operator obtained from $\Lagr^{(0)}(\bk, \eta, \hn)$ in the same fashion as $S^{(0)}_{\partial}$ is obtained from $S^{(0)}$ (c.f. Eq.~\eqref{eq:S_part}).

Using the same steps as in Sec.~\ref{sec:canon}, we thus arrive at the following expression for the spherical-harmonic components of the direct-only part of $\Theta^{(1)}_{\rm hom}$:
\begin{align}
    {\Theta}_{\ell m, \rm hom}^{(1) \rm d} &= 4 \pi (-i)^{\ell} \int Dk~ Y_{\ell m}^*(\hat{k}) \Delta_{\ell, \rm hom}^{(1) \rm d}(k) \zeta(\bk), \label{eq:Theta1_lm_hom} \\
    \Delta_{\ell, \rm hom}^{(1) \rm d}(k) &\equiv \int_0^{\eta_0} d \eta ~g(\eta)~ \Big{[}\overline{\delta}_e(\eta) (\widetilde{\Lagr}^{(0)}_{\partial} + \widetilde{\psi}^{(0)}) \nonumber\\
    &\quad \quad\quad \quad\quad\quad\quad- \overline{\mathcal{D}}_e(\eta) \widetilde{S}^{(0)}_{\partial} \Big{]} j_{\ell}(k \chi), \label{eq:Delta1_lm_hom}
\end{align}
where $\widetilde{\Lagr}^{(0)}_{\partial}(k, \eta)$ and $\widetilde{\psi}^{(0)}(k, \eta)$ are the transfer functions of $\Lagr^{(0)}_{\partial}(\bk, \eta)$ and $\psi^{(0)}(\bk, \eta)$. 

We may now compute the perturbation to the angular power spectrum. To linear order in $\overline{\delta}_e \ll 1$, we have $C_{\ell} = C_{\ell}^{(0)} + {C}_{\ell, \rm hom}^{(1)}$, where we defined $2\braket{{\Theta}^{(1)}_{\ell m, \rm hom}\Theta^{*(0)}_{\ell'm'}}\equiv \delta_{\ell \ell'}\delta_{m m'}{{C}^{(1)}_{\ell, \rm hom}}$. We find that the direct contribution to ${C}_{\ell, \rm hom}^{(1)}$ is then
\begin{align}
{C}_{\ell, \rm hom}^{(1)\rm d} = 8 \pi \int Dk ~P_{\zeta}(k) \Delta_\ell(k) \Delta_{\ell, \rm hom}^{(1) \rm d}(k). \label{eq:Cl1_hom}
\end{align}

\begin{figure}[htb]
\includegraphics[trim={0cm 0.5cm 0.1cm 1cm},width=1\columnwidth,clip]{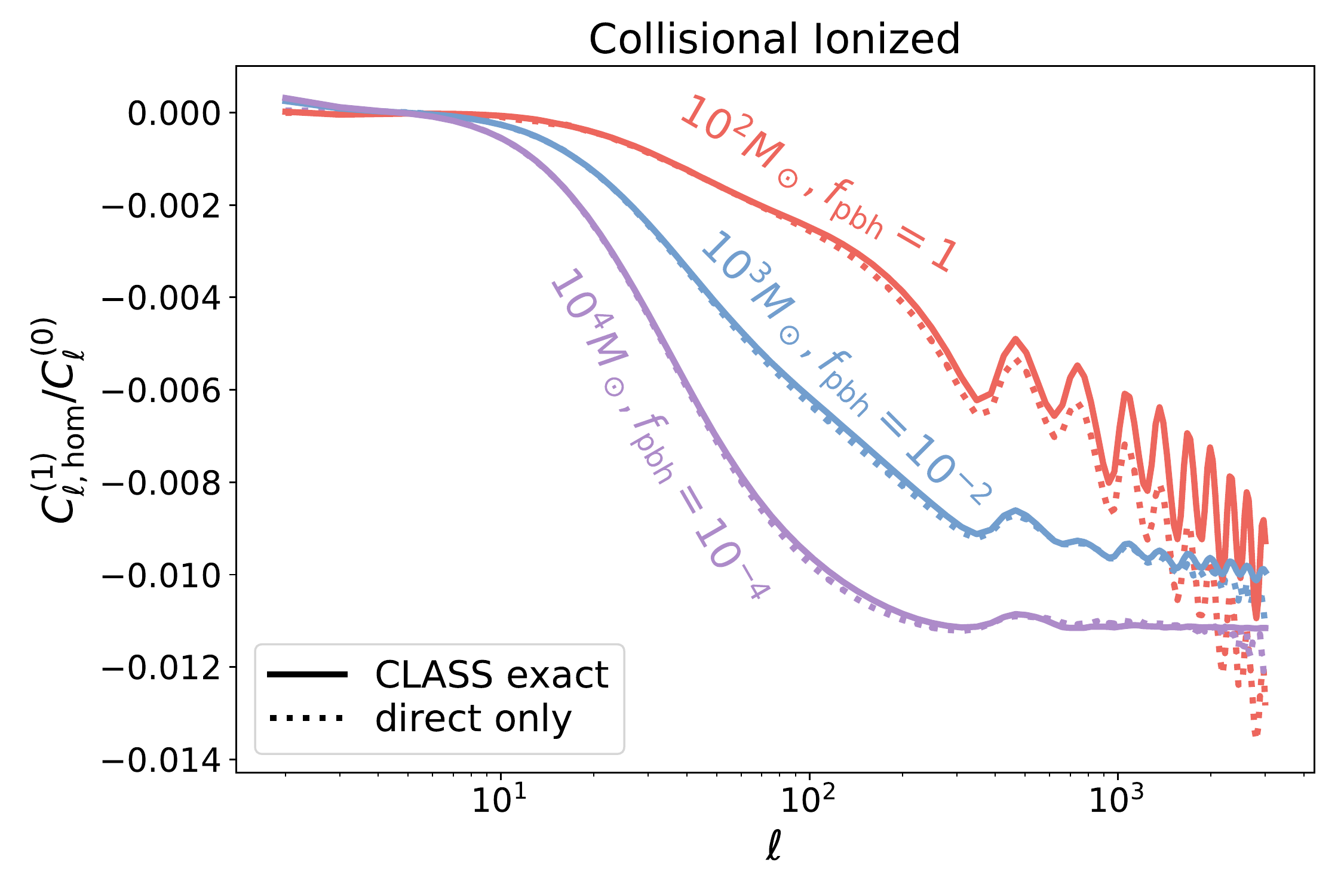}
\caption{\label{fig:class_v_comm} Fractional change to the temperature anisotropy power spectrum from the homogeneous perturbation to the free-electron fraction, $\overline{\delta_e}(\eta)$, for various PBH masses and abundances. We compare the exact non-perturbative effect extracted from \texttt{CLASS} to the perturbative solution including only the ``direct" source term discussed in Sec.~\ref{sec:homo_de}. Our approximation of neglecting the ``feedback'' term is reasonably accurate and we assume this carries over for the inhomogeneous free-electron fraction case.}
\end{figure}

We computed ${C}_{\ell, \rm hom}^{(1) \rm d}$ using the homogeneous part of the free-electron perturbation sourced by accreting PBHs, as calculated in AK17. We compare this result against the exact ${C}_{\ell, \rm hom}^{(1)}$ obtained from \texttt{CLASS} in Fig.~\ref{fig:class_v_comm}. We see that neglecting the feedback term $S^{(1) \rm f}_{\rm hom}$ leads to an order $\sim 10\%$ relative error on ${C}_{\ell,\rm hom}^{(1)}$ for relevant black hole masses, indicating that the term is subdominant. While there is no guarantee that this subdominance carries over in general at higher-order statistics, it still gives us some confidence that neglecting $S^{(1) \rm f}$ is a reasonable approximation, at least as a first step, and especially considering the large theoretical uncertainty in the PBH accretion model. 

In what follows, we will therefore approximate $S^{(1)} \approx S^{(1) \rm d} = \delta_e * \mathcal{C}^{(0)}$, and no longer indicate that we use the direct-term only by a label ``d".

\subsection{Alternative calculation of  \texorpdfstring{$C_{\ell, \rm hom}^{(1) \rm d}$}{Cl1hom}}

Before moving to the full calculation of $\Theta^{(1)}$, including ionization fraction inhomogeneities, we present an alternative calculation of $\Delta_{\ell, \rm hom}^{(1) \rm d}(k)$, required for the cross-power spectrum $C_{\ell, \rm hom}^{(1)}$. This approach relies on intermediate quantities also used for the trispectrum calculation, and provides a useful cross check of our numerical methods.

For any quantity $X(k, \hn \cdot \hk)$, we define its Legendre multipole moments $X_\ell(k)$ as usual through 
\begin{align}
X_\ell(k) \equiv \frac{i^{\ell}}2 \int_{-1}^1 d \mu ~ P_\ell(\mu) X(k, \mu),
\end{align}
such that
\begin{align}
X(k, \hk \cdot \hn) &= \sum_\ell (-i)^{\ell} (2 \ell +1) X_\ell(k) P_\ell(\hn \cdot \hk) \\
&= 4 \pi \sum_{\ell m} (-i)^{\ell} X_\ell(k) Y_{\ell m}(\hk) Y_{\ell m}^*(\hn). \label{eq:multipole_def}
\end{align}
We denote by $\widetilde{\mathcal{C}}^{(0)}(\eta, k, \hat{k} \cdot \hn)$ the transfer function of $\mathcal{C}^{(0)}(\eta, \bk, \hn)$ (i.e.~such that $\mathcal{C}^{(0)} = \widetilde{\mathcal{C}}^{(0)} \zeta$), and define $\mathcal{J}(\eta, k, \hn \cdot \hk) \equiv e^{- i \chi \bk \cdot \hn} \widetilde{\mathcal{C}}^{(0)}(\eta, k, \hat{k} \cdot \hn)$. 

Substituting $S^{(1)}(\bk, \eta, \hn) e^{- i \bk \cdot \hn \chi} = \overline{\delta}_e \mathcal{J}(\eta, k, \hn \cdot \hk) \zeta(\bk)$ and inserting the spherical-harmonic expansion of $\mathcal{J}$ into Eq.~\eqref{eq:los_source}, we then arrive again at Eq.~\eqref{eq:Theta1_lm_hom}, with 
\begin{align}
  \Delta_{\ell, \rm hom}^{(1) \rm d}(k) \equiv \int_0^{\eta_0} d \eta ~ g(\eta)  ~\overline{\delta}_e(\eta) \mathcal{J}_\ell(\eta, k).\label{eq:Delta1_lm_hom_alternative}
\end{align}
Using the plane-wave expansion \eqref{eq:plane-wave} and the Legendre expansion of the product of two Legendre polynomials, we may relate the coefficients $\mathcal{J}_\ell$ to the Legendre coefficients of $\widetilde{\mathcal{C}}^{(0)}$ as follows:
\begin{equation}
\mathcal{J}_\ell(\eta, k) = \frac{4 \pi}{2 \ell +1} \sum_{\ell_1 \ell_2} i^{\ell - \ell_1 - \ell_2} (g_{\ell_1 \ell_2 \ell})^2 j_{\ell_1}(k \chi)\widetilde{\mathcal{C}}^{(0)}_{\ell_2}(\eta, k), \label{eq:mathcalJ_main}
\end{equation}
where $g_{\ell_1 \ell_2 \ell}$ is proportional to a three-J symbol, and is defined in Eq.~\eqref{eq:g_sym}. Since this coefficient is nonvanishing only if $\ell_1 + \ell_2 + \ell$ is even, we may substitute $i^{\ell - \ell_1 - \ell_2} = (-1)^{(\ell -\ell_1 - \ell_2)/2} = (-1)^{(\ell + 3 \ell_1 + 3 \ell_2)/2}$. The Legendre coefficients of the collision operator are given explicitly by
\begin{align}\label{eq:collision_ell}
   \widetilde{\mathcal{C}}^{(0)}_{\ell}=&\frac13 \widetilde{v}_{b \gamma}^{(0)}\delta_{\ell 1} + \frac15 \widetilde{\Pi}^{(0)} \delta_{\ell 2} -\widetilde{\Theta}_{\ell}^{(0)}( 1- \delta_{\ell 0} - \delta_{\ell 1}).
\end{align}
The sums over $\ell_1$ and $\ell_2$ in Eq.~\eqref{eq:mathcalJ_main} are formally infinite, and must be truncated in practice. Since the higher $\ell_2$-modes from the collision term are induced after the peak of the visibility function, we choose to truncate the $\ell_2$ sum at some finite $\ell_{\rm cut}$. This automatically renders the double sum finite, since for a given $\ell_2$, $\ell_1$ is bounded by the triangle condition, $|\ell - \ell_2| \leq \ell_1 \leq \ell + \ell_2$.

\begin{figure}[htb]
\includegraphics[trim={0cm 1cm 0.5cm 1cm},width=.95\columnwidth]{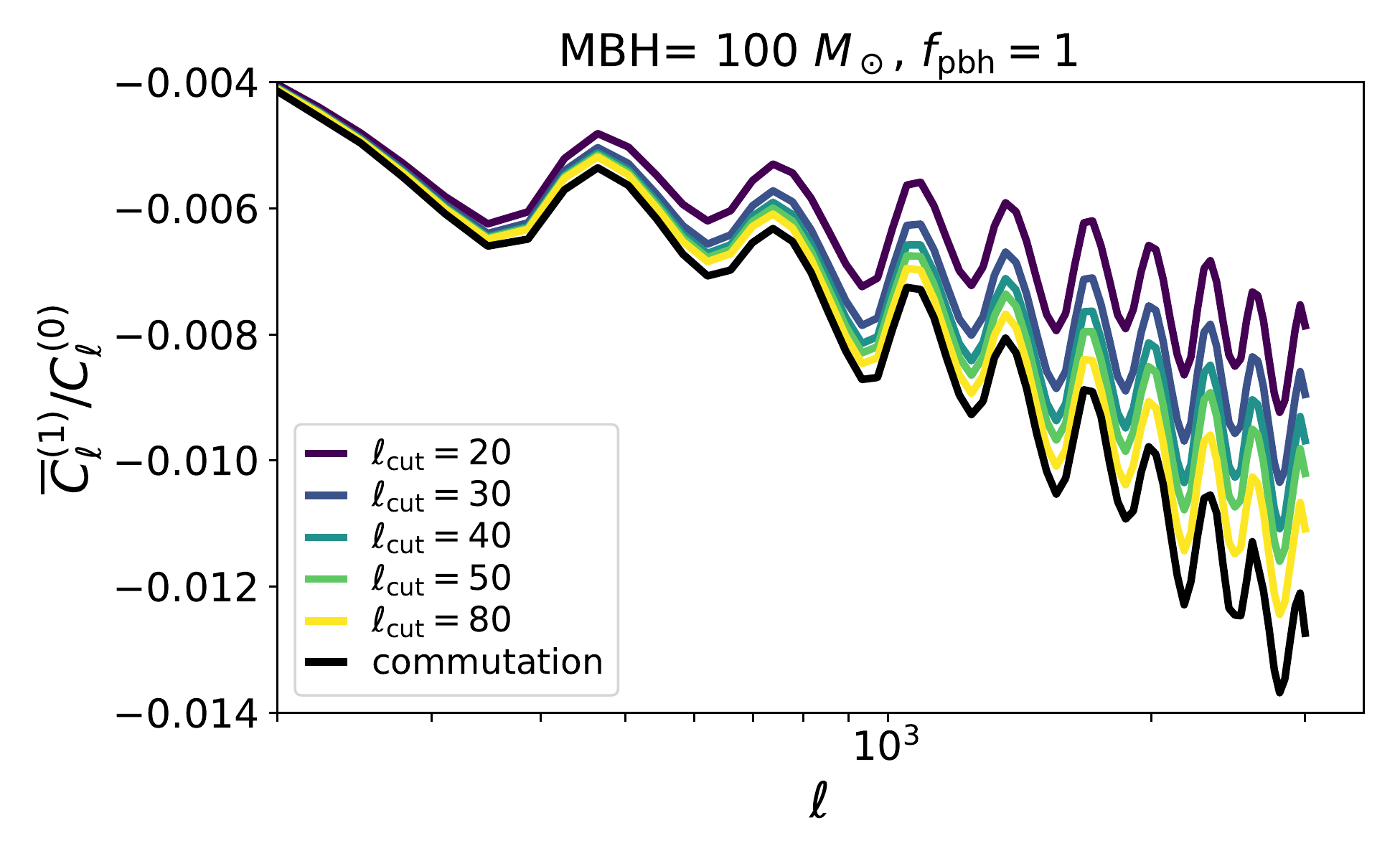}
\caption{\label{fig:homo_CTT} Fractional change to the temperature anisotropy power spectrum from the homogeneous perturbation to the free-electron fraction, $\overline{\delta_e}(\eta)$, due to 100 $M_\odot$ accreting PBHs, computed with the ``commutation" method, using Eq.~\eqref{eq:Delta1_lm_hom}, or with the ``direct summation method", using Eq.~\eqref{eq:Delta1_lm_hom_alternative}, where $\mathcal{J}_\ell$ is given by the sum \eqref{eq:mathcalJ_main}, truncated at $\ell_2 \leq \ell_{\rm cut}$. We see that the direct summation result converges to the ``commutation" result as $\ell_{\rm cut}$ is increased, as it should.}
\end{figure}

We compute $\Delta_{\ell, \rm hom}^{(1) d}$ as given by Eq.~\eqref{eq:Delta1_lm_hom_alternative} and use it to obtain $C_{\ell, \rm hom}^{(1)d}$ from Eq.~\eqref{eq:Cl1_hom}. We show the results in Fig.~\ref{fig:homo_CTT}, for various $\ell_{\rm cut}$, and compare them to the result obtained with the line-of-sight commutation method described in Sec.~\ref{sec:homo_de}. We see that the former converges to the latter as $\ell_{\rm cut}$ is increased, as it should, giving us confidence in the robustness of our numerical methods and results.

\section{Perturbed temperature anisotropy statistics due to inhomogeneously-accreting PBHs}\label{sec:tempstats}

\subsection{Temperature anisotropy transfer functions}\label{sec:transf}

\subsubsection{Definitions}

Neglecting lensing and other nonlinearities, the standard temperature perturbation is linearly related to the primordial curvature perturbation, through (c.f. Eq~\eqref{eq:thet0})
\begin{align}
    \Theta_{\ell m}^{(0)} &= \int Dk ~ T^{(0)}_{\ell m}(\bk) ~ \zeta(\bk),\\
    T_{\ell m}^{(0)}(\bk) &\equiv 4 \pi (-i)^{\ell} \Delta_\ell(k) Y_{\ell m}^*(\hat{k}). \label{eq:T0_lm}
\end{align}
Approximating the free-electron fraction perturbation due to accreting PBHs as quadratic in the initial conditions (c.f.~Eq.~\eqref{eq:Te-def}), the corresponding temperature anisotropy perturbation due to accreting PBHs is \emph{cubic} in the initial curvature perturbation. The goal of this section is to derive an explicit expression for the cubic transfer function $T_{\ell m}^{(1)}(\bk_1, \bk_2, \bk_3)$, defined through
\begin{align}
    \Theta_{\ell m, \rm inh}^{(1)} = \fbh \int D(k_1 k_2 k_3) T_{\ell m}^{(1)}(\bk_1, \bk_2, \bk_3)\nonumber\\
    \times \zeta(\bk_1) \zeta(\bk_2) \zeta(\bk_3), \label{eq:T(1)-def}
\end{align}
where the label ``inh" indicates that here we focus on the inhomogeneous-$\delta_e$ contribution to $\Theta^{(1)}$, recalling that it also has a piece $\Theta^{(1)}_{\rm hom}$ due to the homogeneous $\overline{\delta}_e$, which we computed in Sec.~\ref{sec:homo_de}, so that the total $\Theta^{(1)} = \Theta^{(1)}_{\rm hom} + \Theta^{(1)}_{\rm inh}$.

In addition, we shall derive the harmonic coefficients of this cubic transfer function, defined as 
\begin{align}\label{eq:T_multipoles}
T_{\ell m}^{(1)}(\bk_1, \bk_2, \bk_3) &= (4 \pi)^3 \sum_{\ell_1 \ell_2 \ell_3}(-i)^{\ell_1 + \ell_2 + \ell_3}\nonumber\\
&\times \sum_{m_1 m_2 m_3} T_{\ell_1 \ell_2\ell_3; \ell}^{m_1 m_2 m_3; m}(k_1, k_2, k_3)\nonumber\\
&\times Y_{\ell_1 m_1}(\hat{k}_1)  Y_{\ell_2 m_2}(\hat{k}_2) Y_{\ell_3 m_3}(\hat{k}_3). 
\end{align}

\subsubsection{Calculation}

Neglecting the ``feedback'' term, the source term for the line-of-sight solution of $\Theta^{(1)}_{\rm inh}$ is the convolution between the collision term and inhomogeneous part of the free-electron fraction, $S^{(1)} = \delta_{e, \rm inh} * \mathcal{C}^{(0)}$. Using Eqs.~\eqref{eq:Te-def} and \eqref{eq:dele_pbh}, we can write explicitly
\begin{align}
    S^{(1)}(\eta, \bk, \hat{n})& = \fbh \int ~D(k_1 k_2 k_3) \sld(\bk_1 + \bk_2 +\bk_3 - \bk) \nonumber\\
    &\quad\times (\hk_1 \cdot \hk_2) \Delta_e(\eta,k_1)\Delta_e(\eta, k_2)\nonumber\\
    &\quad \times \widetilde{\mathcal{C}}^{(0)}(\eta, k_3, \hat{k}_3 \cdot \hat{n}) \zeta(\bk_1) \zeta(\bk_2) \zeta(\bk_3),
\end{align}
where again $\widetilde{\mathcal{C}}^{(0)}(\eta, k, \hat{k} \cdot \hn)$ is the transfer function of $\mathcal{C}^{(0)}(\eta, \bk, \hn)$. Taking the harmonic transform of the line-of-sight solution for $\Theta^{(1)}$, Eq.~\eqref{eq:los_source}, we then find 
\begin{align}
    T_{\ell m}^{(1)}(\bk_1, \bk_2, \bk_3) = \int_0^{\eta_0} d \eta ~ g(\eta) \int d^2 \hn ~Y_{\ell m}^*(\hn) \nonumber\\
    \times (\hk_1 \cdot \hk_2) \Delta_e(\eta,k_1)\Delta_e(\eta, k_2)\nonumber\\
    \times \widetilde{\mathcal{C}}^{(0)}(\eta, k_3, \hat{k}_3 \cdot \hat{n}) e^{-i\chi\hat{n}\cdot(\bk_1 + \bk_2 + \bk_3)} . \label{eq:T1_lm}
\end{align}
Note that this function is symmetric under exchange of $\bk_1$ and $\bk_2$. Let us recall, also, that the expression above only holds for $\bk_1 + \bk_2 \neq 0$, and that $T_{\ell m}^{(1)}(\bk_1, - \bk_1, \bk_3) = 0$, since we are only considering the inhomogeneous part (with zero mean) of the free-electron fraction perturbation.

To obtain the harmonic coefficients of $T_{\ell m}^{(1)}$, we first rewrite  (denoting $\bs{\chi} \equiv \chi \hat{n}$),
\begin{align}
(\bk_1 \cdot \bk_2) e^{-i \bs{\chi} \cdot (\bk_1 + \bk_2)} &= -\left[\bs{\nabla}_{\bs{\chi}}  e^{-i \bs{\chi} \cdot \bk_1}\right] \cdot  \left[\bs{\nabla}_{\bs{\chi}}  e^{-i \bs{\chi} \cdot \bk_2}\right] \nonumber\\
&=  -\partial_\chi\left(e^{-i \bs{\chi} \cdot \bk_1}\right) \partial_\chi \left(e^{-i \bs{\chi} \cdot \bk_2}\right)\nonumber\\
& \quad -  \frac1{\chi^2} \left[\bs{\nabla}_{\hat{n}} e^{-i \bs{\chi} \cdot \bk_1}\right] \cdot  \left[\bs{\nabla}_{\hat{n}}  e^{-i \bs{\chi} \cdot \bk_2}\right],
\end{align}
where $\bs{\nabla}_{\bs{\chi}}$ is the gradient with respect to $\bs{\chi}$, which we have split into its radial part $\hat{n} \partial_\chi$ and its angular part $\frac1{\chi} \bs{\nabla}_{\hat{n}}$. Using the plane-wave expansion \eqref{eq:plane-wave}, we thus have
\begin{align}
(\hk_1 \cdot \hk_2 ) e^{-i \chi \hn \cdot (\bk_1 + \bk_2)} = -(4 \pi)^2 \sum_{\ell_1 \ell_2} (-i)^{\ell_1 + \ell_2} \nonumber\\
\sum_{m_1 m_2}Y_{\ell_1 m_1}(\hk_1) Y_{\ell_2 m_2}(\hk_2)\nonumber\\
\times \Big{[}j_{\ell_1}'(\chi k_1) j_{\ell_2}'(\chi k_2) Y_{\ell_1 m_1}^*(\hn) Y_{\ell_2 m_2}^*(\hn)\nonumber\\
+ \frac{j_{\ell_1}(\chi k_1)}{\chi k_1} \frac{j_{\ell_1}(\chi k_2)}{\chi k_2} \bs{\nabla}_{\hn} Y_{\ell_1 m_1}^*(\hn)  \cdot \bs{\nabla}_{\hn} Y_{\ell_2 m_2}^*(\hn) \Big{]}.
\end{align}
Combining this result with the Legendre-expansion of $e^{-i \chi \hn \cdot \bk_3} \widetilde{\mathcal{C}}^{(0)}(\eta, k_3, \hk_3 \cdot \hn)$ [Eq.~\eqref{eq:mathcalJ_main}], we are now in the position to compute $T_{\ell_1 \ell_2 \ell_3; \ell}^{m_1 m_2 m_3; m}$ defined in Eq.~\eqref{eq:T_multipoles}:
\begin{align}
   T_{\ell_1 \ell_2 \ell_3; \ell}^{m_1 m_2 m_3; m}(k_1, k_2, k_3) &= A_{\ell_1 \ell_2, \ell_3}(k_1, k_2, k_3) Q_{\ell_1 \ell_2 \ell_3 \ell}^{m_1 m_2 m_3 m} \nonumber\\
   &+ B_{\ell_1 \ell_2, \ell_3}(k_1, k_2, k_3) \widetilde{Q}_{\ell_1 \ell_2, \ell_3 \ell}^{m_1 m_2, m_3 m}, \label{eq:Tmult-final}
\end{align}
where the rotationally-invariant coefficients $A_{\ell_1 \ell_2, \ell_3}$ and $B_{\ell_1 \ell_2, \ell_3}$ are given by
\begin{align}\label{eq:A}
A_{\ell_1 \ell_2, \ell_3}(k_1, k_2, k_3) \equiv&-\int\!\! d \eta ~ g(\eta) j_{\ell_1}'(\chi k_1)\Delta_e (\eta, k_1)\nonumber\\
&\quad\times j_{\ell_2}'(\chi k_2)\Delta_e(\eta, k_2) \mathcal{J}_{\ell_3}(\eta, k_3), \\
B_{\ell_1 \ell_2, \ell_3}(k_1, k_2, k_3) \equiv&-\int \!\!d \eta ~ g(\eta) \frac{j_{\ell_1}(\chi k_1)}{\chi k_1}\Delta_e (\eta, k_1)\nonumber\\
&\quad\times \frac{j_{\ell_2}(\chi k_2)}{\chi k_2} \Delta_e(\eta, k_2) \mathcal{J}_{\ell_3}(\eta, k_3),\label{eq:B}
\end{align}
and the purely geometric terms $Q_{\ell_1 \ell_2 \ell_3 \ell_4}^{m_1 m_2 m_3 m_4}$ and $\widetilde{Q}_{\ell_1 \ell_2, \ell_3 \ell_4}^{m_1 m_2, m_3 m_4}$ are integrals of the product of four spherical harmonics or their gradients:
\begin{align}
&Q_{\ell_1 \ell_2 \ell_3\ell_4}^{m_1 m_2 m_3 m_4} \!\equiv\nonumber\\
&\quad\quad\!\int\! d^2 \hat{n} ~Y^*_{\ell_1 m_1}(\hat{n}) Y^*_{\ell_2 m_2}(\hat{n}) Y^*_{\ell_3 m_3}(\hat{n}) Y_{\ell_4 m_4}^*(\hat{n}), \label{eq:Q_sym}\\
&\widetilde{Q}_{\ell_1 \ell_2, \ell_3 \ell_4}^{m_1 m_2, m_3 m_4} \equiv\nonumber\\
&\quad\quad\int d^2 \hat{n} ~\bs{\nabla}_{\hat{n}}Y^*_{\ell_1 m_1}(\hat{n})\cdot \bs{\nabla}_{\hat{n}} Y^*_{\ell_2 m_2}(\hat{n}) Y^*_{\ell_3 m_3}(\hat{n}) Y_{\ell_4 m_4}^*(\hat{n}). \label{eq:Qt_sym}
\end{align}
Note that we have separated the groups of indices on which the functions depend fully symmetrically: $A_{\ell_1 \ell_2, \ell_3}(k_1, k_2, k_3)$ and $B_{\ell_1 \ell_2, \ell_3}(k_1, k_2, k_3)$ are symmetric under exchange of $(\ell_1, k_1)$ with $(\ell_2, k_2)$, $Q_{\ell_1 \ell_2 \ell_3\ell_4}^{m_1 m_2 m_3 m_4}$ is symmetric under exchange of any two $(\ell, m)$ pairs, and $\widetilde{Q}_{\ell_1 \ell_2, \ell_3\ell_4}^{m_1 m_2, m_3 m_4}$ is symmetric under exchange of $(\ell_1, m_1)$ with $(\ell_2, m_2)$, as well as under exchange of $(\ell_3, m_3)$ with $(\ell_4, m_4)$.

\subsection{Perturbed temperature angular power spectrum}\label{sec:powerspec}
We have derived all the required transfer functions and are now equipped to compute statistical properties of $\Theta^{(1)}_{\rm inh}$. Because the perturbed temperature anisotropy is cubic in the primordial curvature perturbation, it has a non-vanishing cross-correlation with the standard temperature anisotropy. Using Eqs.~\eqref{eq:thet0} and \eqref{eq:T(1)-def}, we have
\begin{align}
    \langle \Theta_{\ell m, \rm inh}^{(1)} \Theta_{\ell' m'}^{*(0)} \rangle &= \fbh \int\!\! D(k_1 k_2 k_3 k') ~   T^{(1)}_{\ell m}(\bk_1, \bk_2, \bk_3) \nonumber\\
    &\quad\times T^{*(0)}_{\ell' m'}(\bk') \langle\zeta(\bk_1) \zeta(\bk_2) \zeta(\bk_3) \zeta^*(\bk') \rangle. 
    \end{align}
Using Wick's theorem, and recalling that $T^{(1)}_{\ell m}(\bk_1, - \bk_1, \bk_3) = 0$, and that $T^{(1)}_{\ell m}$ is symmetric in its first two arguments, we get
\begin{align}
   \langle \Theta_{\ell m, \rm inh}^{(1)} \Theta_{\ell' m'}^{*(0)} \rangle =& 2 \fbh  \int D(k k')   T^{(1)}_{\ell m}(\bk', -\bk, \bk) \nonumber\\
&\quad\times T^{*(0)}_{\ell' m'}(\bk') P_{\zeta}(k) P(k'), \label{eq:Cl1_inh_int}
\end{align}
From Eq.~\eqref{eq:T1_lm}, we have
\begin{align}
    T^{(1)}_{\ell m}(\bk', -\bk, \bk) = -(\hk' \cdot \hk) \int_0^{\eta_0} d \eta ~g(\eta) \Delta_e(\eta, k') \Delta_e(\eta, k) \nonumber\\
    \times \int d^2 \hn ~Y_{\ell m}^*(\hn) \widetilde{\mathcal{C}}^{(0)}(\eta, k, \hk \cdot \hn) e^{-i \chi \hn \cdot \bk'}.  
\end{align}
Averaging over the direction $\hk$, we then obtain,
\begin{align}
   \int \frac{d^2 \hk}{4 \pi} T^{(1)}_{\ell m}(\bk', -\bk, \bk) \nonumber\\
   = i \int_0^{\eta_0} d \eta ~g(\eta) \Delta_e(\eta, k') \Delta_e(\eta, k)  \widetilde{\mathcal{C}}^{(0)}_1(\eta, k)\nonumber\\
   \times \int d^2 \hn ~Y_{\ell m}^*(\hn) (\hk' \cdot \hn) e^{-i \chi \hn \cdot \bk'},
\end{align}
where $\widetilde{\mathcal{C}}^{(0)}_1(\eta, k) \equiv i \frac12 \int_{-1}^1 d\mu P_1(\mu) \widetilde{\mathcal{C}}^{(0)}(\eta, k, \mu)$ is the order-1 Legendre coefficient of $\widetilde{\mathcal{C}}^{(0)}$, which is proportional to the baryon-photon relative velocity (or baryon-photon slip):
\begin{align}
    \widetilde{\mathcal{C}}^{(0)}_1(\eta, k) = \frac13 \widetilde{v}_{b \gamma}(\eta, k),
\end{align}
where we defined $\bs{v}_{b \gamma}(\bk) \equiv (\bs{v}_b - \bs{v}_\gamma)(\bk) = -i \hk \widetilde{v}_{b \gamma}(k) \zeta(\bk)$.

Using the plane-wave expansion Eq.~\eqref{eq:plane-wave}, this expression further simplifies to
\begin{align}
    \int \frac{d^2 \hk}{4 \pi} T^{(1)}_{\ell m}(\bk', -\bk, \bk) \nonumber\\
   = -\frac{4 \pi}3 (-i)^{\ell} \int_0^{\eta_0} d \eta ~g(\eta) \Delta_e(\eta, k') \Delta_e(\eta, k)\nonumber\\
   \times \widetilde{v}_{b \gamma}(\eta, k) j_{\ell}'(k' \chi) Y_{\ell m}^*(\hk'). \label{eq:T1kk-av}
\end{align}
Finally inserting Eqs.~\eqref{eq:T0_lm} and \eqref{eq:T1kk-av} into Eq.~\eqref{eq:Cl1_inh_int}, we arrive at the following simple result
\begin{align}
\langle \Theta_{\ell m, \rm inh}^{(1)}\Theta_{\ell' m'}^{*(0)} \rangle = \frac12 \delta_{\ell \ell'} \delta_{m m'} C_{\ell, \rm inh}^{(1)},  
\end{align}
where the cross-power spectrum is given by the conformal time integral
\begin{align}
  C_{\ell, \rm inh}^{(1)} &= - \frac{16 \pi}3 \fbh \int_0^{\eta} d\eta ~g(\eta) \gamma(\eta) \mu_\ell(\eta) , \label{eq:C1_inh}
\end{align}
where we have defined
\begin{align}
 \gamma(\eta) &\equiv \int Dk ~ P_\zeta(k) \Delta_e(\eta, k) \widetilde{v}_{b \gamma}(\eta, k)\label{eq:gamma}, \\
 \mu_\ell(\eta) &\equiv \int Dk ~P_\zeta(k) \Delta_e(\eta, k) \Delta_\ell(k)  j_\ell'(k \chi) \label{eq:mu_l}.
\end{align}

We see that the factorization of the free-electron perturbation transfer function has allowed us to obtain a very simple expression for $C_{\ell, \rm inh}^{(1)}$: it only requires pre-computing tables of $\mu_\ell(\eta)$ and $\gamma(\eta)$, and then computing a one-dimensional integral. It can equivalently be rewritten in the same form as Eq.~\eqref{eq:Cl1_hom}:
\begin{eqnarray}
   C_{\ell, \rm inh}^{(1)} &=& 8\pi \int Dk ~P_\zeta(k) \Delta_\ell(k) \Delta_{\ell, \rm inh}^{(1)}(k), \label{eq:C1_inh_2}\\
   \Delta_{\ell, \rm inh}^{(1)}(k) &\equiv& - \frac23 f_{\rm pbh} \int_0^{\eta_0} d\eta ~g(\eta)\gamma(\eta) \Delta_e(k , \eta) j_{\ell}'(k \chi).~~~
\end{eqnarray}
Equation \eqref{eq:C1_inh}, or the equivalent form \eqref{eq:C1_inh_2}, constitute one of the main results of this work.

\begin{figure*}[ht]
\includegraphics[trim={0 0 0 0},width=\columnwidth]{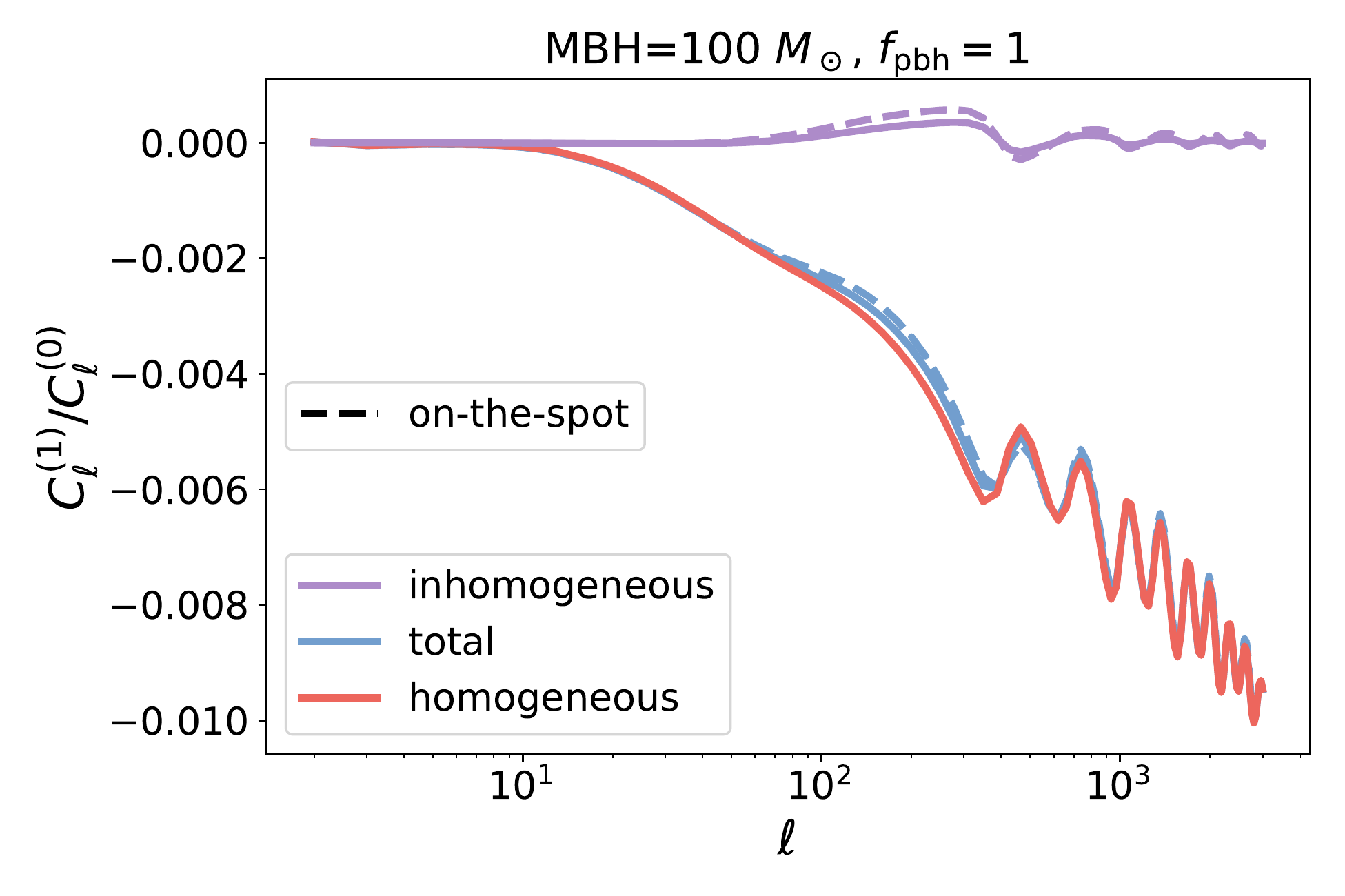}
\includegraphics[trim={0 0 0 0},width=\columnwidth]{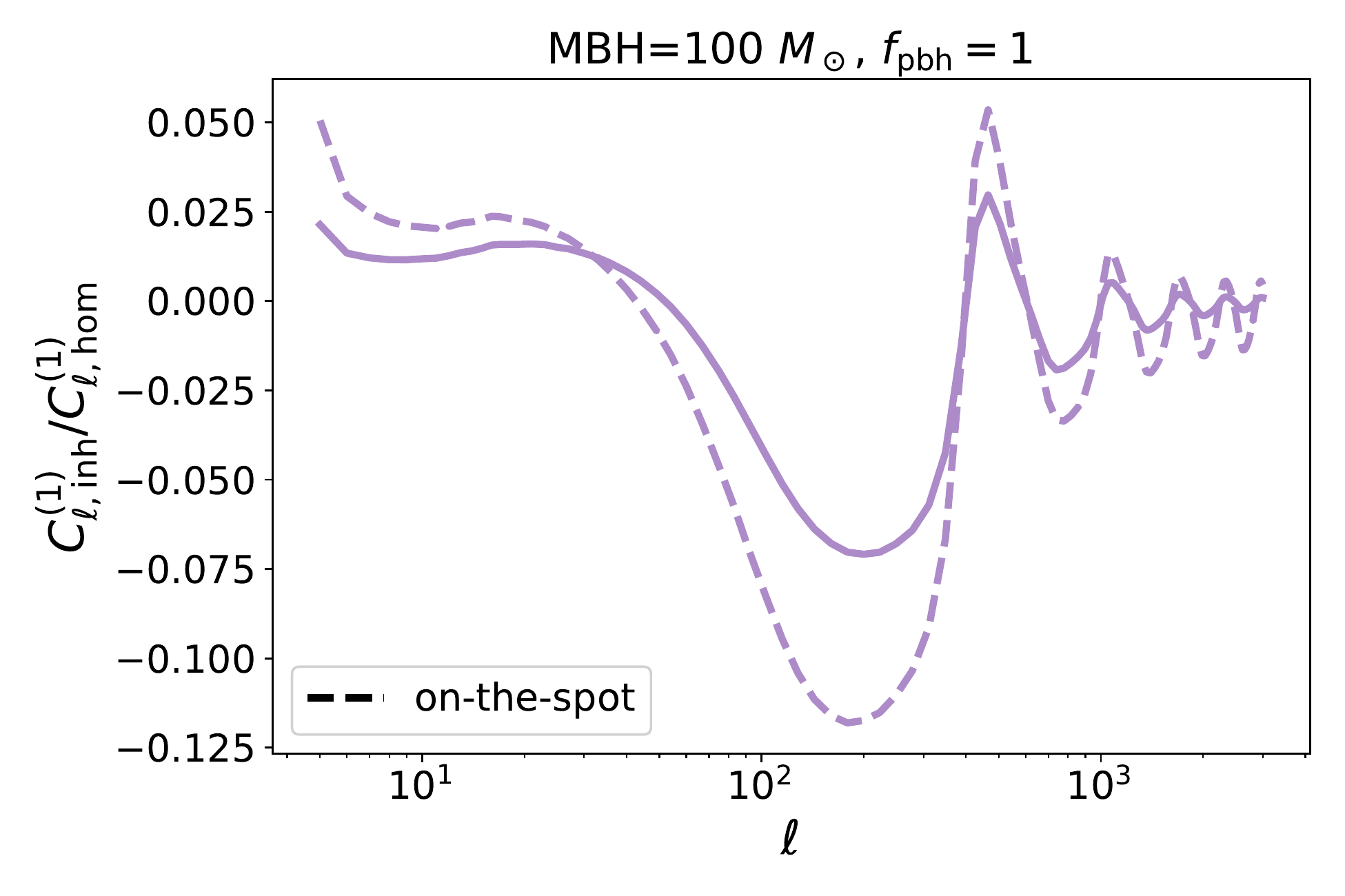}
\caption{\label{fig:AK17_jen} Fractional change to the temperature anisotropy power spectrum due to accreting PBHs of 100 $M_\odot$ comprising all the dark matter. \emph{Left:} Comparison of the contribution due to the inhomogeneous part of the ionization fraction perturbations, $C_{\ell, \rm inh}^{(1)}$, calculated for the first time in this work, with the one arising from the homogeneous part of the free-electron fraction, $C_{\ell, \rm hom}^{(1)}$, previously computed in AK17; we also overlay the total change to the temperature power spectrum from both. \emph{Right}: the ratio between $C_{\ell, \rm inh}^{(1)}$ and $C_{\ell, \rm hom}^{(1)}$. Although one would expect $C_{\ell, \rm inh}^{(1)}$ to be of the same order of magnitude as $C_{\ell, \rm hom}^{(1)}$ \emph{a priori}, we find in practice that the former is $\sim 10-100$ times smaller that the latter. In both cases, dashed curves correspond to the spatial on-the-spot approximation, which neglects the spatial smearing of energy deposition due to the finite propagation of injected photons.}
\end{figure*}

We computed $C_{\ell, \rm inh}^{(1)}$ from Eq.~\eqref{eq:C1_inh}, and checked that Eq.~\eqref{eq:C1_inh_2} gives the same result. We show the result in Fig.~\ref{fig:AK17_jen}, where we compare this term to its counterpart $C_{\ell, \rm hom}^{(1)}$ sourced by the homogeneous part of the free-electron fraction perturbation, for 100-$M_{\odot}$ accreting PBHs. Even though these two contributions should in principle be comparable, given that $\langle \delta_e^2\rangle^{1/2} \sim \overline{\delta}_e$ (see Paper I), we find that $C_{\ell, \rm inh}^{(1)}$ is suppressed by a factor $\sim 10-100$, depending on scale, relative to $C_{\ell, \rm hom}^{(1)}$ for all black hole masses. This turns out to be due to both a poor correlation between $\Theta^{(1)}_{\rm inh}$ and $\Theta^{(0)}$, and a suppression of the characteristic amplitude of $\Theta^{(1)}_{\rm inh}$ itself. We expound on this point in Appendix~\ref{app:auto}.

\subsection{Temperature trispectrum}\label{sec:trispec}

We now compute the connected four-point correlation function of temperature anisotropy,
\begin{align}
    \langle \Theta_1 \Theta_2 \Theta_3 \Theta_4 \rangle_c &\equiv \langle \Theta_1 \Theta_2 \Theta_3 \Theta_4 \rangle  -\langle\Theta_1\Theta_2\rangle\langle\Theta_3\Theta_4\rangle\nonumber\\
    &- \langle\Theta_1\Theta_3\rangle\langle\Theta_2\Theta_4\rangle - \langle\Theta_1\Theta_4\rangle\langle\Theta_2\Theta_3\rangle, \label{eq:4pt-c}
\end{align}
where the numbered subscripts index both $\ell$ and $m$, $\Theta_1\equiv\Theta_{\ell_1 m_1}$, and $c$ denotes subtracting out the unconnected parts of the trispectrum. Recalling that $\Theta = \Theta^{(0)} + \Theta^{(1)}_{\rm hom} + \Theta^{(1)}_{\rm inh}$, and that $\Theta^{(0)}$ and $\Theta^{(1)}_{\rm hom}$ are both linear in the initial Gaussian curvature perturbation, to lowest order in electron density perturbations, the trispectrum is given by
\begin{align}\label{eq:4pt}
    \langle \Theta_1 \Theta_2 \Theta_3 \Theta_4\rangle_c &= \langle \Theta_{1, \rm inh}^{(1)} \Theta_2^{(0)} \Theta_3^{(0)} \Theta_4^{(0)} \rangle_c \nonumber\\
    &+  \langle \Theta_1^{(0)} \Theta_{2, \rm inh}^{(1)} \Theta_3^{(0)} \Theta_4^{(0)} \rangle_c \nonumber\\
    &+  \langle \Theta_1^{(0)} \Theta_2^{(0)} \Theta_{3, \rm inh}^{(1)} \Theta_4^{(0)} \rangle_c \nonumber\\
    &+  \langle \Theta_1^{(0)} \Theta_2^{(0)} \Theta_3^{(0)} \Theta_{4, \rm inh}^{(1)} \rangle_c.
\end{align}
We may now compute each term using Eq.~\eqref{eq:T(1)-def} for $\Theta_{\ell m, \rm inh}^{(1)}$. For instance, the last term is
\begin{align}
    \langle \Theta_{1}^{(0)} \Theta_2^{(0)} \Theta_3^{(0)} \Theta_{4, \rm inh}^{(1)} \rangle_c = \fbh \int D(k k' k'') T_4^{(1)}(\bk, \bk', \bk'') \nonumber\\
    \times \Big{[}\langle \zeta(\bk) \zeta(\bk') \zeta(\bk'') \Theta^{(0)}_1 \Theta^{(0)}_2 \Theta^{(0)}_3 \rangle \nonumber\\
    - \langle \zeta(\bk) \zeta(\bk') \zeta(\bk'') \Theta^{(0)}_1\rangle \langle \Theta_2^{(0)} \Theta_3^{(0)} \rangle\nonumber\\
    - \langle \zeta(\bk) \zeta(\bk') \zeta(\bk'') \Theta^{(0)}_2\rangle \langle \Theta_1^{(0)} \Theta_3^{(0)} \rangle\nonumber\\
    - \langle \zeta(\bk) \zeta(\bk') \zeta(\bk'') \Theta^{(0)}_3\rangle \langle \Theta_1^{(0)} \Theta_2^{(0)} \rangle\Big{]},
\end{align}
where we used the explicit definition of the connected 4-point function, Eq.~\eqref{eq:4pt-c}. Using Wick's theorem to compute the 6-point and 4-point functions of Gaussian fields appearing in the integrand above, simplifying, and renaming dummy integration variables, we arrive at
\begin{align}
\langle \Theta_{1}^{(0)} \Theta_2^{(0)} \Theta_3^{(0)} \Theta_{4, \rm inh}^{(1)} \rangle_c  = \fbh \int D(k k' k'')  \nonumber\\
\times 
\langle \zeta(\bk) \Theta_1^{(0)}\rangle\langle \zeta(\bk') \Theta_2^{(0)}\rangle\langle \zeta(\bk'') \Theta_3^{(0)}\rangle \nonumber\\
\times \left[T_4^{(1)}(\bk, \bk', \bk'') + 5  \textrm{ perms.} \right]
\end{align}
where the 5 permutations involve all other possible permutations of $\bk, \bk', \bk''$. The relevant two-point functions are easily computed with the line-of-sight expression for $\Theta_{\ell m}^{(0)}$, Eq.~\eqref{eq:Theta_lm^0}, and we obtain
\begin{align}
   \langle \zeta(\bk) \Theta_{\ell m}^{(0)}\rangle = 4 \pi (-i)^\ell Y_{\ell m}^*(\hk) \Delta_\ell(k) P_\zeta(k). 
\end{align}
Integrating over the wavenumbers' directions, and using the harmonic decomposition of $T^{(1)}$ given in Eq.~\eqref{eq:T_multipoles}, we thus arrive at
\begin{align}
    \langle \Theta_{\ell_1 m_1}^{(0)} \Theta_{\ell_2 m_2}^{(0)} \Theta_{\ell_3 m_3}^{(0)} \Theta_{\ell_4 m_4, \rm inh}^{(1)} \rangle_c 
    = (4 \pi)^3 \fbh \int D(k_1 k_2 k_3) \nonumber\\
    \times P_\zeta(k_1) P_\zeta(k_2) P_\zeta(k_3) \Delta_{\ell_1}(k_1) \Delta_{\ell_2}(k_2) \Delta_{\ell_3}(k_3) \nonumber\\
   \times \Big{[} T_{\ell_1 \ell_2 \ell_3; \ell_4}^{m_1 m_2 m_3; m_4}(k_1, k_2, k_3) + 5 \textrm{~perms.} \Big{]},
\end{align}
where the 5 permutations involve all other possible permutations of $k_1, k_2, k_3$ simultaneously with the corresponding permutation of the indices $\ell_i, m_i, i = 1, 2, 3$, i.e.~such that the position of the index $\ell_i, m_i$ always corresponds to the position of $k_i$.

We now take advantage of the factorized form of $T_{\ell_1 \ell_2 \ell_3; \ell_4}^{m_1 m_2 m_3; m_4}(k_1, k_2, k_3)$, given in Eqs.~\eqref{eq:Tmult-final}-\eqref{eq:B}. In addition to the function $\mu_\ell(\eta)$ defined in Eq.~\eqref{eq:mu_l}, we define the following functions of time and multipole:
\begin{align}
    \nu_\ell(\eta) &\equiv \int Dk ~P_\zeta(k) \Delta_e(\chi, k) \Delta_\ell(k)  \frac{j_\ell(k \chi )}{k \chi}, \label{eq:nu_l} \\
\lambda_\ell(\eta) &\equiv \int Dk ~P_\zeta(k) \Delta_\ell(k) \mathcal{J}_\ell(\eta, k). \label{eq:lambda_l}
\end{align}
We then define the following one-dimensional integrals:
\begin{align}
    \mathcal{A}_{\ell_1 \ell_2, \ell_3} \equiv&- 2 (4 \pi)^3 \int d\eta ~g(\eta)~ \mu_{\ell_1}(\eta) \mu_{\ell_2}(\eta) \lambda_{\ell_3}(\eta),\\
\mathcal{B}_{\ell_1 \ell_2, \ell_3} \equiv& -2 (4 \pi)^3 \int d\eta ~g(\eta)~ \nu_{\ell_1}(\eta) \nu_{\ell_2}(\eta) \lambda_{\ell_3}(\eta),
\end{align}
which are symmetric in their first two arguments. We then find, using the symmetry of $T^{(1)}$ in its first two arguments, and the symmetries of the $Q$ and $\widetilde{Q}$ symbols defined in Eqs.~\eqref{eq:Q_sym}, \eqref{eq:Qt_sym}:
\begin{align}
    &\frac1{\fbh}\langle \Theta_{\ell_1 m_1}^{(0)} \Theta_{\ell_2 m_2}^{(0)} \Theta_{\ell_3 m_3}^{(0)} \Theta_{\ell_4 m_4, \rm inh}^{(1)} \rangle_c \nonumber\\
    &= \mathcal{A}_{(\ell_1 \ell_2 \ell_3)} Q_{\ell_1 \ell_2 \ell_3 \ell_4}^{m_1 m_2 m_3 m_4} +\mathcal{B}_{\ell_1\ell_2,\ell_3}\widetilde{Q}_{ \ell_1 \ell_2,\ell_3\ell_4}^{m_1 m_2, m_3 m_4} \nonumber\\
    &+ \mathcal{B}_{\ell_2\ell_3,\ell_1}\widetilde{Q}_{ \ell_2 \ell_3,\ell_1\ell_4}^{m_2 m_3, m_1 m_4}+\mathcal{B}_{\ell_3\ell_1,\ell_2}\widetilde{Q}_{ \ell_3 \ell_1,\ell_2\ell_4}^{m_3 m_1, m_2 m_4},
\end{align}
where\footnote{Note that we do not use the standard symmetrization notation, i.e.~do not divide by the number of terms, in order to avoid the proliferation of numerical prefactors.} 
\begin{align}
   \mathcal{A}_{(\ell_1 \ell_2 \ell_3)} \equiv \mathcal{A}_{\ell_1 \ell_2, \ell_3} + \mathcal{A}_{\ell_2 \ell_3, \ell_1} + \mathcal{A}_{\ell_3 \ell_1, \ell_2}.
\end{align}

Finally, summing over the four permutations in Eq.~\eqref{eq:4pt}, we arrive at the main result of this work, which is the temperature trispectrum sourced by accreting PBHs:
\begin{align}
    &\langle \Theta_{\ell_1 m_1} \Theta_{\ell_2 m_2} \Theta_{\ell_3 m_3} \Theta_{\ell_4 m_4}\rangle_c = \fbh \left(\mathcal{T}_{\rm pbh}\right)_{\ell_1 \ell_2 \ell_3 \ell_4}^{m_1 m_2 m_3 m_4}, \label{eq:Tpbh-general}\\
    &\left(\mathcal{T}_{\rm pbh}\right)_{\ell_1 \ell_2 \ell_3 \ell_4}^{m_1 m_2 m_3 m_4} \equiv  
    \mathcal{A}_{(\ell_1 \ell_2 \ell_3 \ell_4)} Q_{\ell_1 \ell_2 \ell_3 \ell_4}^{m_1 m_2 m_3 m_4}\nonumber\\
&+ \mathcal{B}_{\ell_1 \ell_2, (\ell_3 \ell_4)} \widetilde{Q}_{\ell_1 \ell_2, \ell_3 \ell_4}^{m_1 m_2, m_3 m_4} + \mathcal{B}_{\ell_1 \ell_3, (\ell_2 \ell_4)}\widetilde{Q}_{\ell_1 \ell_3, \ell_2 \ell_4}^{m_1 m_3, m_2 m_4} \nonumber\\
&+ \mathcal{B}_{\ell_1 \ell_4, (\ell_2 \ell_3)}\widetilde{Q}_{\ell_1 \ell_4, \ell_2 \ell_3}^{m_1 m_4, m_2 m_3}+ \mathcal{B}_{\ell_2 \ell_3, (\ell_1 \ell_4)} \widetilde{Q}_{\ell_2 \ell_3, \ell_1 \ell_4}^{m_2 m_3, m_1 m_4} \nonumber\\
 & + \mathcal{B}_{\ell_2 \ell_4, (\ell_1 \ell_3)} \widetilde{Q}_{\ell_2 \ell_4, \ell_1 \ell_3}^{m_2 m_4, m_1 m_3}+ \mathcal{B}_{\ell_3 \ell_4, (\ell_1 \ell_2)} \widetilde{Q}_{\ell_3 \ell_4, \ell_1 \ell_2}^{m_3 m_4, m_1 m_2}, \label{eq:Trispec-final}
\end{align}
where we have defined the symmetrized coefficients
\begin{align}
    \mathcal{A}_{(\ell_1 \ell_2 \ell_3 \ell_4)}&\equiv \mathcal{A}_{(\ell_1 \ell_2 \ell_3)}+ \mathcal{A}_{(\ell_2 \ell_3 \ell_4)}\nonumber\\
    &\quad+ \mathcal{A}_{(\ell_3 \ell_4 \ell_1)}+  \mathcal{A}_{(\ell_4 \ell_1 \ell_2)},\\
    \mathcal{B}_{\ell_1 \ell_2, (\ell_3 \ell_4)} &\equiv  \mathcal{B}_{\ell_1 \ell_2, \ell_3} + \mathcal{B}_{\ell_1 \ell_2, \ell_4}.
\end{align}

\section{Trispectrum constraints and sensitivity forecasts}\label{sec:forecast}

In this section we compute and present trispectrum constraints and sensitivity forecasts on the fraction of dark matter made of PBHs, $f_{\rm pbh}$. A full trispectrum analysis of the \emph{Planck} satellite temperature data would be very challenging and is well beyond the scope of this work. Instead, we compute the overlap of the PBH-induced trispectrum with the local-type primordial non-Gaussianity (PNG) trispectrum template, in order to extract an indirect limit on the PBH abundance, given Planck's limits on $g_{\rm NL}^{\rm loc}$ \citep{planck20c}. In addition, we forecast Planck's sensitivity to the trispectrum induced by accreting PBHs. For the scope of this paper, we ignore biases that may arise due to lensing or other nonlinear effects, but they should of course be accounted for in a full data analysis.

\subsection{General equations}

Given that the trispectrum induced by accreting PBHs is approximately linear in $f_{\rm pbh}$, as given by Eq.~\eqref{eq:Tpbh-general}, one can build an optimal quartic estimator $\widehat{f}_{\rm pbh}$ for $f_{\rm pbh}$ \cite{smith15a, regan10a}. Its precise expression will not be needed here, and is given in Eq.~(24) of Ref.~\cite{smith15a}. The inverse variance of this estimator is given by Eq.~(25) in Ref.~\cite{smith15a}. Approximating the noise covariance matrix as diagonal in $\ell$, the variance of the estimator is given by
\begin{align}
\sigma_{f_{\rm pbh}}^2 = \left \langle \mathcal{T}_{\rm pbh} \cdot  \mathcal{T}_{\rm pbh} \right \rangle^{-1},  \label{eq:var_f_pbh}
\end{align}
where for any two trispectra $\mathcal{T}_A, \mathcal{T}_B$, we define their inverse-noise weighted dot product as
\begin{align}
\left \langle \mathcal{T}_A \cdot  \mathcal{T}_B \right  \rangle & \equiv\frac{f_{\rm sky}}{4!} \sum_{\ell's}\frac{1}{C'_{\ell_1} C'_{\ell_2} C'_{\ell_3} C'_{\ell_4}}\nonumber\\
&\times\sum_{m's}  \left(\mathcal{T}_A\right)_{\ell_1 \ell_2 \ell_3 \ell_4}^{m_1 m_2 m_3 m_4} \left(\mathcal{T}_B\right)_{\ell_1 \ell_2 \ell_3 \ell_4}^{m_1 m_2 m_3 m_4}, \label{eq:dotprod}
\end{align}
where $C_{\ell}' \equiv C_\ell + N_\ell$ is the total variance of the observed CMB temperature, including both the cosmological signal $C_\ell$ and instrumental noise $N_\ell$, $f_{\rm sky}$ is the fraction of the sky covered by the experiment, and the sums carry over all four indices. 

Primordial non-Gaussianity also generates a CMB temperature trispectrum, proportional to a non-Gaussianity parameter $g_{\rm NL}$:
\beq
\langle \Theta_{\ell_1 m_1} \Theta_{\ell_2 m_2} \Theta_{\ell_3 m_3} \Theta_{\ell_4 m_4} \rangle_c = g_{\rm NL} ~\left(\mathcal{T}_{\rm png}\right)_{\ell_1 \ell_2 \ell_3 \ell_4}^{m_1 m_2 m_3 m_4}.
\eeq
One can build an optimal estimator $\widehat{g}_{\rm NL}$ for $g_{\rm NL}$ in the same way as $f_{\rm pbh}$. The non-Gaussianity sourced by inhomogeneously accreting PBHs would lead to a systematic bias in this estimator, even in the absence of primordial non-Gaussianity. This bias is linear in $f_{\rm pbh}$:
\begin{align}
\langle \Delta \widehat{g}_{\rm NL}\rangle_{\rm pbh} &= f_{\rm pbh}~ \mathcal{R}, \label{eq:bias}\\
\mathcal{R} &\equiv \sigma_{g_{\rm NL}}^2 \left \langle \mathcal{T}_{\rm pbh} \cdot \mathcal{T}_{\rm png} \right \rangle, 
\end{align}
where $\sigma_{g_{\rm NL}}^2$ is the variance of the quadratic estimator $\widehat{g}_{\rm NL}$, given by
\begin{align}
\sigma_{g_{\rm NL}}^2&\equiv \left \langle \mathcal{T}_{\rm png}  \cdot \mathcal{T}_{\rm png} \right \rangle^{-1}. \label{eq:F_png}
\end{align}
Constraints on the amplitude $g_{\rm NL}$ of primordial non-Gaussianity therefore directly translate into bounds on the PBH abundance $f_{\rm pbh}$. In what follows we will specifically consider the local-type primordial trispectrum, which is most tightly constrained by CMB anisotropy observations, and whose shape is given in \citep{smith15a},
\begin{align}
\left(\mathcal{T}^{\rm loc}_{\rm png}\right)_{\ell_1 \ell_2 \ell_3 \ell_4}^{m_1 m_2 m_3 m_4} &= \mathcal{C}_{(\ell_1 \ell_2 \ell_3 \ell_4)}Q_{\ell_1 \ell_2 \ell_3 \ell_4}^{m_1 m_2 m_3 m_4}, \\
\mathcal{C}_{(\ell_1 \ell_2 \ell_3 \ell_4)} &\equiv \mathcal{C}_{\ell_1 \ell_2 \ell_3, \ell_4} + 3 ~ \textrm{perm}, \\
\mathcal{C}_{\ell_1 \ell_2 \ell_3, \ell_4} &\equiv 6 \int r^2 dr ~ \beta_{\ell_1}(r)\beta_{\ell_2}(r)\beta_{\ell_3}(r) \alpha_{\ell_4}(r),
\end{align}
where $Q_{\ell_1 \ell_2 \ell_3 \ell_4}^{m_1 m_2 m_3 m_4}$ is given by Eq.~\eqref{eq:Q_sym} and we used the standard notation of Refs.~\cite{smith15a, komatsu05a}:
\begin{align}
\alpha_{\ell}(r) &\equiv \frac53 (4 \pi) \int Dk~  \Delta_\ell(k) j_{\ell}(k r) ,\\
\beta_{\ell}(r) &\equiv \frac35 (4 \pi) \int Dk~  \Delta_\ell(k) j_{\ell}(k r) P_{\zeta}(k).
\end{align}

\subsection{Sums over  \texorpdfstring{$m$}{m}'s}

Before proceeding with the numerical evaluation of Eqs.~\eqref{eq:var_f_pbh} and \eqref{eq:bias}, we first simplify the sums over $m$'s, which involve purely geometric quantities. Specifically, we define
\begin{align} \label{eq:Qstart}
    (\mathcal{Q}^2)_{\ell_1 \ell_2 \ell_3 \ell_4} \equiv& \sum_{m's} \left(Q_{\ell_1 \ell_2 \ell_3 \ell_4}^{m_1 m_2 m_3 m_4}\right)^2,\\
    (\mathcal{Q \widetilde{Q}})_{\ell_1 \ell_2, \ell_3 \ell_4} \equiv& \sum_{m's} {Q^*}_{\ell_1 \ell_2 \ell_3 \ell_4}^{m_1 m_2 m_3 m_4} \widetilde{Q}_{\ell_1 \ell_2, \ell_3 \ell_4}^{m_1 m_2, m_3 m_4}, \\
 (\mathcal{\widetilde{Q}}^2)_{\ell_1 \ell_2, \ell_3 \ell_4} \equiv& \sum_{m's} \left( \widetilde{Q}_{\ell_1 \ell_2, \ell_3 \ell_4}^{m_1 m_2, m_3 m_4}\right)^2, \\
 (\mathcal{\widetilde{Q} \widetilde{Q}}^{\rm T})_{\ell_1 \ell_2, \ell_3 \ell_4} \equiv& \sum_{m's} \widetilde{Q^*}_{\ell_1 \ell_2, \ell_3 \ell_4}^{m_1 m_2, m_3 m_4} \widetilde{Q}_{\ell_3 \ell_4, \ell_1 \ell_2}^{m_3 m_4, m_1 m_2},\\ (\mathcal{\widetilde{Q} \widetilde{Q}}^{\rm S})_{\ell_1, \ell_2 \ell_3, \ell_4} \equiv& \sum_{m's} \widetilde{Q^*}_{\ell_1 \ell_2, \ell_3 \ell_4}^{m_1 m_2, m_3 m_4} \widetilde{Q}_{\ell_1 \ell_3, \ell_2 \ell_4}^{m_1 m_3, m_2 m_4},\label{eq:Qend}
\end{align}
where ``T" stands for transpose and ``S" for ``scrambled". The same symmetry rules apply where each set of indices divided by or surrounded by commas are symmetric. We simplify these quantities in Appendix~\ref{app:Q-sums}, where we reduce them to a single sum of products of 3-J symbols. 

Inserting Eq.~\eqref{eq:Trispec-final} into Eq.~\eqref{eq:var_f_pbh}, carrying out the sums over $m$'s, and simplifying, the inverse variance of the estimator $\widehat{f}_{\rm pbh}$ becomes
\begin{align}\label{eq:inv_fpbh}
    \left(\sigma_{f_{\rm pbh}}^2\right)^{-1}&=\frac{f_{\rm sky}}{4!}\sum_{\ell's} \frac1{C'_{\ell_1}C'_{\ell_2} C'_{\ell_3} C'_{\ell_4}} \nonumber\\
    &\times \Big{[}~~~~~\left(\mathcal{A}_{(\ell_1 \ell_2 \ell_3 \ell_4)}\right)^2 (\mathcal{Q}^2)_{\ell_1 \ell_2 \ell_3 \ell_4}\nonumber\\  
    &~~~~ + 12 ~\mathcal{A}_{(\ell_1 \ell_2 \ell_3 \ell_4)}\mathcal{B}_{\ell_1 \ell_2, (\ell_3 \ell_4)} (\mathcal{Q \widetilde{Q}})_{\ell_1 \ell_2, \ell_3 \ell_4} \nonumber\\
    & ~~~~ + 6 ~\left(\mathcal{B}_{\ell_1 \ell_2, (\ell_3 \ell_4)}\right)^2 (\mathcal{\widetilde{Q}}^2)_{\ell_1 \ell_2, \ell_3 \ell_4}  \nonumber\\
    &~~~~ +6~\mathcal{B}_{\ell_1 \ell_2, (\ell_3 \ell_4)}\mathcal{B}_{\ell_3 \ell_4, (\ell_1 \ell_2)} (\mathcal{\widetilde{Q} \widetilde{Q}}^{\rm T})_{\ell_1 \ell_2, \ell_3 \ell_4} \nonumber\\
&~~~~ + 24~\mathcal{B}_{\ell_1 \ell_2, (\ell_3 \ell_4)}\mathcal{B}_{\ell_1 \ell_3, (\ell_2 \ell_4)}(\mathcal{\widetilde{Q} \widetilde{Q}}^{\rm S})_{\ell_1, \ell_2 \ell_3, \ell_4} \Big{]}.
\end{align}
Similarly, the bias on local-type non-Gaussianity due to accreting PBHs simplifies to
\begin{align}
    \langle \Delta \widehat{g}^{\rm loc}_{\rm NL}\rangle_{\rm pbh}&= f_{\rm pbh}\times \frac{f_{\rm sky}}{4!} \sigma_{g_{\rm NL}^{\rm loc}}^2 \sum_{\ell's} \frac{\mathcal{C}_{(\ell_1 \ell_2 \ell_3 \ell_4)} }{C'_{\ell_1} C'_{\ell_2} C'_{\ell_3} C'_{\ell_4}}\nonumber\\
    &\quad\times\Big{[}~~~~\mathcal{A}_{(\ell_1 \ell_2 \ell_3 \ell_4)} (\mathcal{Q}^2)_{\ell_1 \ell_2 \ell_3 \ell_4}  \nonumber\\
    &\quad\quad + 6~\mathcal{B}_{\ell_1 \ell_2, (\ell_3 \ell_4)} (\mathcal{Q \widetilde{Q}})_{\ell_1 \ell_2, \ell_3 \ell_4}  \Big{]},
\end{align}
where the inverse variance of $\widehat{g}_{\rm NL}^{\rm loc}$ is given by
\begin{align}\label{eq:FPNG}
\left(\sigma_{g_{\rm NL}^{\rm loc}}^2\right)^{-1} &= \frac{f_{\rm sky}}{4!} \sum_{\ell's} \frac{\left(\mathcal{C}_{(\ell_1 \ell_2 \ell_3 \ell_4)}\right)^2}{C'_{\ell_1} C'_{\ell_2} C'_{\ell_3} C'_{\ell_4}} (\mathcal{Q}^2)_{\ell_1 \ell_2 \ell_3 \ell_4}.
\end{align}

\subsection{Application to Planck data}

We now apply the above results to the Planck experiment \citep{planck20a,planck20b,planck20c}. The relevant fraction of sky coverage is $f_{\rm sky}=0.78$ \citep{planck20c}, and the instrumental noise $N_\ell$ is obtained from combining the noises of the 100, 143 and 217 GHz frequency channels, 
\begin{align}
    N_\ell=\left[\sum_c N^{-1}_{\ell,c}\right]^{-1},
\end{align}
where, for each channel $c$, the noise is modelled as a Gaussian with variance per pixel $\sigma^2_c$ and beam size $\theta_{{\rm FWHM},c}$:
\begin{align}
    N_{\ell,c}=\left(\frac{\sigma_c~\theta_{{\rm FWHM},c}}{T_0}\right)\exp\left[\frac{\ell(\ell+1)\theta^2_{{\rm FWHM},c}}{8\ln 2}\right],
\end{align}
where $T_0=2.73$ K is the CMB monopole. The respective parameters for each channel are\footnote{\url{https://wiki.cosmos.esa.int/planckpla/index.php/Main_Page}} 
\begin{center}
\begin{tabular}{ c c c }
 $\nu_c$ & $\theta_{\rm FWHM, c}$ & $\sigma_c$ \\ 
 \hline
  100 GHz & 9.66$'$ & 10.77 $\mu$K\\ 
  143 GHz & 7.27$'$ & 6.40 $\mu$K\\ 
  217 GHz & 5.01$'$ & 12.48 $\mu$K\\ 
\end{tabular}
\end{center}
The \textit{Planck} 2018 limits on $g^{\rm loc}_{\rm NL}$ are given by \citep{planck20c}:
\begin{align}
g_{\rm NL}^{\rm loc} &= \left(- 5.8 \pm 6.5 \right) \times 10^4 \ \ \ (68\%~~ \textrm{confidence}) \\
&\equiv \widehat{g_{\rm NL}^{\rm loc}} \pm \sigma_{g_{\rm NL}^{\rm loc}}
\end{align}
As a cross check of our numerical code, we compared the standard deviation of the local-type trispectrum estimator that we obtain from Eq.~\eqref{eq:FPNG} to the one reported by the Planck collaboration, and given above. We find that they agree within 5\%.

To derive an indirect bound on $f_{\rm pbh}$ from the Planck constraint on $g_{\rm NL}^{\rm loc}$, we proceed as follows. Using Bayes' theorem, and assuming the estimator for $g_{\rm NL}^{\rm loc}$ has a Gaussian distribution, the un-normalized posterior probability distribution for $f_{\rm pbh}$ is given by
\begin{align}
    \mathcal{P}(f_{\rm pbh}) \propto \exp\left[-\frac12 \frac{\left(\mathcal{R} f_{\rm pbh} - \widehat{g_{\rm NL}^{\rm loc}} \right)^2}{\sigma_{g_{\rm NL}^{\rm loc}}^2} \right] H(f_{\rm pbh}),
\end{align}
where in this context $H(x)$ designates the Heaviside function, enforcing a positive prior on $f_{\rm pbh}$. The $(1 - \epsilon)$-confidence upper limit on $f_{\rm pbh}$, is then obtained from solving the implicit equation
\beq
\int_{f_{\rm pbh}}^\infty d f ~\mathcal{P}(f) = \epsilon ~ \int_0^{\infty} d f ~\mathcal{P}(f).
\eeq

\subsection{Results and discussion}

We are now fully equipped to compute upper limits on $f_{\rm pbh}$ indirectly from the Planck constraint on $g_{\rm NL}^{\rm loc}$, and to forecast Planck's sensitivity to $f_{\rm pbh}$ from the temperature trispectrum. We shall compare these limits and forecasts to Planck power-spectra limits on $f_{\rm pbh}$. We obtain the latter with exactly the same procedure as in AK17, but using Planck 2018 data \cite{planck20a} (instead of 2015). Specifically, we use the foreground-marginalized Plik-lite log-likelihood for $C_\ell$'s at $\ell \geq 30$, which we Taylor-expand near the Planck best-fit cosmology, and account approximately for the low-$\ell$ data by imposing a Gaussian prior on the optical depth to reionization. For the joint $TT,TE,EE$ limits, we use the modified version of \texttt{HYREC} and \texttt{CLASS} as implemented by AK17; in particular we use their approximate homogeneous injection-to-deposition Green's function. In addition to the joint $TT,TE,EE$ limits, we also compute a $TT$-only upper limit on $f_{\rm pbh}$ -- we still retain the optical depth prior, however, so our ``$TT$-only" limits are technically temperature + low-$\ell$ polarization limits. For a fair comparison with our $TTTT$ trispectrum limits and forecasts, for the $TT$ limit we compute the effect of accreting PBHs at first order in $f_{\rm pbh}$, including the ``direct" term only, and using our more accurate injection-to-ionization Green's function. We also include the effect of inhomogeneous ionization perturbations on the temperature power spectrum for completeness, but this makes a negligible difference on the results.

We find that the indirect limit on $f_{\rm pbh}$ obtained from Planck's bounds on $g_{\rm NL}^{\rm loc}$ is systematically one order of magnitude weaker than the $TT$-only power spectrum limit, for all PBH masses. This is due to the weak overlap of the trispectrum induced by primordial non-Gaussianity with the one induced by accreting PBHs: we find that the correlation coefficient of the two shapes is less than $10\%$ across all black hole masses (using the dot product defined in Eq.~\eqref{eq:dotprod}). We therefore do not show this limit on our final figure.

\begin{figure*}[ht]
\includegraphics[trim={0 1cm 0 .5cm},width=1.5\columnwidth]{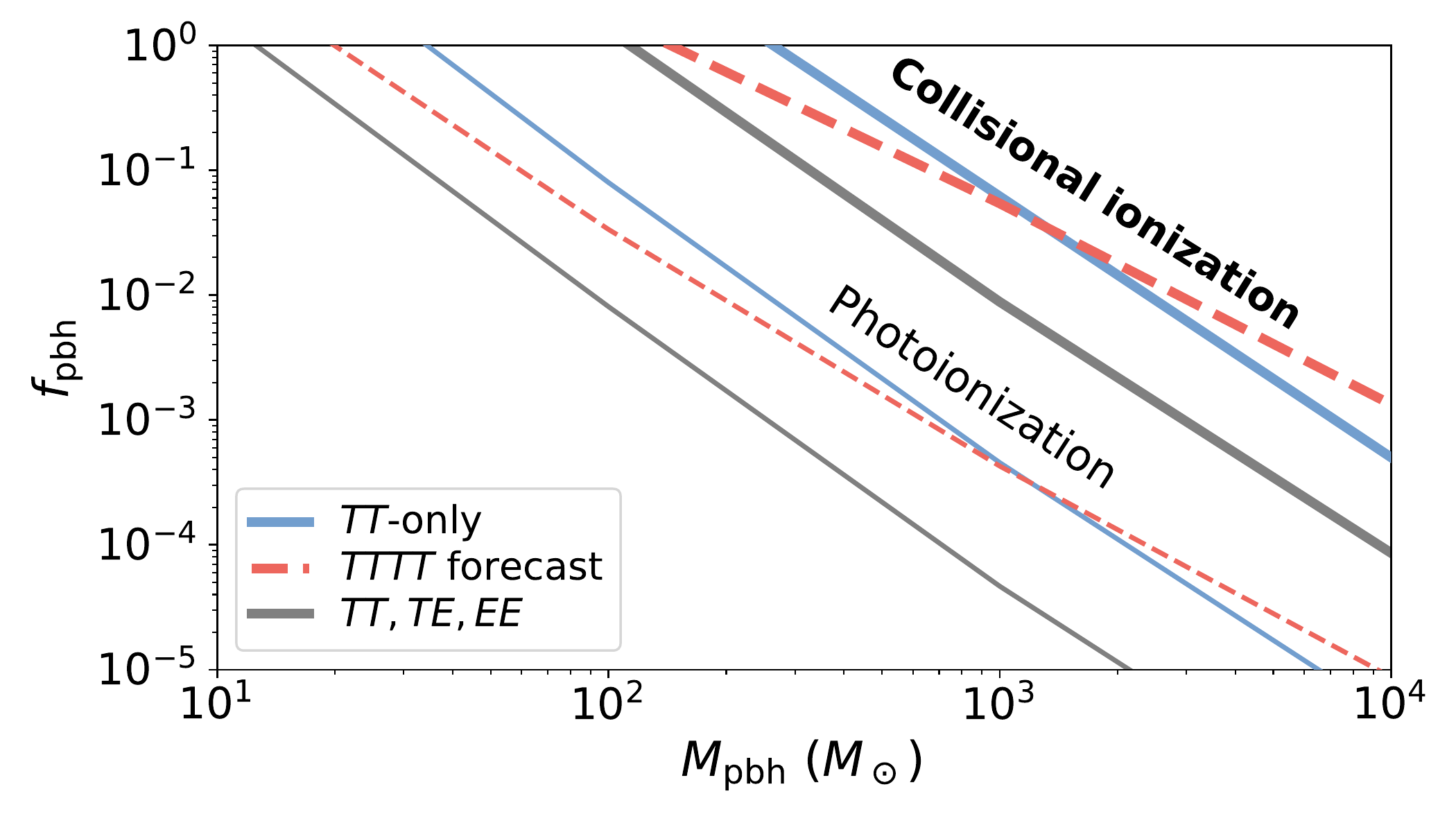}
\caption{\label{fig:const} \emph{Planck} 2018 CMB power spectra constraints (solid lines) and temperature trispectrum forecasted sensitivity (dashed red line) to the fraction of dark matter in PBHs, as a function of PBH mass. Our forecasted sensitivity from the temperature trispectrum is better than $TT$-only constraints for $M_{\rm pbh} \lesssim 10^3 M_{\odot}$ for both the collisional ionization (thick lines) and photoionization (thin lines) limits (see AK17 for details about these different regimes).}
\end{figure*}

We show our forecasted 1-$\sigma$ sensitivity of Planck to the trispectrum of accreting PBHs in Fig.~\ref{fig:const}, alongside current Planck power spectra upper limits on $f_{\rm pbh}$. The upper set of curves correspond to the conservative collisional ionization limit of AK17, while the lower set of curves correspond to the photoionization limit (see AK17 for details). In both cases the qualitative results are the same: we see that the temperature-only trispectrum is not as sensitive as we had expected it to be a priori, as its sensitivity is comparable to current $TT$ upper limits (rather than an order of magnitude better than joint temperature and polarization limits). Nevertheless, the temperature trispectrum is still more sensitive than temperature-only power spectrum constraints for $M_{\rm pbh} \lesssim 10^3 M_{\odot}$. In particular, the temperature-only trispectrum has the potential to probe PBHs lighter by a factor $\sim 2$ than the current reach of temperature-only power spectrum limits.

Interestingly, the mass dependence of the trispectrum sensitivity forecast is shallower than that of the power spectrum constraints. Moreover, we find that making the spatial on-the-spot approximation (as described in Sec.~\ref{sec:approx}) affects trispectrum forecasts by no more than $20\%$. Both of these features can be explained qualitatively by the different redshift dependence of the trispectrum and power spectrum signals, which we explore in Appendix~\ref{app:slope}.

Fig.~\ref{fig:const} also shows the updated Planck joint temperature and polarization power-spectrum constraints ($TT,TE,EE$). We see that these constraints are tighter than the $TT$-only constrains by about an order of magnitude. This stems from the relatively larger effect of recombination perturbations on the polarization signal (see e.g.~Fig.~13 of AK17), indicating a stronger cross-correlation of the perturbed CMB polarization with the unperturbed field. This provides a strong motivation to extend our work to all temperature and $E$-mode polarization trispectra, $TTTE, TTEE, TEEE, EEEE$, which may be significantly more sensitive to accreting PBHs than the temperature-only trispectrum. In addition, the inhomogeneity in the free-electron fraction ought to induce $B$-mode polarization of magnitude comparable with the corresponding $E$-mode polarization, $B^{(1)}_{\rm inh} \sim E^{(1)}_{\rm inh}$. This means that, to linear order in $f_{\rm pbh}$, trispectra involving one $B$ mode ($TTTB, TTEB, TEEB, EEEB$) ought to carry comparable signal to the corresponding 4-point functions involving temperature and $E$-modes only. Importantly, absent primordial tensor modes or accreting PBHs, the primary (unlensed) CMB $B$-mode polarization vanishes. Therefore, after delensing, one can effectively eliminate cosmic variance in the $B$-mode measurement. We thus expect these $B$-mode trispectra to have a significantly enhanced signal-to-noise ratio relative to their $E$-mode counterparts \cite{Meerburg_16}. We defer to a future publication the extension of this work to polarization trispectra.

\section{Conclusions}\label{sec:conc}

This work is the second part of a series of three papers studying the imprints of inhomogeneously-accreting PBHs on CMB anisotropies, in particular their higher-order statistics. The first part, Ref.~\citep{jensen21a}, inspected in detail how inhomogeneous energy injection from non-uniformly accreting PBHs perturbs recombination. In the present analysis, we compute the perturbed temperature anisotropy and its 2-point and 4-point functions. In the upcoming third paper of this series, we will extend this work to polarization.

Our main results can be summarized as follows: \\
$(i)$ The inhomogeneous part of the free-electron perturbation leads to a sub-10\% effect on the perturbation to the CMB temperature power spectrum. In other words, it is sufficient to only account for the average perturbation to the free-electron fraction when computing the effect of accreting PBHs on the CMB temperature power spectrum. This sub-dominant contribution was not expected a priori and is due to the poor correlation of the perturbed CMB temperature field with the standard temperature anisotropy. It is not guaranteed that the same holds true for CMB polarization power spectra.
\\
$(ii)$ We set new constraints on the PBH abundance, obtained indirectly from Planck's upper limits on local-type primordial non-Gaussianity. Indeed, the shape of PBH-induced trispectrum overlaps with that of primoridal non-Gaussianities, although weakly. This weak correlation implies that our new constraints are not competitive with existing CMB temperature power spectrum constraints. Still, they provide a qualitatively different probe of PBH abundance, complementary to the usual 2-point function limits.\\
$(iii)$ We forecast the sensitivity of Planck to the temperature trispectrum induced by inhomogeneously-accreting PBHs. Although our numerical results show a weaker sensitivity than what could have been expected from simple order-of-magnitude estimates, still we find that the temperature trispectrum would be sensitive to PBH abundances lower than current bounds from the CMB temperature-only 2-point function, for $M_{\rm pbh} \lesssim 10^3 M_\odot$. This is our most important result, which demonstrates that the CMB trispectrum is indeed a useful probe of PBHs. 

The calculation of higher-order CMB statistics is quite involved, and we necessarily had to make several approximation to keep it tractable. First, following previous studies of perturbed recombination, we only accounted for the ``direct" piece of the source term for the perturbation to CMB anisotropies, and neglected the ``feedback" piece. Unlike previous studies, however, we explicitly quantified this approximation in the limiting case of homogeneous ionization perturbations, and showed that it is accurate to better than $\sim 20\%$ in that case. Still, a rigorous and definitive calculation of the trispectrum should eventually include the ``feedback" term self-consistently. Second, we made several approximations in order to derive a factorized quadratic transfer function for the free-electron fraction perturbation. In particular, we conservatively approximated the injection-to-ionization Green's function by a factorized form that bounds it from below. This approximation was needed to get a factorized trispectrum, much more manageable computationally than the exact trispectrum would be. In order to quantify the error induced by this approximation, we also considered the limit of spatially-on-the-spot energy deposition, which bounds the injection-to-ionization Green's function from above. We found that all our results are nearly unchanged when considering this limit, thus giving us confidence in their robustness. Third, in our analysis of the primordial non-Gaussianity bias and our trispectrum sensitivity forecast, we neglected non-Gaussianities induced by CMB lensing. An actual analysis of CMB data should of course correct for the lensing bias. 

The most uncertain part of our calculation remains the physics of accretion and radiation. All our numerical results rely on the semi-analytic model of AK17 \cite{yacine17a}, with a simple prescription for the effect of relative velocities. While of course the quantitative results would change with different assumptions about the accretion geometry and radiative efficiency, it seems unavoidable that the PBH accretion luminosity should be strongly modulated by large-scale supersonic relative velocities. We also neglected entirely the effects of non-linear clustering post-recombination \cite{inman19}. We expect relative velocities would also modulate the baryon content of the first halos, hence the accretion rate in these environments. Hence, our results should still be robust qualitatively, regardless of the details of the accretion model, or of the relevance of accretion in non-linear halos. Moreover, the formalism we develop is quite general and could be applied to arbitrary perturbations of recombination spatially modulated by relative velocities, or even more generally quadratic in initial conditions.

Even if the temperature trispectrum is not quite as sensitive to PBHs as we had anticipated from the simple order of magnitude presented in the introduction, our results are still very significant and promising. Indeed, we uncovered a completely new CMB observable to probe PBHs, with a sensitivity comparable to, and in some cases better than, current CMB temperature power-spectrum constraints. Importantly, while several energy injection processes could in principle mimick the effect of accreting PBHs in CMB power spectra, to our knowledge the trispectrum signature studied in this work is unique to them. These considerations provide strong motivation to extend this work and study the polarization signal of inhomogeneously accreting PBHs. In addition to trispectra involving $E$-mode polarization ($TTTE, TTEE, TEEE, EEEE$), we also expect $B$-mode non-Gaussianity, in the form of $TTTB, TTEB, TEEB, EEEB$ trispectra at leading order in PBH abundance. These $B$-mode trispectra ought to have amplitudes comparable to their $E$-mode counterparts, but much lower noise. We defer the computation of these promising observables to the third and last installment of this series of publications.

\section*{Acknowledgements}
YAH is a CIFAR-Azrieli Global scholar and acknowledges funding from CIFAR. This work was supported in part by NSF grant No.~1820861. We thank Daan Meerburg for suggesting to seek a factorized form for the trispectrum, and Colin Hill for comments on this manuscript.

\begin{appendix}

\section{Correlation functions involving a function of relative velocity}\label{app:vbcsq}

In what follows we denote $\bs{v} \equiv \bs{v}_{\rm bc}$ the relative velocity of baryons and dark matter. We need to compute $(N+1)$-point functions of the form
\begin{align}
\braket{F(\bk) \delta_1(\bk_1) \cdots \delta_{N}(\bk_{N})} = P_N(\bk, \bk_1, \cdots, \bk_{N})\nonumber\\
\times (2 \pi)^3 \delta^{(3)}(\bk + \bk_1 + \cdots + \bk_N), \label{eq:BFd1d1-def}
\end{align}
where $F$ depends on position only through the magnitude $v$ of the relative velocity field, i.e.~$F(\bs{x}) = F(v(\bs{x}))$, and has zero mean, $\braket{F} = 0$, and $\delta_1, \cdots, \delta_N$ are scalars also with zero mean linearly related to the primordial curvature perturbation. This $(N+1)$-point function is non-zero only if $N$ is even, given that $F$ is an even function of relative velocity. The $(N+1)$-spectrum $P_N$ is the Fourier transform of the $(N+1)$-point correlation function
\beq
\xi_N(\bx_1, \cdots, \bx_N) \equiv \braket{F(v(\bs{0})) \delta_1(\bx_1) \cdots \delta_N(\bx_N)}.
\eeq
The goal of this appendix is to derive an approximate expression for $\xi_N$, from which one can also approximate $P_N$.

In full generality, provided $\bs{v}, \delta_1, \cdots, \delta_N$ are Gaussian-distributed, we have
\begin{align}
\xi_N(\bx_1, \cdots, \bx_N) = \int d^3 v~ d\delta_1\cdots d \delta_N ~ F(v) \delta_1 \cdots \delta_N \nonumber\\
\times \frac{1}{\sqrt{2 \pi \det(\bs{C})}}  \exp\left[- \frac12 \bs{X}^{\rm T} \cdot \bs{C}^{-1} \cdot \bs{X} \right],
\end{align}
where 
\begin{align}
    \bs{X}^{\rm T} \equiv (\widetilde{\bs{v}}, \widetilde{\bs{\delta}}^{\rm T}) \equiv \left(\frac{\bs{v}}{\sigma_{1d}}, \frac{\delta_1}{\sigma_{\delta_1}}, \cdots,  \frac{\delta_N}{\sigma_{\delta_N}}\right),
\end{align}
    with $\sigma_{1d}^2 \equiv \braket{v^2}/3$ and $\sigma_{\delta_i}^2\equiv \braket{\delta_i^2}$. $\bs{C}$ is the $(N+3)$ by $(N+3)$ normalized covariance matrix of $\widetilde{\bs{v}}(\bs{0}), \widetilde{\delta_1}(\bx_1), \cdots , \widetilde{\delta_N}(\bx_N)$. Explicitly, this matrix is given by $\bs{C} = \bs{C}_0 + \bs{\Delta}$, with
\begin{align}
&\bs{C}_0 \equiv \begin{pmatrix}
  \bigone_{3\times 3}
  & \rvline & \bigzero_{3 \times N} \\[8pt]
\hline
  \bigzero_{N \times 3} & \rvline &
  \bs{C}_{\widetilde{\delta}}
\end{pmatrix}, \\
&\bs{\Delta} \equiv \begin{pmatrix}
  \bigzero_{3 \times 3} & \rvline & 
 \begin{matrix} 
 \bs{\Xi}_1^{\rm T}\\
 \bs{\Xi}_2^{\rm T}\\
 \bs{\Xi}_3^{\rm T}
 \end{matrix} \\
\hline
\begin{matrix}
\bs{\Xi}_1 & \bs{\Xi}_2 & \bs{\Xi}_3
\end{matrix}
& \rvline &  \bigzero_{N \times N}
\end{pmatrix},
\end{align}
where $\bs{C}_{\widetilde{\delta}}$ is the $N\times N$ normalized covariance matrix of the $\widetilde{\delta}$'s, and $\bs{\Xi}_i,  i = 1, 2, 3$, are the $N$-dimensional column vectors
\beq
\bs{\Xi}_i \equiv \begin{pmatrix} \braket{{\widetilde{v}_i(0)\widetilde{\delta}_1(\bx_1)}}\\ \vdots\\ \braket{\widetilde{v}_i(0)\widetilde{\delta}_N(\bx_N)} \end{pmatrix}.
\eeq

 In words, the matrix $\bs{C}_0$ includes all correlations except for the velocity-$\delta$ correlations, which are included in $\bs{\Delta}$.

\begin{figure}[ht]
\includegraphics[trim={0 0 0 0},width=\columnwidth]{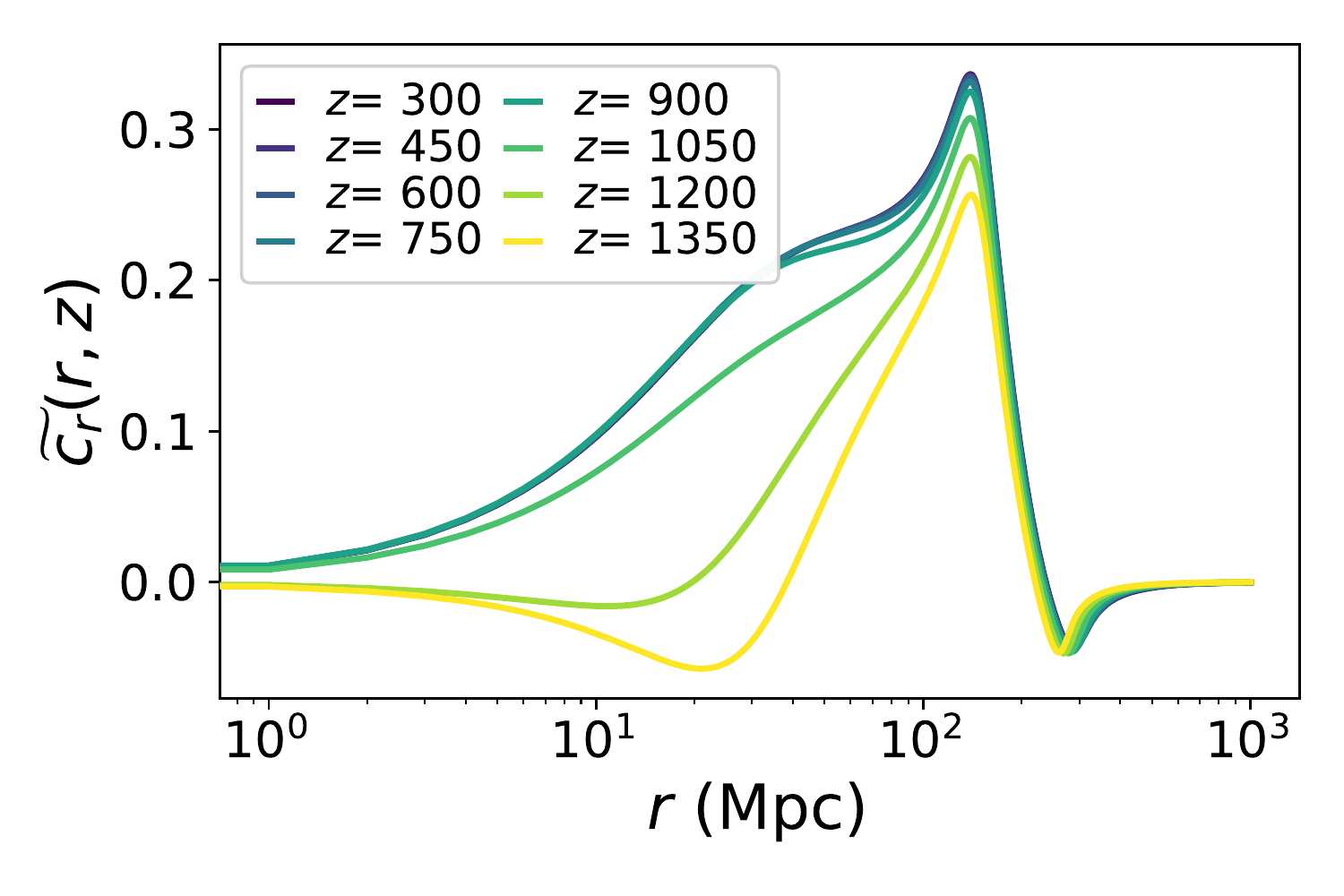}
\caption{\label{fig:v_corr} Correlation function of the monopole terms of the line-of-sight source for $\Theta^{(0)}$ near recombination with the relative velocity between CDM and baryons at various redshifts. Namely we plot the variance normalized correlation function $\widetilde{c}_r$, where $\langle {\bf v}(z){S^{(0)}_0}(z'=1100)\rangle \equiv c_r(r,z)\hat{r}$, and the subscript implies the monopole terms only. Even at intermediate scales it is less than unity and justifies expanding the covariance matrix discussed in Appendix~\ref{app:vbcsq}.}
\end{figure}

So far, these expressions are exact. We expect that, in general, $\bs{\Delta}$ is small for \emph{any} separation. Indeed, this is always true in the large-separation limit. Moreover, statistical isotropy implies that $\braket{\bs{v}(0) \delta(\bs{x})} \rightarrow 0$ when $\bs{x} \rightarrow 0$, since there is no non-null isotropic rank-1 tensor. This can be seen in Fig.~\ref{fig:v_corr} where we correlate $v_i$ with the canonical monopoles of the $\Theta^{(0)}$ line-of-sight source transfer functions of ${S}^{(0)}$ (c.f. the first line of Eq.~\eqref{eq:S0}) for example.

We may therefore expand $\bs{C}^{-1}$ around $\bs{C}_0^{-1}$ to compute $\xi_N$. We'll see that it is required to include terms at second order in $\bs{\Delta}$:
\beq
\bs{C}^{-1} = \bs{C}_0^{-1} - \bs{C}_0^{-1}\bs{\Delta} \bs{C}_0^{-1} + \bs{C}_0^{-1}\bs{\Delta} \bs{C}_0^{-1} \bs{\Delta} \bs{C}_0^{-1} + \mathcal{O}(\Delta^3),
\eeq
from which we get, at second order in $\Delta$, 
\barr
&&\exp\left[- \frac12 \bs{X}^{\rm T}  \bs{C}^{-1}  \bs{X} \right] = \Lambda \times \exp\left[- \frac12 \bs{X}^{\rm T}  \bs{C}_0^{-1}  \bs{X} \right] ,\nonumber\\
&&\Lambda  \equiv 1+ \frac12 \bs{\tilde{X}}^{\rm T}  \bs{\Delta}  \bs{\tilde{X}} \nonumber\\
&&~~ + \frac18 \left(\bs{\tilde{X}}^{\rm T}  \bs{\Delta}  \bs{\tilde{X}} \right)^2 - \frac12 \bs{\tilde{X}}^{\rm T}  \bs{\Delta}\bs{C}_0^{-1} \bs{\Delta} \bs{\tilde{X}},~~~~~~~~~
\earr
with $\bs{\tilde{X}} \equiv \bs{C}_0^{-1} \bs{X}$. With this approximation, we thus have
\beq
\xi_N \approx \sqrt{\frac{\det(\bf{C}_0)}{\det(\bf{C})}}\langle F(v) \delta_1 \cdots \delta_N \Lambda\rangle_0 \approx \langle F(v) \delta_1 \cdots \delta_N \Lambda\rangle_0, \label{eq:xi_N-approx}
\eeq
where the average $\langle \cdots \rangle_0$ is over the ``unperturbed" $(N+3)$-D Gaussian distribution with covariance matrix $\bf{C}_0$, which is the product of two uncorrelated Gaussian distributions: an isotorpic Gaussian distribution for $\bs{v}(\bs{0})$ and a $N$-dimensional Gaussian for $(\delta_1(\bs{x}_1), \cdots, \delta_N(\bs{x}_N))$, with covariance matrix $\bs{C}_\delta$. The second equality is valid to lowest order in $\Delta$.

Upon integrating over velocities, the contribution of the first term in $\Lambda$ (i.e.~1) vanishes, since $\braket{F(v)} = 0$. Let us now compute the other terms. First, let us compute
\beq
\tilde{\bs{X}}^{\rm T}  \bs{\Delta} = \left( \bs{\delta}^{\rm T} \bs{C}_\delta^{-1}   \bs{\Xi}_1, \bs{\delta}^{\rm T}  \bs{C}_\delta^{-1}  \bs{\Xi}_2, \bs{\delta}^{\rm T}  \bs{C}_\delta^{-1}  \bs{\Xi}_3, \frac{v_i}{\sigma_{1 \rm d}^2} \bs{\Xi}_i^{\rm T} \right), 
\eeq
where the first three terms are scalars, and the last term contains an implicit sum over $i$, and is a $N$-dimensional vector. We therefore have
\beq
\Lambda_1 \equiv \bs{\tilde{X}}^{\rm T}  \bs{\Delta}  \bs{\tilde{X}} =  \frac{2v_i}{\sigma_{\rm 1d}^2} \bs{\Xi}_i^{\rm T}  \bs{C}_\delta^{-1}  \bs{\delta} = \frac{2v_i}{\sigma_{\rm 1d}^2}  \bs{\delta}^{\rm T}  \bs{C}_{\delta}^{-1}  \bs{\Xi}_i.
\eeq
The second term $\Lambda_1$ is therefore linear in $v_i$. Therefore $\langle F(v) \delta_1... \delta_N \Lambda_1\rangle_0 = 0$ since $\langle v_i F(v) \rangle = 0$, by isotropy. 

We thus need to only include the third and last terms in $\Lambda$, quadratic in $\Delta$.

Let's start with the third term, proportional to $\Lambda_1^2$. From our previous results, we have
\beq
\Lambda_1^2 = \frac4{\sigma_{\rm 1d}^4} v_i v_j \bs{\Xi}_i^{\rm T} \bs{C}_\delta^{-1} \bs{\delta} \bs{\delta}^{\rm T}  \bs{C}_\delta^{-1}\bs{\Xi}_j,
\eeq
where repeated indices are summed over. Using $\langle F(v) v_i v_j \rangle_0 = \frac13 \delta_{ij} \langle v^2 F(v)\rangle_0$, we thus find
\begin{align}
\Big{\langle} F(v) \delta_1 \cdots \delta_N \Lambda_1^2 \Big{\rangle}_0 = \frac{4}{3 \sigma_{1d}^4} \langle v^2 F(v) \rangle \nonumber\\
\times \bs{\Xi}_i^{\rm T} \bs{C}_{\delta}^{-1} \langle \delta_1 \cdots  \delta_N \bs{\delta} \bs{\delta}^{\rm T} \rangle_0 \bs{C}_{\delta}^{-1} \bs{\Xi}_i.
\end{align}
On the other hand, we have
\barr
\Lambda_2 &\equiv& \bs{\tilde{X}}^{\rm T}  \bs{\Delta}\bs{C}_0^{-1} \bs{\Delta} \bs{\tilde{X}} = \frac{v_i v_j}{\sigma_{1d}^4} \bs{\Xi}_i^{\rm T} \bs{C}_{\delta}^{-1} \bs{\Xi}_j \nonumber\\
&+& \textrm{terms independent of $v$}.
\earr
This implies
\begin{align}
\Big{\langle} F(v) \delta_1 \cdots \delta_N \Lambda_2 \Big{\rangle}_0 = \frac{\braket{v^2 F(v)}}{3 \sigma_{1d}^4}  \langle \delta_1 \cdots \delta_N \rangle \bs{\Xi}_i^{\rm T} \bs{C}_{\delta}^{-1}   \bs{\Xi}_i.
\end{align}
Therefore, combining terms, we obtain
\barr
\xi_N(\bx_1, \cdots, \bx_N) &\approx& \frac18 \Big{\langle} F(v) \delta_1 \cdots \delta_N \Lambda_1^2 \Big{\rangle}_0 \nonumber\\
&-& \frac12 \Big{\langle} F(v) \delta_1 \cdots \delta_N \Lambda_2 \Big{\rangle}_0\nonumber\\
&=& \braket{v^2 F(v)}S(\bx_1, \cdots, \bx_N),
\earr
where we have defined 
\beq
S \equiv \frac1{6 \sigma_{1d}^4} \bs{\Xi}_i^{\rm T} \bs{C}_{\delta}^{-1} \Big{\langle} \delta_1 \cdots \delta_N \left(\bs{\delta \delta}^{\rm T} - \bs{C}_\delta \right)\Big{\rangle} \bs{C}_{\delta}^{-1} \bs{\Xi}_i.
\eeq
We see that in this approximation, the shape of the $N$-point correlation is entirely determined by $S(\bx_1, \cdots, \bx_N)$, regardless of the function $F(v)$. The latter only affects the overall amplitude of the correlation function, and only through its moment $\langle v^2 F(v) \rangle$. 

Therefore, to compute the $N$-point correlation function, one may substitute $F(v)$ with a simpler function $\tilde{F}(v)$, as long as $\langle v^2 \tilde{F}(v) \rangle = \langle v^2 F(v) \rangle$. The simplest such function is $\tilde{F}(v) \equiv b_F \left(\frac{v^2}{3 \sigma_{1d}^2} -1\right)$. It is such that
\beq
\langle v^2 \tilde{F}(v) \rangle  = 2 b_F \sigma_{1 \rm d}^2.   
\eeq
Hence, we may use $\tilde{F}(v)$ instead of $F(v)$ provided the parameter $b_F$ is given by
\beq
b_F = \frac1{2 \sigma_{1d}^2} \braket{v^2 F(v)}.
\eeq

This result was proven in configuration space but also holds in the Fourier domain, where it is most useful: we have proven that, for any $N$-point function involving $N$ scalar functions (provided the $\delta v_i$ correlations are sufficiently small at all separations), we may use $\tilde{F}(v) = b_F \left(\frac{v^2}{3 \sigma_{1d}^2} -1\right) $ in order to compute $N$-point functions. Importantly, this means that the shape we derive for the trispectrum should be relatively insensitive to the details of accretion physics -- the shape still has some dependence on it, as in practice the bias parameter $b_F$ is redshift-dependent, in a way that depends on the details of accretion.

\section{Numerical resolution and convergence}\label{app:conv}

In this appendix we describe our sampling of $\eta, k$ and $\ell$ integrals and sums.

Because each conformal-time integral relevant to the PBH-induced trispectrum includes the visibility function $g(\eta)$, we sample $\eta$ more finely during recombination. Starting from $z_{\rm max} = 1400$, we sample $\eta$ with logarithmic step size $\Delta\ln \eta = 10^{-3}$ until $z_{\rm rec} = 900$, after which we increase the step size to $\Delta \ln \eta = 2\times 10^{-2}$ until $z_{\rm re} = 10$, and then finally we sample linearly in $\eta$ until $z = 0$ with step size $\Delta\eta = 50$ Mpc.

For $k$ integrals, we compute quantities on a grid from $k_{\rm min}=10^{-5}$ Mpc$^{-1}$ up to a maximum wave number $k_{\rm max} = 5000 \eta_0^{-1}$ with a step size $\Delta k={\rm min}(\epsilon k, \kappa_0)$, where $\epsilon = 0.006$ and $\kappa_0 = 10^{-4}$ Mpc$^{-1}$, i.e.~use logarithm spacing for low-$k$ to linear spacing at high-$k$.

Finally, our $\ell$ sampling consists of the floors of an array of real $\ell$ values spaced logarithmically in $2 \leq \ell < 400$ with $\Delta \ln \ell = 0.0225$, and linearly in $400 \leq \ell < \ell_{\max} = 3000$ with $\Delta \ell = 19.5$. Note that these values were chosen to produce an $\ell$ sampling similar to the standard output of \texttt{CLASS}, with almost double the resolution.

We reproduce the standard CMB temperature angular power spectrum, Eq.~\eqref{eq:C_l}, and compare to the output of \texttt{CLASS} \citep{CLASS}. We find a sub-percent fractional difference for all $\ell <\ell_{\rm max}$.
We also recompute all results with increased resolution prescribed via,
\begin{align}
    (\Delta\ln \eta,\,\Delta\eta,\,k_{\rm max}&,\epsilon, \kappa_0)\rightarrow\nonumber\\
    &\left(\frac{2}{3}\Delta\ln \eta,\,\frac{2}{3}\Delta\eta,\,\frac{3}{2}k_{\rm max},\frac{2}{3}\epsilon, \frac{2}{3}\kappa_0\right).
\end{align}
We find, for both the power spectrum and trispectrum calculations, there is a fractional change in the results only at sub-percent level, far below the theoretical uncertainty of the problem at hand.

Lastly, the trispectrum results depend on the intermediate quantity $\mathcal{J}_\ell$, given in Eq.~\eqref{eq:mathcalJ_main} as an infinite double sum. We truncate this sum at a maximum $\ell_2 = \ell_{\rm cut}$ (which automatically truncates the $\ell_1$ sum due to the triangle inequality). We find that our trispectrum results are converged within $0.1\%$ by $\ell_{\rm cut} = 50$.

\section{Spin-weighted spherical harmonics}\label{app:spin}
Spin-weighted spherical harmonics are related to Wigner $D$-matrices. They become regular spherical harmonics when their spin is zero, $(_0Y_{\ell m})=Y_{\ell m}$ and inherit similar orthogonal and completeness relations. They have the familiar property $(_sY_{\ell m})^*=(-1)^{s+m}(_{-s}Y_{\ell -m})$, as well as a product rule similar to the Gaunt relation involving Wigner 3$j$ symbols \citep{Qtheory},
\begin{align}\label{eq:product_rule}
    ~_{s_1}Y_{\ell_1 m_1}(\hat{n}) ~_{s_2}Y_{\ell_2 m_2}(\hat{n}) =\sum_{s_3,\ell_3m_3}g^{-s_1(-s_2)(-s_3)}_{\ell_1\ell_2\ell_3}\nonumber\\
    \times\threej{\ell_1}{\ell_2}{\ell_3}{m_1}{m_2}{m_3} ~_{s_3}Y^*_{\ell_3 m_3}(\hat{n}),
\end{align}
where the $g$-symbols are defined by
\begin{align}\label{eq:g_sym}
    g_{\ell_1 \ell_2 \ell_3}^{s_1 s_2 s_3} &\equiv \sqrt{\frac{(2 \ell_1 +1)(2 \ell_2 +1)(2 \ell_3 +1)}{4 \pi}}  \threej{\ell_1}{\ell_2}{\ell_3}{s_1}{s_2}{s_3},\nonumber\\
    g_{\ell_1 \ell_2 \ell_3}&\equiv g_{\ell_1 \ell_2 \ell_3}^{0\, 0\, 0},
\end{align}
For shorthand we also define the Gaunt coefficient,
\begin{align}\label{def:G}
\mathcal{G}^{\ell_1\ell_2\ell_3}_{m_1 m_2m_3}\equiv\,& g_{\ell_1 \ell_2 \ell_3}
\begin{pmatrix}
\ell_1 & \ell_2 & \ell_3\\
m_1 & m_2 & m_3
\end{pmatrix}.
\end{align}

For the Wigner $3j$ symbols to be nonzero, the $\ell$'s in the first row must be positive and obey the triangle inequality. Likewise, the sum of the bottom row of azimuthal modes ($m_1,m_2,m_3$) must equate to zero, and each must satisfy $-\ell_i\le m_i\le \ell_i$.
The Wigner 3$j$ symbols also have an orthogonality condition that we utilize,
\begin{align}\label{eq:wig_orth}
\sum_{m_1 m_2} \threej{\ell_1}{\ell_2}{\ell_3}{m_1}{m_2}{m_3}& \threej{\ell_1}{\ell_2}{\ell_3'}{m_1}{m_2}{m_3'} =\nonumber\\
&\quad\quad\frac{\delta_{\ell_3 \ell_3'} \delta_{m_3 m_3'}}{2 \ell_3 + 1} \{ \ell_1 \ell_2 \ell_3 \},
\end{align}
where $\{\ell_1 \ell_2 \ell_3 \}$ is 1 if the three $\ell$'s satisfy the triangle inequality and 0 otherwise. 

When summing over the azimuthal modes of the product of spin-weighted spherical harmonics, we have,
\begin{align}\label{eq:d_sum}
\sum_m(_sY_{\ell m}(\hat{n}))(_{s'}Y_{\ell m}(\hat{n}'))^*=(-1)^s\frac{2\ell+1}{4\pi}d^\ell_{ss'}\left(\mu\right),
\end{align}
where we have introduced the Wigner small $d$-functions and $\mu\equiv \hat{n}\cdot\hat{n}'$ \citep{smith15a}. If $s=s'=0$, then the $d$-functions reduce to normal Legendre polynomials. These $d$-functions themselves satisfy the orthogonality condition,
\begin{align}\label{eq:d_orth}
    \int^1_{-1}d\mu~ d^{\ell_{1}}_{ss'}(\mu)d^{\ell_{2}}_{ss'}(\mu)=\frac{2}{2\ell+1}\delta_{\ell_1\ell_2},
\end{align}
and are equipped with the identity,
\begin{align}
    d^\ell_{-s(-s')}(\mu)=d^\ell_{s's}(\mu)=(-1)^{s+s'}d^\ell_{ss'}(\mu).
\end{align}
They also have the property that their product can be expanded via \cite{Qtheory},
\begin{align}\label{eq:d_prod}
    &d^{\ell_{1}}_{s_1s_2'}(\mu)d^{\ell_{2}}_{s_2s_2'}(\mu)=\nonumber\\
    &\quad\quad\sum_{\ell,s,s'}(2\ell+1)\threej{\ell_1}{\ell_2}{\ell}{s_1}{s_2}{s} d^{\ell}_{s s'}(\mu){\threej{\ell_1}{\ell_2}{\ell}{s_1'}{s_2'}{s'}}.
\end{align}

\section{Sums of products of \texorpdfstring{$Q$}{Q} and \texorpdfstring{$\widetilde{Q}$}{Q~} symbols}
\label{app:Q-sums}

Computing the multipoles of the nonlinear perturbation of temperature anisotropy introduces integrals of products of four (spin-weighted) spherical harmonics, denoted as the $Q$-symbols in Eq.~\eqref{eq:Q_sym} and \eqref{eq:Qt_sym}. In this appendix we lay out the math to simplify the sums and products of these $Q$-symbols necessary for the first-order trispectrum calculations. We borrow the tools introduced in Appendix~\ref{app:spin}.

Lest us start with $(\mathcal{Q}^2)_{\ell_1 \ell_2 \ell_3 \ell_4}$ defined in Eq.~\eqref{eq:Qstart}. Given the definitions of $Q_{\ell_1 \ell_2 \ell_3 \ell_4}^{m_1 m_2 m_3 m_4}$, it is given by
\beq
(\mathcal{Q}^2)_{\ell_1 \ell_2 \ell_3 \ell_4} = \sum_{m's} \int d^2 \hat{n} \int d^2 \hat{n}' \prod_i Y_{\ell_i m_i}(\hat{n})Y^*_{\ell_i m_i}(\hat{n}').
\eeq
Let us now use Eq.~\eqref{eq:d_sum}, which reduces to Legendre polynomials in this case:
\beq
\sum_{m_i} Y_{\ell_i m_i}(\hat{n})Y^*_{\ell_i m_i}(\hat{n}') = \frac1{4 \pi} (2 \ell_i +1) P_{\ell_i}(\mu),
\eeq
where $\mu\equiv \hat{n}\cdot\hat{n}'$.
We may carry out one of the angular integrals and get 
\begin{align}
(\mathcal{Q}^2)_{\ell_1 \ell_2 \ell_3 \ell_4} &= \frac1{(4 \pi)^2} \frac12 \prod_{i} (2 \ell_i + 1)\nonumber\\
&\times\int_{-1}^1 d \mu~ P_{\ell_1}(\mu) P_{\ell_2}(\mu) P_{\ell_3}(\mu) P_{\ell_4}(\mu).
\end{align}
Now, recall the product rule for Wigner $d$-functions (or Legendre polynomials in this case), Eq~\eqref{eq:d_prod},
\beq
P_{\ell_1} P_{\ell_2} = \sum_{\ell} (2 \ell + 1) {\threej{\ell_1}{\ell_2}{\ell}{0}{0}{0}}^2 P_{\ell}.
\eeq
Therefore, using the orthogonality relation Eq.~\eqref{eq:d_orth}, we obtain
\begin{align}
(\mathcal{Q}^2)_{\ell_1 \ell_2 \ell_3 \ell_4} =& \frac1{(4 \pi)^2} \prod_{i} (2 \ell_i + 1) \sum_{\ell} (2 \ell +1)\nonumber\\
&\quad\quad\times{\threej{\ell_1}{\ell_2}{\ell}{0}{0}{0}}^2{\threej{\ell}{\ell_3}{\ell_4}{0}{0}{0}}^2 \nonumber\\
=& \sum_{\ell} \frac1{2 \ell +1} (g_{\ell_1 \ell_2 \ell})^2 (g_{\ell \ell_3 \ell_4})^2,
\end{align}
with a result that should be independent of the grouping of the two pairs of $\ell$'s. 

To generalize this to sums involving $\widetilde{Q}$ is relatively straightforward, but requires general Wigner small $d$-functions. We note that the angular derivatives present in $\widetilde{Q}$ can be expressed in terms of spin-1 spin-weighted spherical harmonics \citep{smith15a}. That is,
\begin{align}
\bs{\nabla}_{\hat{n}}Y_{\ell_1 m_1} \cdot \bs{\nabla}_{\hat{n}} Y_{\ell_2 m_2} &= -\frac12 \sqrt{\ell_1(\ell_1 +1) \ell_2 (\ell_2 + 1)}\nonumber\\
\quad &\times \sum_{s = \pm 1} ~_s Y_{\ell_1 m_1} ~_{-s}Y_{\ell_2 m_2},
\end{align}
such that,
\begin{align}\label{eq:Qtil_sym}
&\widetilde{Q}_{\ell_1 \ell_2, \ell_3 \ell_4}^{m_1 m_2, m_3 m_4} = - \frac12 \sqrt{\ell_1(\ell_1 +1) \ell_2 (\ell_2 + 1)}\nonumber\\
&\quad\sum_{s = \pm 1}\int d^2 \hat{n} ~_sY^*_{\ell_1 m_1}(\hat{n})~_{-s}Y^*_{\ell_2 m_2}(\hat{n}) Y^*_{\ell_3 m_3}(\hat{n}) Y_{\ell_4 m_4}^*(\hat{n}).
\end{align}
For short we define
\beq
\widetilde{g}_{\ell_1 \ell_2, \ell} \equiv \sum_{s = \pm 1} g_{\ell_1, \ell_2, \ell}^{s, -s, 0},
\eeq
which is symmetric in its first two indices. Using the properties of $d$-functions outlined in Appendix~\ref{app:spin}, we obtain for Eq.~\eqref{eq:Qstart}~--~\eqref{eq:Qend},
\begin{align}
&(\mathcal{Q \widetilde{Q}})_{\ell_1 \ell_2, \ell_3 \ell_4} = - \frac12 \sqrt{\ell_1 (\ell_1 +1) \ell_2 (\ell_2 +1)} \nonumber\\
&\quad\quad\times\sum_{\ell}\frac1{2 \ell +1} g_{\ell_1 \ell_2 \ell}~\widetilde{g}_{\ell_1 \ell_2, \ell} (g_{\ell_3 \ell_4 \ell})^2, \\
&(\mathcal{\widetilde{Q}}^2)_{\ell_1 \ell_2, \ell_3 \ell_4} = \frac14 \ell_1 (\ell_1 +1) \ell_2 (\ell_2 +1)  \nonumber\\
&\quad\quad\times\sum_{\ell} \frac1{2 \ell +1} (\widetilde{g}_{\ell_1 \ell_2, \ell})^2 (g_{\ell \ell_3 \ell_4})^2, \\
&(\mathcal{\widetilde{Q} \widetilde{Q}}^{\rm T})_{\ell_1 \ell_2, \ell_3 \ell_4} =  \frac14 \sqrt{\ell_1 (\ell_1 +1) \ell_2 (\ell_2 +1)} \nonumber\\
&\quad\quad\times\sqrt{\ell_3 (\ell_3 +1) \ell_4 (\ell_4 +1)}  \nonumber\\
&\quad\quad\times\sum_{\ell} \frac1{2 \ell +1} (g_{\ell_1 \ell_2 \ell} ~\widetilde{g}_{\ell_1 \ell_2, \ell}) (g_{\ell_3 \ell_4 \ell} ~\widetilde{g}_{\ell_3 \ell_4, \ell}), \\
 &(\mathcal{\widetilde{Q} \widetilde{Q}}^{\rm S})_{\ell_1, \ell_2 \ell_3, \ell_4} = - \frac14 \ell_1 (\ell_1 +1) \sqrt{\ell_2(\ell_2 +1) \ell_3 (\ell_3 + 1)}  \nonumber\\
&\quad\times \sum_{s = \pm 1} \sum_{\ell} \frac1{2 \ell +1}\widetilde{g}_{\ell_1 \ell_2, \ell} g_{\ell, \ell_1, \ell_2}^{s, -s, 0} ~g_{\ell, \ell_3 ,\ell_4}^{-s, s, 0}~g_{\ell \ell_3 \ell_4}.
\end{align}

\section{Perturbed temperature anisotropy auto-power-spectrum}\label{app:auto}
\begin{figure}[htb]
\includegraphics[trim={0cm 1cm 0.5cm .5cm},width=.9\columnwidth]{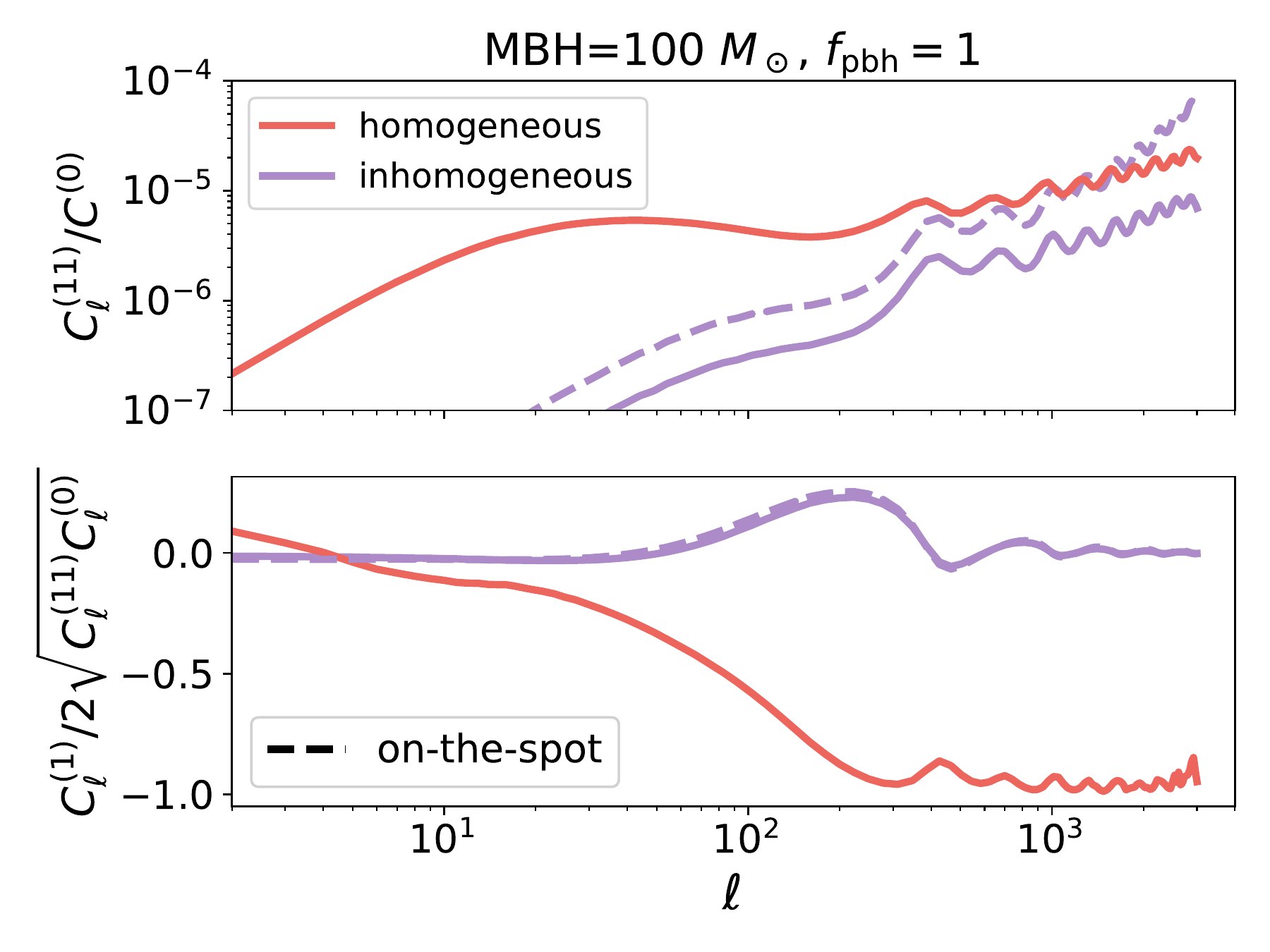}
\caption{\label{fig:auto} Top: auto-power-spectrum of the perturbed temperature anisotropy due to accreting PBHs defined as $\langle\Theta^{(1)}_{\ell m}\Theta^{*(1)}_{\ell' m'}\rangle\equiv \delta_{\ell\ell'}\delta_{mm'}C_{\ell}^{(11)}$, normalized by the standard angular power spectrum. Bottom: correlation coefficients between $\Theta^{(1)}$ and $\Theta^{(0)}$. In both cases we assume 100-$M_\odot$ PBHs comprising all the dark matter, but the qualitative trends are general for all PBH masses. The suppressed amplitude in the auto-power-spectrum of $\Theta^{(1)}_{\rm inh}$ (purple) compared to $\Theta^{(1)}_{\rm hom}$ (red) and the poor correlation explains the large difference in amplitude for the computed power spectra in Sec.~\ref{sec:powerspec}.}
\end{figure}

In this paper we found that the temperature-only trispectrum induced by accreting PBHs was not as sensitive as we expected a priori. Additionally, the amplitude of the power spectrum perturbation sourced by inhomogeneities in the free-electron fraction, $C^{(1)}_{\ell,\rm inh} \equiv 2 \langle \Theta^{(1)}_{\ell m, \rm inh} \Theta^{(0)*}_{\ell m} \rangle$ is up to two orders of magnitude smaller than its counterpart $C^{(1)}_{\ell,\rm hom} \equiv 2 \langle \Theta^{(1)}_{\ell m, \rm hom} \Theta^{(0)*}_{\ell m} \rangle$, as revealed in Fig.~\ref{fig:AK17_jen}. In this appendix, we show this is due to a combination of a poor correlation between $\Theta^{(1)}_{\rm inh}$ and the standard CMB temperature anisotropy $\Theta^{(0)}$, and a suppression of the characteristic amplitude of $\Theta^{(1)}_{\rm inh}$ itself, relative to its counterpart $\Theta^{(1)}_{\rm hom}$. We do so by computing and comparing the \textit{auto}-power-spectra of $\Theta^{(1)}_{\rm hom}$ and $\Theta^{(1)}_{\rm inh}$. The results are shown in Fig.~\ref{fig:auto}.

From Eq.~\eqref{eq:Theta1_lm_hom}, the auto-power-spectrum of $\Theta^{(1)}_{\rm hom}$ is trivially 
\begin{align}\label{eq:homo_auto}
    C_{\ell,\rm hom}^{(11)}=4\pi\int Dk~ P_\zeta(k) \left[\Delta_{\ell, \rm hom}^{(1) \rm d}(k)\right]^2,
\end{align}
where $\langle\Theta^{(1)}_{\ell m,\rm hom}\Theta^{*(1)}_{\ell' m',\rm hom}\rangle\equiv \delta_{\ell\ell'}\delta_{mm'}C_{\ell,\rm hom}^{(11)}$.

The auto-power-spectrum of the inhomogeneous-ionization counterpart, $\langle\Theta^{(1)}_{\ell m,\rm inh}\Theta^{*(1)}_{\ell' m',\rm inh}\rangle\equiv \delta_{\ell\ell'}\delta_{mm'}C_{\ell,\rm inh}^{(11)}$, is much more involved. In what follows, we denote the integral operator,
\begin{align}
\int D(k_1 k_2 k_3)P_\zeta(k_1)P_\zeta(k_2)P_\zeta(k_3)\equiv \int \mathcal{D}^3\!P.
\end{align}
We begin similarly as we did for the trispectrum calculation in Sec.~\ref{sec:trispec}. Starting with Eq.~\eqref{eq:T(1)-def}, using Wick's theorem, and exploiting the fact that $T^{(1)}_{\ell m}(\bk_1, - \bk_1, \bk_3) = 0$ and $T^{(1)}_{\ell m}$ is symmetric in its first two ${\bm k}$ arguments, we find,
\begin{align}
    \langle&\Theta_{\ell m, \rm inh}^{(1)}\Theta_{\ell' m', \rm inh}^{*(1)}\rangle =\nonumber\\
    &\quad\quad\fbh^2\int \mathcal{D}^3\!P\left\{4T_{\ell m}^{(1)}(\bk_1, \bk_2, -\bk_2)T_{\ell' m'}^{*(1)}(\bk_1, \bk_3, -\bk_3)\right.\nonumber\\
    &\quad\quad\quad\quad+2T_{\ell m}^{(1)}(\bk_1, \bk_2, \bk_3)T_{\ell' m'}^{*(1)}(\bk_1, \bk_2, \bk_3)\nonumber\\
    &\quad\quad\quad\quad+4\left.T_{\ell m}^{(1)}(\bk_1, \bk_2, \bk_3)T_{\ell' m'}^{*(1)}(\bk_3, \bk_2, \bk_1)\right\}.
\end{align}
The first term can be solved with the same method as for the inhomogeneous power spectrum in Sec.~\ref{sec:powerspec}. That is, 
\begin{align}
    \int \mathcal{D}^3\!P &~T_{\ell m}^{(1)}(\bk_1, \bk_2, -\bk_2)T_{\ell' m'}^{*(1)}(\bk_1, \bk_3, -\bk_3)=\nonumber\\
    &\quad\delta_{\ell\ell'}\delta_{m m'}\frac{16\pi}9\int_0^{\eta_0} d \eta\int_0^{\eta_0} d \eta' g(\eta)g(\eta')\nonumber\\
    &\quad \quad \quad\quad\quad\quad\quad\quad\quad\times\mathcal{A}_\ell(\eta,\eta')\gamma(\eta)\gamma(\eta'),
\end{align}
where $\gamma(\eta)$ is defined in Eq.~\eqref{eq:gamma} and,
\begin{align}
    \mathcal{A}_\ell(\eta,\eta')&\equiv \int\! Dk P_\zeta(k) \Delta_e(\eta, k) j_{\ell}'(k\chi)\Delta_e(\eta', k) j_{\ell}'(k\chi').
\end{align}

The remaining two terms are not as simple. Using Eq.~\eqref{eq:T_multipoles} and integrating over all three $\hat{k}$'s (but restoring them for notational convenience by absorbing factors of $4\pi$), the second term can be written as,
\begin{align}
    &\int \mathcal{D}^3\!P~T_{\ell m}^{(1)}(\bk_1, \bk_2, \bk_3)T_{\ell' m'}^{*(1)}(\bk_1, \bk_2, \bk_3)=\nonumber\\
    &\quad\quad (4\pi)^3\sum_{m_i,\ell_i}\int \mathcal{D}^3\!P ~T_{\ell_1 \ell_2\ell_2; \ell}^{m_1 m_2 m_3; m}T_{\ell_1 \ell_2\ell_3; \ell'}^{*m_1 m_2 m_3; m'},
\end{align}
where we have suppressed the $k$ dependence in $T_{\ell_1 \ell_2 \ell_3; \ell_4}^{m_1 m_2 m_3; m_4}$, defined in Eq.~\eqref{eq:Tmult-final}, and the sum is over $\ell_i$ and $m_i$, with $i = 1, 2, 3$. Without having to expand the terms with spherical harmonics, we can exploit the fact that the Universe is statistically isotropic and instead compute, 
\begin{align}
    \langle\Theta_{\ell m, \rm inh}^{(1)}\Theta_{\ell' m', \rm inh}^{*(1)}\rangle=\frac{\delta_{\ell\ell'}\delta_{mm'}}{2\ell+1}\sum_{m''}\langle\Theta_{\ell m'', \rm inh}^{(1)}\Theta_{\ell m'', \rm inh}^{*(1)}\rangle.
\end{align}
This enables us to use the machinery we derived in Appendix~\ref{app:Q-sums} and write the second term as,
\begin{align}
    \int \mathcal{D}^3\!P~T_{\ell m}^{(1)}(\bk_1, \bk_2,& \bk_3)T_{\ell' m'}^{*(1)}(\bk_1, \bk_2, \bk_3)=\nonumber\\
    (4\pi)^3\frac{\delta_{\ell\ell'}\delta_{mm'}}{2\ell+1}\!\!\sum_{m_i,\ell_i}\!&\int \!\!\mathcal{D}^3\!P\!\left\{\vphantom{\widetilde{Q}^2}A^2_{\ell_1 \ell_2, \ell_3}(k_1, k_2, k_3) (\mathcal{Q}^2)_{\ell_1 \ell_2 \ell_3 \ell}\right.\nonumber\\
    & + 2AB_{\ell_1 \ell_2, \ell_3}(k_1, k_2, k_3)(\mathcal{Q \widetilde{Q}})_{\ell_1 \ell_2, \ell_3 \ell}\nonumber\\
    &\left.+B^2_{\ell_1 \ell_2, \ell_3}(k_1, k_2, k_3)(\mathcal{\widetilde{Q}}^2)_{\ell_1 \ell_2, \ell_3 \ell}\right\}\nonumber.
\end{align}

We apply the same logic to the third term. We then take advantage of the factorized forms of Eqs.~\eqref{eq:Tmult-final}-\eqref{eq:B} to write a computationally manageable final solution as,
\begin{align}\label{eq:inh_auto}
    C_{\ell,\rm inh}^{(11)}=f_{\rm pbh}^2\Bigr[4\mathfrak{A}_\ell+\sum_{\ell_1\ell_2\ell_3}\left(2\mathfrak{B}_{\ell_1\ell_2,\ell_3;\ell}+4\mathfrak{C}_{\ell_1,\ell_2,\ell_3;\ell}\right)\Bigr],
\end{align}
where
\begin{align}
    &\mathfrak{A}_\ell\equiv\!\frac{4\pi}9\int_0^{\eta_0}\!\!\! d \eta\int_0^{\eta_0} \!\!\!d \eta' g(\eta)g(\eta')\mathcal{A}_\ell(\eta,\eta')\beta(\eta)\beta(\eta'),\\  
    &\mathfrak{B}_{\ell_1\ell_2,\ell_3;\ell}\equiv\frac{(4\pi)^3}{2\ell+1}\int d \eta\int d \eta' g(\eta)g(\eta')\nonumber\\
    &\times\left\{\mathcal{A}_{\ell_1}(\eta,\eta')\mathcal{A}_{\ell_2}(\eta,\eta') J_{\ell_3}(\eta,\eta')(\mathcal{Q}^2)_{\ell_1 \ell_2 \ell_3 \ell}\right.\nonumber\\
    &\quad+2\mathcal{K}_{\ell_1}(\eta,\eta')\mathcal{K}_{\ell_2}(\eta,\eta') J_{\ell_3}(\eta,\eta')(\mathcal{Q \widetilde{Q}})_{\ell_1 \ell_2, \ell_3 \ell}\nonumber\\
    &\quad\left.+\mathcal{B}_{\ell_1}(\eta,\eta')\mathcal{B}_{\ell_2}(\eta,\eta') J_{\ell_3}(\eta,\eta')(\mathcal{\widetilde{Q}}^2)_{\ell_1 \ell_2, \ell_3 \ell}\right\},
    \end{align}
    \begin{align}
    &\mathfrak{C}_{\ell_1,\ell_2,\ell_3;\ell}\equiv\frac{(4\pi)^3}{2\ell+1}\int d \eta \int d\eta' g(\eta)g(\eta')\nonumber\\
    &\times\left\{\widetilde{\mathcal{A}}_{\ell_1}(\eta,\eta')\widetilde{\mathcal{A}}_{\ell_2} (\eta',\eta)\mathcal{A}_{\ell_3}(\eta,\eta')(\mathcal{Q}^2)_{\ell_1 \ell_2 \ell_3 \ell}\right.\nonumber\\
    &\quad\!\!\!+2\widetilde{\mathcal{A}}_{\ell_1}(\eta,\eta'){\mathcal{K}}_{\ell_2}(\eta,\eta') \widetilde{\mathcal{B}}_{\ell_3}(\eta',\eta)(\mathcal{Q \widetilde{Q}})_{\ell_3 \ell_2, \ell_1 \ell}\nonumber\\
    &\quad\!\!\!\left.+\widetilde{\mathcal{B}}_{\ell_1}(\eta,\eta')\widetilde{\mathcal{B}}_{\ell_2}(\eta',\eta) \mathcal{B}_{\ell_3}(\eta,\eta')(\mathcal{\widetilde{Q} \widetilde{Q}}^{\rm S})_{\ell_1, \ell_2 \ell_3, \ell} \right\}\!,
\end{align}
with
\begin{align}
    \mathcal{B}_{\ell}(\eta,\eta')&\equiv\!\int\!\! D(k)P_\zeta(k) \frac{j_{\ell}(\chi k)}{\chi k}\Delta_e (\eta, k)\frac{j_{\ell}(\chi' k)}{\chi' k}\Delta_e (\eta', k),\nonumber\\
    \mathcal{K}_{\ell}(\eta,\eta')&\equiv\!\int\!\! D(k) P_\zeta(k)j_{\ell}'(\chi k)\Delta_e (\eta, k)\frac{j_{\ell}(\chi' k)}{\chi' k}\Delta_e (\eta', k),\nonumber\\
    J_{\ell}(\eta,\eta')&\equiv\!\int\!\! D(k) P_\zeta(k)\mathcal{J}_{\ell}(\eta, k)\mathcal{J}_{\ell}(\eta', k),\nonumber\\
    \widetilde{\mathcal{A}}_{\ell}(\eta,\eta')&\equiv\!\int\!\! D(k) P_\zeta(k)j_{\ell}'(\chi k)\Delta_e (\eta, k)\mathcal{J}_{\ell}(\eta', k),\nonumber\\
    \widetilde{\mathcal{B}}_{\ell}(\eta,\eta')&\equiv\!\int\!\! D(k)P_\zeta(k) \frac{j_{\ell}(\chi k)}{\chi k}\Delta_e (\eta, k)\mathcal{J}_{\ell}(\eta', k).
\end{align}

We plot the results, Eq.~\eqref{eq:inh_auto} and Eq.~\eqref{eq:homo_auto}, in Fig.~\ref{fig:auto}. We see there is ultimately an order of magnitude difference between the amplitudes of the inhomogeneous and homogeneous temperature perturbation auto-power-spectrum. We also see that the correlation between our newly computed inhomogeneous temperature perturbation and the standard CMB temperature anisotropy is very poor. Both these facts are likely the culprits behind both the unexpected sensitivity from the forecast on the trispectrum \textit{and} the two orders of magnitude difference in the power spectra amplitudes we observe in Fig.~\ref{fig:AK17_jen}. Additionally, it can be seen that, if it were not for the very small correlation at large $\ell$, the scale suppression due to photon propagation would have a much bigger effect in both the inhomogeneous power spectrum and trispectrum results.

\section{Redshift dependence of the temperature trispectrum induced by accreting PBHs}\label{app:slope}

In this appendix we inspect the redshift dependence of the temperature trispectrum from accreting PBHs, by reproducing the forecast analysis of Sec.~\ref{sec:forecast}, but artificially imposing that the free-electron fraction perturbation vanishes outside of redshift bins of size $\Delta z = 50$. In a given redshift bin, we compute the signal-to-noise, $S/N$, assuming a Planck-like experiment for both the temperature-only trispectrum and power spectrum. That is, for the trispectrum we compute $(S/N)_{\rm tri}\equiv 1/\sigma_{f_{\rm pbh}}$ from Eq.~\eqref{eq:inv_fpbh}. For the power spectrum we compute the similar forecasted quantity,
\begin{align}
    (S/N)_{\rm ps}\equiv \left[\frac{f_{\rm sky}}{2}\sum_\ell (2\ell+1)\left(\frac{C_{\ell}^{(1)}}{C_\ell'}\right)^2\right]^{1/2},
\end{align}
where $C_{\ell}^{(1)}=C_{\ell,\rm hom}^{(1)}+C_{\ell, \rm inh}^{(1)}$ is the total (c.f. Fig.~\ref{fig:AK17_jen}) perturbed $TT$ power spectrum due to accreting PBHs (considering only the ``direct'' term discussed in Sec.~\ref{sec:homo_de}). Note that a rigorous treatment would properly account for correlations between different redshift bins, and involve a principal component analysis. Still, the simple estimation of $S/N$ should give us a reasonable qualitative understanding of the redshift dependence of the signal. 

\begin{figure}[htb]
\includegraphics[trim={0cm 1cm 0.5cm .25cm},width=.95\columnwidth]{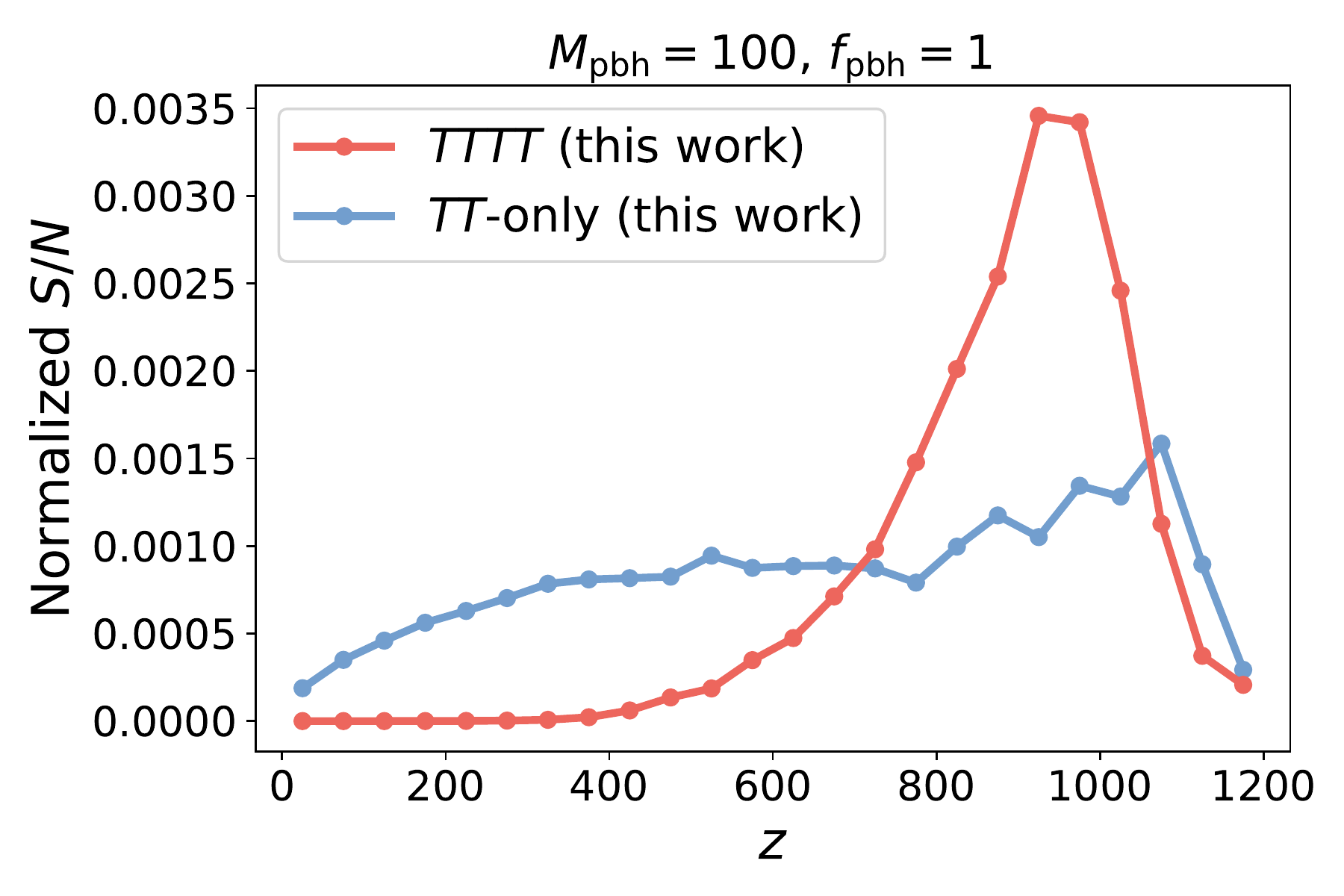}
\caption{\label{fig:SN} Forecasted Planck signal-to-noise ratio for the $TT$-only power spectrum and $TTTT$ trispectrum induced by accreting PBHs. For ease of comparison we normalize the curves such that they integrate to unity over redshift. Each point is computed assuming the perturbed free-electron fraction is only nonzero in redshift bins of size $\Delta z=50$.}
\end{figure}

We compare the two $S/N$ as a function of redshift in Fig.~\ref{fig:SN} for 100-$M_\odot$ PBHs. We see that the temperature trispectrum $S/N$ is rather sharply peaked around $z \sim 900-1000$, in contrast with the temperature power spectrum signal, which receives comparable contributions from a broad range of redshifts $200 \lesssim z \lesssim 1200$. 

This is consistent with the the following observations. First, by inspecting the on-the-spot energy deposition limit discussed in Sec.~\ref{sec:approx}, we found that the trispectrum is negligibly affected by photon propagation that is more suppressive at late times. Namely the strongest spatial fluctuations for the accreting PBHs due to relative velocities occur at a few $10$'s of Mpc scales as shown in Paper~I Fig.~13, but this is not noticeably suppressed until $z\approx 800$ when inspecting Fig.~\ref{fig:G_e}. Second, we find that the trispectrum constraints converge much more quickly than compared to the power spectrum when varying the max multipole on the zeroth-order collision term present in the line-of-sight source. Namely, the trispectrum is unaffected by higher order multipoles of zeroth-order temperature anisotropy which are induced at later times. Thirdly and more subtly, the $f_{\rm pbh}$-$M_{\rm pbh}$ powerlaw dependence is weaker for the trispectrum constraints than it is for the power spectrum constraints. As found in AK17, the luminosity of a spherically accreting PBH is proportional to $M^3$ at all times when excluding their radiative efficiency. The radiative efficiency, however, turns out to have an inverse dependence on black hole mass whose power depends on redshift. This power converges to zero at late times, and implies that the mass dependence on $f_{\rm pbh}$ constraints is weaker if the signal receives support from earlier redshifts. This can be seen directly in Fig.~8 of AK17 where they plot the mean luminosity as a function of redshift for various $M_{\rm pbh}$.

\end{appendix}

\bibliography{main}

\providecommand{\noopsort}[1]{}\providecommand{\singleletter}[1]{#1}%
\begin{thebibliography}{29}%
\makeatletter
\providecommand \@ifxundefined [1]{%
 \@ifx{#1\undefined}
}%
\providecommand \@ifnum [1]{%
 \ifnum #1\expandafter \@firstoftwo
 \else \expandafter \@secondoftwo
 \fi
}%
\providecommand \@ifx [1]{%
 \ifx #1\expandafter \@firstoftwo
 \else \expandafter \@secondoftwo
 \fi
}%
\providecommand \natexlab [1]{#1}%
\providecommand \enquote  [1]{``#1''}%
\providecommand \bibnamefont  [1]{#1}%
\providecommand \bibfnamefont [1]{#1}%
\providecommand \citenamefont [1]{#1}%
\providecommand \href@noop [0]{\@secondoftwo}%
\providecommand \href [0]{\begingroup \@sanitize@url \@href}%
\providecommand \@href[1]{\@@startlink{#1}\@@href}%
\providecommand \@@href[1]{\endgroup#1\@@endlink}%
\providecommand \@sanitize@url [0]{\catcode `\\12\catcode `\$12\catcode
  `\&12\catcode `\#12\catcode `\^12\catcode `\_12\catcode `\%12\relax}%
\providecommand \@@startlink[1]{}%
\providecommand \@@endlink[0]{}%
\providecommand \url  [0]{\begingroup\@sanitize@url \@url }%
\providecommand \@url [1]{\endgroup\@href {#1}{\urlprefix }}%
\providecommand \urlprefix  [0]{URL }%
\providecommand \Eprint [0]{\href }%
\providecommand \doibase [0]{http://dx.doi.org/}%
\providecommand \selectlanguage [0]{\@gobble}%
\providecommand \bibinfo  [0]{\@secondoftwo}%
\providecommand \bibfield  [0]{\@secondoftwo}%
\providecommand \translation [1]{[#1]}%
\providecommand \BibitemOpen [0]{}%
\providecommand \bibitemStop [0]{}%
\providecommand \bibitemNoStop [0]{.\EOS\space}%
\providecommand \EOS [0]{\spacefactor3000\relax}%
\providecommand \BibitemShut  [1]{\csname bibitem#1\endcsname}%
\let\auto@bib@innerbib\@empty
\bibitem [{\citenamefont {{Bean}}\ and\ \citenamefont
  {{Magueijo}}(2002)}]{bean02a}%
  \BibitemOpen
  \bibfield  {author} {\bibinfo {author} {\bibfnamefont {R.}~\bibnamefont
  {{Bean}}}\ and\ \bibinfo {author} {\bibfnamefont {J.}~\bibnamefont
  {{Magueijo}}},\ }\href {\doibase 10.1103/PhysRevD.66.063505} {\bibfield
  {journal} {\bibinfo  {journal} {\prd}\ }\textbf {\bibinfo {volume} {66}},\
  \bibinfo {eid} {063505} (\bibinfo {year} {2002})},\ \Eprint
  {http://arxiv.org/abs/astro-ph/0204486} {arXiv:astro-ph/0204486} \BibitemShut
  {NoStop}%
\bibitem [{\citenamefont {{Bird}}\ \emph {et~al.}(2016)\citenamefont {{Bird}}
  \emph {et~al.}}]{bird16a}%
  \BibitemOpen
  \bibfield  {author} {\bibinfo {author} {\bibfnamefont {S.}~\bibnamefont
  {{Bird}}} \emph {et~al.},\ }\href {\doibase 10.1103/PhysRevLett.116.201301}
  {\bibfield  {journal} {\bibinfo  {journal} {\prl}\ }\textbf {\bibinfo
  {volume} {116}},\ \bibinfo {eid} {201301} (\bibinfo {year} {2016})},\ \Eprint
  {http://arxiv.org/abs/1603.00464} {arXiv:1603.00464} \BibitemShut {NoStop}%
\bibitem [{\citenamefont {{Carr}}\ \emph {et~al.}(2021)\citenamefont {{Carr}},
  \citenamefont {{Kohri}}, \citenamefont {{Sendouda}},\ and\ \citenamefont
  {{Yokoyama}}}]{carr21a}%
  \BibitemOpen
  \bibfield  {author} {\bibinfo {author} {\bibfnamefont {B.}~\bibnamefont
  {{Carr}}}, \bibinfo {author} {\bibfnamefont {K.}~\bibnamefont {{Kohri}}},
  \bibinfo {author} {\bibfnamefont {Y.}~\bibnamefont {{Sendouda}}}, \ and\
  \bibinfo {author} {\bibfnamefont {J.}~\bibnamefont {{Yokoyama}}},\ }\href
  {\doibase 10.1088/1361-6633/ac1e31} {\bibfield  {journal} {\bibinfo
  {journal} {Reports on Progress in Physics}\ }\textbf {\bibinfo {volume}
  {84}},\ \bibinfo {eid} {116902} (\bibinfo {year} {2021})},\ \Eprint
  {http://arxiv.org/abs/2002.12778} {arXiv:2002.12778} \BibitemShut {NoStop}%
\bibitem [{\citenamefont {{Ricotti}}\ \emph {et~al.}(2008)\citenamefont
  {{Ricotti}}, \citenamefont {{Ostriker}},\ and\ \citenamefont
  {{Mack}}}]{ricotti08a}%
  \BibitemOpen
  \bibfield  {author} {\bibinfo {author} {\bibfnamefont {M.}~\bibnamefont
  {{Ricotti}}}, \bibinfo {author} {\bibfnamefont {J.~P.}\ \bibnamefont
  {{Ostriker}}}, \ and\ \bibinfo {author} {\bibfnamefont {K.~J.}\ \bibnamefont
  {{Mack}}},\ }\href {\doibase 10.1086/587831} {\bibfield  {journal} {\bibinfo
  {journal} {\apj}\ }\textbf {\bibinfo {volume} {680}},\ \bibinfo {pages} {829}
  (\bibinfo {year} {2008})},\ \Eprint {http://arxiv.org/abs/0709.0524}
  {arXiv:0709.0524} \BibitemShut {NoStop}%
\bibitem [{\citenamefont {{Ali-Ha{\"i}moud}}\ and\ \citenamefont
  {{Kamionkowski}}(2017)}]{yacine17a}%
  \BibitemOpen
  \bibfield  {author} {\bibinfo {author} {\bibfnamefont {Y.}~\bibnamefont
  {{Ali-Ha{\"i}moud}}}\ and\ \bibinfo {author} {\bibfnamefont {M.}~\bibnamefont
  {{Kamionkowski}}},\ }\href {\doibase 10.1103/PhysRevD.95.043534} {\bibfield
  {journal} {\bibinfo  {journal} {\prd}\ }\textbf {\bibinfo {volume} {95}},\
  \bibinfo {eid} {043534} (\bibinfo {year} {2017})},\ \Eprint
  {http://arxiv.org/abs/1612.05644} {arXiv:1612.05644} \BibitemShut {NoStop}%
\bibitem [{\citenamefont {{Poulin}}\ \emph {et~al.}(2017)\citenamefont
  {{Poulin}}, \citenamefont {{Serpico}}, \citenamefont {{Calore}},
  \citenamefont {{Clesse}},\ and\ \citenamefont {{Kohri}}}]{poulin17a}%
  \BibitemOpen
  \bibfield  {author} {\bibinfo {author} {\bibfnamefont {V.}~\bibnamefont
  {{Poulin}}}, \bibinfo {author} {\bibfnamefont {P.~D.}\ \bibnamefont
  {{Serpico}}}, \bibinfo {author} {\bibfnamefont {F.}~\bibnamefont {{Calore}}},
  \bibinfo {author} {\bibfnamefont {S.}~\bibnamefont {{Clesse}}}, \ and\
  \bibinfo {author} {\bibfnamefont {K.}~\bibnamefont {{Kohri}}},\ }\href
  {\doibase 10.1103/PhysRevD.96.083524} {\bibfield  {journal} {\bibinfo
  {journal} {\prd}\ }\textbf {\bibinfo {volume} {96}},\ \bibinfo {eid} {083524}
  (\bibinfo {year} {2017})},\ \Eprint {http://arxiv.org/abs/1707.04206}
  {arXiv:1707.04206} \BibitemShut {NoStop}%
\bibitem [{\citenamefont {{Senatore}}\ \emph {et~al.}(2009)\citenamefont
  {{Senatore}}, \citenamefont {{Tassev}},\ and\ \citenamefont
  {{Zaldarriaga}}}]{senatore09a}%
  \BibitemOpen
  \bibfield  {author} {\bibinfo {author} {\bibfnamefont {L.}~\bibnamefont
  {{Senatore}}}, \bibinfo {author} {\bibfnamefont {S.}~\bibnamefont
  {{Tassev}}}, \ and\ \bibinfo {author} {\bibfnamefont {M.}~\bibnamefont
  {{Zaldarriaga}}},\ }\href {\doibase 10.1088/1475-7516/2009/09/038} {\bibfield
   {journal} {\bibinfo  {journal} {\jcap}\ }\textbf {\bibinfo {volume}
  {2009}},\ \bibinfo {eid} {038} (\bibinfo {year} {2009})},\ \Eprint
  {http://arxiv.org/abs/0812.3658} {arXiv:0812.3658} \BibitemShut {NoStop}%
\bibitem [{\citenamefont {{Khatri}}\ and\ \citenamefont
  {{Wandelt}}(2009)}]{khatri09a}%
  \BibitemOpen
  \bibfield  {author} {\bibinfo {author} {\bibfnamefont {R.}~\bibnamefont
  {{Khatri}}}\ and\ \bibinfo {author} {\bibfnamefont {B.~D.}\ \bibnamefont
  {{Wandelt}}},\ }\href {\doibase 10.1103/PhysRevD.79.023501} {\bibfield
  {journal} {\bibinfo  {journal} {\prd}\ }\textbf {\bibinfo {volume} {79}},\
  \bibinfo {eid} {023501} (\bibinfo {year} {2009})},\ \Eprint
  {http://arxiv.org/abs/0810.4370} {arXiv:0810.4370} \BibitemShut {NoStop}%
\bibitem [{\citenamefont {{Dvorkin}}\ \emph {et~al.}(2013)\citenamefont
  {{Dvorkin}}, \citenamefont {{Blum}},\ and\ \citenamefont
  {{Zaldarriaga}}}]{dvorkin13a}%
  \BibitemOpen
  \bibfield  {author} {\bibinfo {author} {\bibfnamefont {C.}~\bibnamefont
  {{Dvorkin}}}, \bibinfo {author} {\bibfnamefont {K.}~\bibnamefont {{Blum}}}, \
  and\ \bibinfo {author} {\bibfnamefont {M.}~\bibnamefont {{Zaldarriaga}}},\
  }\href {\doibase 10.1103/PhysRevD.87.103522} {\bibfield  {journal} {\bibinfo
  {journal} {\prd}\ }\textbf {\bibinfo {volume} {87}},\ \bibinfo {eid} {103522}
  (\bibinfo {year} {2013})},\ \Eprint {http://arxiv.org/abs/1302.4753}
  {arXiv:1302.4753} \BibitemShut {NoStop}%
\bibitem [{\citenamefont {{Jensen}}\ and\ \citenamefont
  {{Ali-Ha{\"\i}moud}}(2021)}]{jensen21a}%
  \BibitemOpen
  \bibfield  {author} {\bibinfo {author} {\bibfnamefont {T.~W.}\ \bibnamefont
  {{Jensen}}}\ and\ \bibinfo {author} {\bibfnamefont {Y.}~\bibnamefont
  {{Ali-Ha{\"\i}moud}}},\ }\href {\doibase 10.1103/PhysRevD.104.063534}
  {\bibfield  {journal} {\bibinfo  {journal} {\prd}\ }\textbf {\bibinfo
  {volume} {104}},\ \bibinfo {eid} {063534} (\bibinfo {year} {2021})},\ \Eprint
  {http://arxiv.org/abs/2106.10266} {arXiv:2106.10266} \BibitemShut {NoStop}%
\bibitem [{\citenamefont {{Tseliakhovich}}\ and\ \citenamefont
  {{Hirata}}(2010)}]{Tseliakhovich_10}%
  \BibitemOpen
  \bibfield  {author} {\bibinfo {author} {\bibfnamefont {D.}~\bibnamefont
  {{Tseliakhovich}}}\ and\ \bibinfo {author} {\bibfnamefont {C.}~\bibnamefont
  {{Hirata}}},\ }\href {\doibase 10.1103/PhysRevD.82.083520} {\bibfield
  {journal} {\bibinfo  {journal} {\prd}\ }\textbf {\bibinfo {volume} {82}},\
  \bibinfo {eid} {083520} (\bibinfo {year} {2010})},\ \Eprint
  {http://arxiv.org/abs/1005.2416} {arXiv:1005.2416} \BibitemShut {NoStop}%
\bibitem [{\citenamefont {{Novosyadlyj}}(2006)}]{Novosyadlyj_06}%
  \BibitemOpen
  \bibfield  {author} {\bibinfo {author} {\bibfnamefont {B.}~\bibnamefont
  {{Novosyadlyj}}},\ }\href {\doibase 10.1111/j.1365-2966.2006.10593.x}
  {\bibfield  {journal} {\bibinfo  {journal} {\mnras}\ }\textbf {\bibinfo
  {volume} {370}},\ \bibinfo {pages} {1771} (\bibinfo {year} {2006})},\ \Eprint
  {http://arxiv.org/abs/astro-ph/0603674} {arXiv:astro-ph/0603674} \BibitemShut
  {NoStop}%
\bibitem [{\citenamefont {{Huang}}\ and\ \citenamefont
  {{Vernizzi}}(2013)}]{Huang_13}%
  \BibitemOpen
  \bibfield  {author} {\bibinfo {author} {\bibfnamefont {Z.}~\bibnamefont
  {{Huang}}}\ and\ \bibinfo {author} {\bibfnamefont {F.}~\bibnamefont
  {{Vernizzi}}},\ }\href {\doibase 10.1103/PhysRevLett.110.101303} {\bibfield
  {journal} {\bibinfo  {journal} {\prl}\ }\textbf {\bibinfo {volume} {110}},\
  \bibinfo {eid} {101303} (\bibinfo {year} {2013})},\ \Eprint
  {http://arxiv.org/abs/1212.3573} {arXiv:1212.3573} \BibitemShut {NoStop}%
\bibitem [{\citenamefont {{Planck
  Collaboration}}(2020{\natexlab{a}})}]{planck20c}%
  \BibitemOpen
  \bibfield  {author} {\bibinfo {author} {\bibnamefont {{Planck
  Collaboration}}},\ }\href {\doibase 10.1051/0004-6361/201935891} {\bibfield
  {journal} {\bibinfo  {journal} {\aap}\ }\textbf {\bibinfo {volume} {641}},\
  \bibinfo {eid} {A9} (\bibinfo {year} {2020}{\natexlab{a}})},\ \Eprint
  {http://arxiv.org/abs/1905.05697} {arXiv:1905.05697} \BibitemShut {NoStop}%
\bibitem [{\citenamefont {{Seljak}}\ and\ \citenamefont
  {{Zaldarriaga}}(1996)}]{seljak96a}%
  \BibitemOpen
  \bibfield  {author} {\bibinfo {author} {\bibfnamefont {U.}~\bibnamefont
  {{Seljak}}}\ and\ \bibinfo {author} {\bibfnamefont {M.}~\bibnamefont
  {{Zaldarriaga}}},\ }\href {\doibase 10.1086/177793} {\bibfield  {journal}
  {\bibinfo  {journal} {\apj}\ }\textbf {\bibinfo {volume} {469}},\ \bibinfo
  {pages} {437} (\bibinfo {year} {1996})},\ \Eprint
  {http://arxiv.org/abs/astro-ph/9603033} {arXiv:astro-ph/9603033} \BibitemShut
  {NoStop}%
\bibitem [{\citenamefont {{Khatri}}\ and\ \citenamefont
  {{Wandelt}}(2010)}]{khatri10a}%
  \BibitemOpen
  \bibfield  {author} {\bibinfo {author} {\bibfnamefont {R.}~\bibnamefont
  {{Khatri}}}\ and\ \bibinfo {author} {\bibfnamefont {B.~D.}\ \bibnamefont
  {{Wandelt}}},\ }\href {\doibase 10.1103/PhysRevD.81.103518} {\bibfield
  {journal} {\bibinfo  {journal} {\prd}\ }\textbf {\bibinfo {volume} {81}},\
  \bibinfo {eid} {103518} (\bibinfo {year} {2010})},\ \Eprint
  {http://arxiv.org/abs/0903.0871} {arXiv:0903.0871} \BibitemShut {NoStop}%
\bibitem [{\citenamefont {{Smith}}\ \emph {et~al.}(2015)\citenamefont
  {{Smith}}, \citenamefont {{Senatore}},\ and\ \citenamefont
  {{Zaldarriaga}}}]{smith15a}%
  \BibitemOpen
  \bibfield  {author} {\bibinfo {author} {\bibfnamefont {K.~M.}\ \bibnamefont
  {{Smith}}}, \bibinfo {author} {\bibfnamefont {L.}~\bibnamefont {{Senatore}}},
  \ and\ \bibinfo {author} {\bibfnamefont {M.}~\bibnamefont {{Zaldarriaga}}},\
  }\href@noop {} {\bibfield  {journal} {\bibinfo  {journal} {arXiv e-prints}\ }
  (\bibinfo {year} {2015})},\ \Eprint {http://arxiv.org/abs/1502.00635}
  {arXiv:1502.00635} \BibitemShut {NoStop}%
\bibitem [{\citenamefont {{Lee}}\ and\ \citenamefont
  {{Ali-Ha{\"\i}moud}}(2020)}]{hyrec2}%
  \BibitemOpen
  \bibfield  {author} {\bibinfo {author} {\bibfnamefont {N.}~\bibnamefont
  {{Lee}}}\ and\ \bibinfo {author} {\bibfnamefont {Y.}~\bibnamefont
  {{Ali-Ha{\"\i}moud}}},\ }\href {\doibase 10.1103/PhysRevD.102.083517}
  {\bibfield  {journal} {\bibinfo  {journal} {\prd}\ }\textbf {\bibinfo
  {volume} {102}},\ \bibinfo {eid} {083517} (\bibinfo {year} {2020})},\ \Eprint
  {http://arxiv.org/abs/2007.14114} {arXiv:2007.14114} \BibitemShut {NoStop}%
\bibitem [{\citenamefont {{Ali-Ha{\"\i}moud}}\ and\ \citenamefont
  {{Hirata}}(2011)}]{yacine11a}%
  \BibitemOpen
  \bibfield  {author} {\bibinfo {author} {\bibfnamefont {Y.}~\bibnamefont
  {{Ali-Ha{\"\i}moud}}}\ and\ \bibinfo {author} {\bibfnamefont {C.~M.}\
  \bibnamefont {{Hirata}}},\ }\href {\doibase 10.1103/PhysRevD.83.043513}
  {\bibfield  {journal} {\bibinfo  {journal} {\prd}\ }\textbf {\bibinfo
  {volume} {83}},\ \bibinfo {eid} {043513} (\bibinfo {year} {2011})},\ \Eprint
  {http://arxiv.org/abs/1011.3758} {arXiv:1011.3758} \BibitemShut {NoStop}%
\bibitem [{\citenamefont {{Ali-Ha{\"\i}moud}}\ and\ \citenamefont
  {{Hirata}}(2010)}]{YAH_10}%
  \BibitemOpen
  \bibfield  {author} {\bibinfo {author} {\bibfnamefont {Y.}~\bibnamefont
  {{Ali-Ha{\"\i}moud}}}\ and\ \bibinfo {author} {\bibfnamefont {C.~M.}\
  \bibnamefont {{Hirata}}},\ }\href {\doibase 10.1103/PhysRevD.82.063521}
  {\bibfield  {journal} {\bibinfo  {journal} {\prd}\ }\textbf {\bibinfo
  {volume} {82}},\ \bibinfo {eid} {063521} (\bibinfo {year} {2010})},\ \Eprint
  {http://arxiv.org/abs/1006.1355} {arXiv:1006.1355} \BibitemShut {NoStop}%
\bibitem [{\citenamefont {{Ma}}\ and\ \citenamefont
  {{Bertschinger}}(1995)}]{ma95a}%
  \BibitemOpen
  \bibfield  {author} {\bibinfo {author} {\bibfnamefont {C.-P.}\ \bibnamefont
  {{Ma}}}\ and\ \bibinfo {author} {\bibfnamefont {E.}~\bibnamefont
  {{Bertschinger}}},\ }\href {\doibase 10.1086/176550} {\bibfield  {journal}
  {\bibinfo  {journal} {\apj}\ }\textbf {\bibinfo {volume} {455}},\ \bibinfo
  {pages} {7} (\bibinfo {year} {1995})},\ \Eprint
  {http://arxiv.org/abs/astro-ph/9506072} {arXiv:astro-ph/9506072} \BibitemShut
  {NoStop}%
\bibitem [{\citenamefont {{Blas}}\ \emph {et~al.}(2011)\citenamefont {{Blas}},
  \citenamefont {{Lesgourgues}},\ and\ \citenamefont {{Tram}}}]{CLASS}%
  \BibitemOpen
  \bibfield  {author} {\bibinfo {author} {\bibfnamefont {D.}~\bibnamefont
  {{Blas}}}, \bibinfo {author} {\bibfnamefont {J.}~\bibnamefont
  {{Lesgourgues}}}, \ and\ \bibinfo {author} {\bibfnamefont {T.}~\bibnamefont
  {{Tram}}},\ }\href {\doibase 10.1088/1475-7516/2011/07/034} {\bibfield
  {journal} {\bibinfo  {journal} {\jcap}\ }\textbf {\bibinfo {volume} {2011}},\
  \bibinfo {eid} {034} (\bibinfo {year} {2011})},\ \Eprint
  {http://arxiv.org/abs/1104.2933} {arXiv:1104.2933} \BibitemShut {NoStop}%
\bibitem [{\citenamefont {{Regan}}\ \emph {et~al.}(2010)\citenamefont
  {{Regan}}, \citenamefont {{Shellard}},\ and\ \citenamefont
  {{Fergusson}}}]{regan10a}%
  \BibitemOpen
  \bibfield  {author} {\bibinfo {author} {\bibfnamefont {D.~M.}\ \bibnamefont
  {{Regan}}}, \bibinfo {author} {\bibfnamefont {E.~P.~S.}\ \bibnamefont
  {{Shellard}}}, \ and\ \bibinfo {author} {\bibfnamefont {J.~R.}\ \bibnamefont
  {{Fergusson}}},\ }\href {\doibase 10.1103/PhysRevD.82.023520} {\bibfield
  {journal} {\bibinfo  {journal} {\prd}\ }\textbf {\bibinfo {volume} {82}},\
  \bibinfo {eid} {023520} (\bibinfo {year} {2010})},\ \Eprint
  {http://arxiv.org/abs/1004.2915} {arXiv:1004.2915} \BibitemShut {NoStop}%
\bibitem [{\citenamefont {{Komatsu}}\ \emph {et~al.}(2005)\citenamefont
  {{Komatsu}}, \citenamefont {{Spergel}},\ and\ \citenamefont
  {{Wandelt}}}]{komatsu05a}%
  \BibitemOpen
  \bibfield  {author} {\bibinfo {author} {\bibfnamefont {E.}~\bibnamefont
  {{Komatsu}}}, \bibinfo {author} {\bibfnamefont {D.~N.}\ \bibnamefont
  {{Spergel}}}, \ and\ \bibinfo {author} {\bibfnamefont {B.~D.}\ \bibnamefont
  {{Wandelt}}},\ }\href {\doibase 10.1086/491724} {\bibfield  {journal}
  {\bibinfo  {journal} {\apj}\ }\textbf {\bibinfo {volume} {634}},\ \bibinfo
  {pages} {14} (\bibinfo {year} {2005})},\ \Eprint
  {http://arxiv.org/abs/astro-ph/0305189} {arXiv:astro-ph/0305189} \BibitemShut
  {NoStop}%
\bibitem [{\citenamefont {{Planck
  Collaboration}}(2020{\natexlab{b}})}]{planck20a}%
  \BibitemOpen
  \bibfield  {author} {\bibinfo {author} {\bibnamefont {{Planck
  Collaboration}}},\ }\href {\doibase 10.1051/0004-6361/201936386} {\bibfield
  {journal} {\bibinfo  {journal} {\aap}\ }\textbf {\bibinfo {volume} {641}},\
  \bibinfo {eid} {A5} (\bibinfo {year} {2020}{\natexlab{b}})},\ \Eprint
  {http://arxiv.org/abs/1907.12875} {arXiv:1907.12875} \BibitemShut {NoStop}%
\bibitem [{\citenamefont {{Planck
  Collaboration}}(2020{\natexlab{c}})}]{planck20b}%
  \BibitemOpen
  \bibfield  {author} {\bibinfo {author} {\bibnamefont {{Planck
  Collaboration}}},\ }\href {\doibase 10.1051/0004-6361/201833910} {\bibfield
  {journal} {\bibinfo  {journal} {\aap}\ }\textbf {\bibinfo {volume} {641}},\
  \bibinfo {eid} {A6} (\bibinfo {year} {2020}{\natexlab{c}})},\ \Eprint
  {http://arxiv.org/abs/1807.06209} {arXiv:1807.06209} \BibitemShut {NoStop}%
\bibitem [{\citenamefont {{Meerburg}}\ \emph {et~al.}(2016)\citenamefont
  {{Meerburg}}, \citenamefont {{Meyers}}, \citenamefont {{van Engelen}},\ and\
  \citenamefont {{Ali-Ha{\"\i}moud}}}]{Meerburg_16}%
  \BibitemOpen
  \bibfield  {author} {\bibinfo {author} {\bibfnamefont {P.~D.}\ \bibnamefont
  {{Meerburg}}}, \bibinfo {author} {\bibfnamefont {J.}~\bibnamefont
  {{Meyers}}}, \bibinfo {author} {\bibfnamefont {A.}~\bibnamefont {{van
  Engelen}}}, \ and\ \bibinfo {author} {\bibfnamefont {Y.}~\bibnamefont
  {{Ali-Ha{\"\i}moud}}},\ }\href {\doibase 10.1103/PhysRevD.93.123511}
  {\bibfield  {journal} {\bibinfo  {journal} {\prd}\ }\textbf {\bibinfo
  {volume} {93}},\ \bibinfo {eid} {123511} (\bibinfo {year} {2016})},\ \Eprint
  {http://arxiv.org/abs/1603.02243} {arXiv:1603.02243} \BibitemShut {NoStop}%
\bibitem [{\citenamefont {{Inman}}\ and\ \citenamefont
  {{Ali-Ha{\"\i}moud}}(2019)}]{inman19}%
  \BibitemOpen
  \bibfield  {author} {\bibinfo {author} {\bibfnamefont {D.}~\bibnamefont
  {{Inman}}}\ and\ \bibinfo {author} {\bibfnamefont {Y.}~\bibnamefont
  {{Ali-Ha{\"\i}moud}}},\ }\href {\doibase 10.1103/PhysRevD.100.083528}
  {\bibfield  {journal} {\bibinfo  {journal} {\prd}\ }\textbf {\bibinfo
  {volume} {100}},\ \bibinfo {eid} {083528} (\bibinfo {year} {2019})},\ \Eprint
  {http://arxiv.org/abs/1907.08129} {arXiv:1907.08129} \BibitemShut {NoStop}%
\bibitem [{\citenamefont {{Varshalovich}}\ \emph {et~al.}(1988)\citenamefont
  {{Varshalovich}}, \citenamefont {{Moskalev}},\ and\ \citenamefont
  {{Khersonskii}}}]{Qtheory}%
  \BibitemOpen
  \bibfield  {author} {\bibinfo {author} {\bibfnamefont {D.~A.}\ \bibnamefont
  {{Varshalovich}}}, \bibinfo {author} {\bibfnamefont {A.~N.}\ \bibnamefont
  {{Moskalev}}}, \ and\ \bibinfo {author} {\bibfnamefont {V.~K.}\ \bibnamefont
  {{Khersonskii}}},\ }\href {\doibase 10.1142/0270} {\emph {\bibinfo {title}
  {{Quantum Theory of Angular Momentum}}}}\ (\bibinfo  {publisher} {WORLD
  SCIENTIFIC},\ \bibinfo {year} {1988})\BibitemShut {NoStop}%
\end{thebibliography}%

\end{document}